\documentclass[iop]{emulateapj}

\usepackage{natbib}  
\usepackage{rotating}
\usepackage{placeins}
\usepackage{amsmath}
\usepackage{url}
\newcommand{\msun}{\ensuremath{M_{\odot}}}			
\newcommand{\hh}{\ensuremath{\textrm{H}_{2}}}			
\newcommand{\kms}{\textrm{km~s}\ensuremath{^{-1}}}	
\newcommand{\perkmspc}{\textrm{per~km~s}\ensuremath{^{-1}}\textrm{pc}\ensuremath{^{-1}}}	
\newcommand{\kmspc}{\textrm{km~s}\ensuremath{^{-1}}\textrm{pc}\ensuremath{^{-1}}}	
\newcommand{\perkms}{\textrm{per~km~s}\ensuremath{^{-1}}}	
\newcommand{\percc}{\ensuremath{\textrm{cm}^{-3}}}
\newcommand{\persc}{\ensuremath{\textrm{cm}^{-2}}}

\newcommand{\um}{\ensuremath{\mu m}}    

\newcommand{\formaldehyde}{\ensuremath{\textrm{H}_2\textrm{CO}}}
\newcommand{\ortho}{\ensuremath{\textrm{o-H}_2\textrm{CO}}}
\newcommand{\oneone}{\ensuremath{1_{10}-1_{11}}}
\newcommand{\twotwo}{\ensuremath{2_{11}-2_{12}}}
\newcommand{\threethree}{\ensuremath{3_{12}-3_{13}}}

\newcommand{\thirteenco}{\ensuremath{^{13}\textrm{CO}}}

\def\ee#1{\ensuremath{\times10^{#1}}}
\newcommand{\degree}{\ensuremath{^{\circ}}}

\newcommand{\uchii}{UC\ion{H}{2}}

\def\Figure#1#2#3#4#5{
\begin{figure*}[htp]
\epsscale{#4}
\includegraphics[scale=#4,angle=#5,width=7in]{#1}
\caption{#2}
\label{#3}
\end{figure*}
}

\def\OneColFigure#1#2#3#4#5{
\begin{figure}[htpb]
\epsscale{#4}
\includegraphics[scale=#4,angle=#5,width=3in]{#1}
\caption{#2}
\label{#3}
\end{figure}
}

\def\FigureTwo#1#2#3#4#5{
\begin{figure*}[htpb]
\epsscale{#5}
\plottwo{#1}{#2}
\caption{#3}
\label{#4}
\end{figure*}
}

\def\OneColFigureTwo#1#2#3#4#5{
\begin{figure}[htp]
\epsscale{#5}
\plottwo{#1}{#2}
\caption{#3}
\label{#4}
\end{figure}
}

\def\Table#1#2#3#4#5#6{
\begin{deluxetable*}{#1}
\tablewidth{0pt}
\tabletypesize{\footnotesize}
\tablecaption{#2}
\tablehead{#3}
\startdata
\label{#4}
#5
\enddata
#6
\end{deluxetable*}
}

    \shorttitle{\formaldehyde\ toward \uchii s}
\shortauthors{Ginsburg et al}

\begin{document}

\title{Galactic \formaldehyde\ Densitometry I: Pilot Survey of Ultracompact
\ion{H}{2} Regions and Methodology}

\author{Adam Ginsburg}
\author{Jeremy Darling}
\author{Cara Battersby}
\author{Ben Zeiger}
\author{John Bally}
\email{Adam.Ginsburg@colorado.edu}
\affil{Center for Astrophysics and Space Astronomy, 
Department of Astrophysical and Planetary Sciences,
University of Colorado 389 UCB, Boulder, CO 80309-0389}

\begin{abstract}

  We present a pilot survey of 21 lines of sight towards ultracompact \ion{H}{2}
  (\uchii) regions and three towards continuum-free lines of sight in the
  formaldehyde (\formaldehyde)\ \oneone\ (6 cm) and \twotwo\ (2 cm)
  transitions, using the \formaldehyde\ centimeter lines as a molecular gas densitometer.
  Using Arecibo and Green Bank beam-matched observations, we measure the
  density of 51 detected \formaldehyde\ line pairs and present upper limits on
  density for an additional 24 detected \oneone\ lines.  We analyze the
  systematic uncertainties in the \formaldehyde\ densitometer, achieving \hh\ density measurements
  with accuracies $\sim0.1-0.3$ dex.  The densities measured are not correlated with distance,
  implying that it is possible to make accurate density measurements throughout the galaxy without 
  a distance bias.  We confirm that
  ultracompact HII regions are associated with, and possibly embedded in, gas
  at densities $n(\hh)\gtrsim10^{5}$ \percc.  The densities measured in
  line-of-sight molecular clouds suggest that they consist of low
  volume filling factor ($f\sim10^{-2}$) gas at high ($n(\hh)>10^4$ \percc) density, which is inconsistent
  with purely supersonic turbulence and requires high-density clumping greater than
  typically observed in gravo-turbulent simulations.  We observe complex line
  morphologies that indicate density variations
  with velocity around \uchii\ regions, and we classify a subset of the \uchii\ molecular envelopes
  as collapsing or expanding.  We compare these measurements to Bolocam Galactic Plane
  Survey 1.1 mm observations, and note
  that most \uchii\ regions have 1.1 mm emission consisting of
  significant (5-70\%) free-free emission and are therefore not necessarily
  dominated by optically thin dust emission as is often assumed when computing
  clump masses.  A comparison of our data with the \citet{Mangum2008}
  starburst sample shows that the area filling factor of dense
  ($n(\hh)\sim10^5$ \percc) molecular gas in typical starburst galaxies is
  $\lesssim0.01$, but in extreme starburst galaxies like Arp 220, is $\sim0.1$,
  suggesting that Arp 220 is physically similar to an oversized \uchii\ region.
    

\end{abstract}


\bibliographystyle{apj_w_etal}

\section{Introduction}

Massive stars are known to form preferentially in clustered environments
\citep{DeWit2005}.  They therefore likely form from ``clumps,'' collections of
gas and dust more dense and compact than Giant Molecular Clouds (GMCs) but larger and
more diffuse than typical low-mass protostellar cores.  ``Clumps'' have been
observed with masses ranging from $10-10^6$\msun\ (but more typically
$10^2-10^3\msun$) and with beam-averaged densities in the range $10^3 \lesssim n(\hh) \lesssim
10^5$ \percc\ and sizes $\sim1$ pc \citep[e.g., ][]{Rosolowsky2010,Dunham2010}.
While giant molecular clouds  in the Galaxy have been surveyed
\citep[e.g., ][]{Jackson2006}, the process by which these clouds condense into
clumps and cores and the mechanisms by which they are dispersed are not
understood. 

It is still not known what sets the final mass of massive stars, but it
is thought that they must ignite while still accreting
\citep{mckee2007}.  Hot O and B stars emitting strongly in the ultraviolet will
ionize their surroundings, creating density-bounded \ion{H}{2} regions.  They
progress from hypercompact through ultracompact and compact and finally diffuse
\ion{H}{2} region phases, during which they either dissociate or blow out their
surrounding medium \citep{Churchwell2002,keto2007}.  The brightest sources in the
Galactic plane in both the free-free continuum in the cm-wavelength regime and
the dust continum in the sub-mm to mm-wavelength regime generally host \uchii\
regions.

While the gas within \uchii\ regions is hot and ionized, the surrounding gas is
initially molecular.  At the interface between the molecular cloud and
the ionization front, a photon-dominated or photodissociation region appears
\citep{Roshi2005}.  \citet{Churchwell2010} observed HCO$^+$ towards a
sample of \uchii\ regions and noted both infall and outflow motions in
molecular tracers towards these objects.  It should be possible to determine
whether the \uchii\ regions still have collapsing envelopes (infall signatures)
or only disks (outflow signatures) and thereby determine relative evolutionary
states of the regions.

Two centimeter transitions of formaldehyde, \ortho\ \oneone\ (6 cm) and
\twotwo\ (2 cm)\footnote{All references to \formaldehyde\ in this paper,
except where otherwise noted, are to the ortho \ortho\ population, as no para p-\formaldehyde\ 
lines were observed}, have been used to measure the density of molecular
clouds in massive-star-forming regions \citep[e.g., ][]{Dickel1986,Dickel1987},
high-latitude Galactic clouds \citep[e.g., ][]{Turner1989}, the Galactic Center
\citep[e.g., ][]{Zylka1992} starburst galaxies \citep[e.g., ][]{Mangum2008}, and molecular clouds
in a gravitational lens \citep[e.g., ][]{Zeiger2010}.  Studies similar to our own have
been performed by \citet{Wadiak1988} and \citet{Henkel1983}, in which bright
continuum sources were observed in the same transitions with
(approximately) beam-matched telescopes at $\sim$2\arcmin\ resolution. Our
study delves deeper into the spectral line profiles and systematic uncertainties of
\formaldehyde\ densitometry and is performed at higher spatial resolution than
past work.

This paper presents a pilot study as a proof-of-concept for a much larger
ongoing survey\footnote{GBT project code GBT10B-019} towards 400 lines of sight
and the methodology applicable to the larger survey.

In section \ref{sec:observations} we present the new observations and describe
other data sets used in our analysis.  Section \ref{sec:models} describes the
modeling procedure used to derive density from the \formaldehyde\ line
observations.  Section \ref{sec:analysis} presents detailed discussion of the
modeling and derivation of physical parameters and their uncertainties.
Section \ref{sec:results} describes the derived and measured values. 
Section \ref{sec:discussion} discusses the larger implications of our results.
We conclude with a brief summary of important results.

\section{Observations and Data}
\label{sec:observations}
\subsection{Source Selection} 
The observed lines-of-sight included 21 sources selected from the \citet{Araya2002} \uchii\
sample and 3 from the \citet{Araya2004} ``massive-star forming candidate''
sample.  The sources were selected primarily on the basis of having been
previously observed with Arecibo\footnote{The Arecibo Observatory is part of
the National Astronomy and Ionosphere Center, which is operated by Cornell
University under a cooperative agreement with the National Science Foundation.
} in the \oneone\ transition of \formaldehyde\
with the intent of demonstrating the densitometry method within the Galaxy
rather than making systematic observations of any source class.  Nonetheless,
the \citet{Araya2002} sample includes the majority of the bright \uchii\
regions accessible to Arecibo.  Additionally, there are many detected GMCs
along the line of sight to these \uchii\ regions.

The \citet{Araya2004} observations included 15 pointings towards infrared dark cloud
(IRDC) candidates and High-Mass Protostellar Object (HMPO) candidates.  The
sources we selected from this sample include two sources classified as IRDCs
based on MSX data and one HMPO candidate.  The selection of these sources was
arbitrary; we were only able to observe 24 lines-of-sight in our 4 hour
observation block.  The remaining sources will be discussed in a later paper.
The observed lines of sight are listed in Table \ref{tab:h2comeasured_a}.

\LongTables
\Table{lcccccccc}{Measured \formaldehyde\ \oneone\ line properties}
{\colhead{Source Name\tablenotemark{a}}&\colhead{$l$}&\colhead{$b$}&\colhead{6cm Continuum}&\colhead{Peak}&\colhead{Center}&\colhead{FWHM }&\colhead{RMS}&\colhead{Channel Width}\\
\colhead{           }&\colhead{\degree}&\colhead{\degree}&\colhead{(Jy)}&\colhead{(Jy)}&\colhead{(\kms)}&\colhead{(\kms)}&\colhead{(Jy)}&\colhead{(\kms)}\\ }
{tab:h2comeasured_a}{
       G32.80+0.19 0&              0.1904&             32.7968&         2.18 (0.01)&      -0.393 (0.008)&        15.39 (0.05)&         6.57 (0.06)&              0.0049&              1.1374\\
       G32.80+0.19 1&              0.1904&             32.7968&         2.18 (0.01)&      -0.092 (0.008)&        11.45 (0.26)&        10.25 (0.65)&              0.0049&              1.1374\\
       G32.80+0.19 2&              0.1904&             32.7968&         2.18 (0.01)&      -0.063 (0.008)&        80.63 (0.13)&         2.49 (0.36)&              0.0049&              1.1374\\
       G32.80+0.19 3&              0.1904&             32.7968&         2.18 (0.01)&      -0.254 (0.008)&        84.61 (0.02)&         1.37 (0.06)&              0.0049&              1.1374\\
       G32.80+0.19 4&              0.1904&             32.7968&         2.18 (0.01)&      -0.090 (0.008)&        88.66 (0.09)&         3.21 (0.31)&              0.0049&              1.1374\\
       G33.13-0.09 0&             -0.0949&             33.1297&         0.49 (0.00)&      -0.192 (0.007)&        75.92 (0.05)&         3.80 (0.12)&              0.0045&              1.1374\\
       G33.13-0.09 1&             -0.0949&             33.1297&         0.49 (0.00)&      -0.023 (0.007)&        81.62 (0.35)&         2.49 (0.88)&              0.0045&              1.1374\\
       G33.13-0.09 2&             -0.0949&             33.1297&         0.49 (0.00)&      -0.040 (0.007)&       101.50 (0.40)&        11.30 (0.80)&              0.0045&              1.1374\\
       G33.13-0.09 3&             -0.0949&             33.1297&         0.49 (0.00)&      -0.039 (0.007)&        10.39 (0.08)&         2.04 (0.24)&              0.0045&              1.1374\\
       G33.92+0.11 0&              0.1112&              33.914&         0.83 (0.00)&      -0.081 (0.008)&       107.28 (0.18)&         6.62 (0.34)&               0.005&              1.1374\\
       G33.92+0.11 1&              0.1112&              33.914&         0.83 (0.00)&      -0.079 (0.008)&       106.03 (0.06)&         2.41 (0.23)&               0.005&              1.1374\\
       G33.92+0.11 2&              0.1112&              33.914&         0.83 (0.00)&      -0.160 (0.030)&        57.30 (0.40)&        10.60 (0.80)&               0.005&              1.1374\\
       G34.26+0.15 0&              0.1538&             34.2572&         5.57 (0.01)&      -1.828 (0.015)&        60.24 (0.01)&         3.80 (0.03)&              0.0063&              1.1374\\
       G34.26+0.15 1&              0.1538&             34.2572&         5.57 (0.01)&      -0.160 (0.015)&        26.69 (0.08)&         1.04 (0.22)&              0.0063&              1.1374\\
       G34.26+0.15 2&              0.1538&             34.2572&         5.57 (0.01)&      -0.099 (0.015)&        11.25 (0.19)&         2.01 (0.40)&              0.0063&              1.1374\\
       G34.26+0.15 3&              0.1538&             34.2572&         5.57 (0.01)&      -0.126 (0.015)&        51.70 (2.00)&         4.20 (1.00)&              0.0063&              1.1374\\
       G34.26+0.15 4&              0.1538&             34.2572&         5.57 (0.01)&      -0.047 (0.015)&        48.20 (2.00)&         1.80 (1.00)&              0.0063&              1.1374\\
       G35.20-1.74 0&             -1.7409&             35.1997&         5.17 (0.00)&      -1.018 (0.008)&        43.37 (0.01)&         3.67 (0.02)&              0.0051&              1.1374\\
       G35.20-1.74 1&             -1.7409&             35.1997&         5.17 (0.00)&      -0.147 (0.008)&        36.67 (0.10)&         1.49 (0.27)&              0.0051&              1.1374\\
       G35.20-1.74 2&             -1.7409&             35.1997&         5.17 (0.00)&      -0.324 (0.008)&        14.08 (0.01)&         0.93 (0.03)&              0.0051&              1.1374\\
       G35.20-1.74 3&             -1.7409&             35.1997&         5.17 (0.00)&      -0.039 (0.008)&        50.59 (0.53)&         4.92 (1.31)&              0.0051&              1.1374\\
       G35.57-0.03 0&             -0.0306&             35.5779&         0.47 (0.00)&      -0.064 (0.009)&        52.10 (0.10)&         4.60 (0.30)&              0.0053&              1.1374\\
       G35.57-0.03 1&             -0.0306&             35.5779&         0.47 (0.00)&      -0.021 (0.009)&        45.60 (0.30)&         1.90 (0.60)&              0.0053&              1.1374\\
       G35.57-0.03 2&             -0.0306&             35.5779&         0.47 (0.00)&      -0.019 (0.009)&        57.60 (0.50)&         2.90 (0.97)&              0.0053&              1.1374\\
       G35.57-0.03 3&             -0.0306&             35.5779&         0.47 (0.00)&      -0.031 (0.009)&        12.80 (0.20)&         1.84 (0.41)&              0.0053&              1.1374\\
       G35.57-0.03 4&             -0.0306&             35.5779&         0.47 (0.00)&      -0.031 (0.008)&        29.04 (0.11)&         0.82 (0.25)&              0.0053&              1.1374\\
       G35.58+0.07 0&              0.0657&             35.5801&         0.53 (0.01)&      -0.146 (0.004)&        49.37 (0.21)&         5.33 (0.34)&              0.0048&              1.1374\\
       G35.58+0.07 1&              0.0657&             35.5801&         0.53 (0.01)&      -0.049 (0.013)&        53.13 (0.25)&         2.98 (0.64)&              0.0048&              1.1374\\
       G35.58+0.07 2&              0.0657&             35.5801&         0.53 (0.01)&      -0.025 (0.004)&        58.12 (0.29)&         3.63 (0.74)&              0.0048&              1.1374\\
       G35.58+0.07 3&              0.0657&             35.5801&         0.53 (0.01)&      -0.034 (0.004)&        13.24 (0.17)&         2.80 (0.39)&              0.0048&              1.1374\\
       G37.87-0.40 0&             -0.3993&              37.873&         4.40 (0.01)&      -0.531 (0.006)&        60.23 (0.11)&         8.73 (0.35)&              0.0069&              1.1374\\
       G37.87-0.40 1&             -0.3993&              37.873&         4.40 (0.01)&      -0.124 (0.014)&        53.27 (0.19)&         4.03 (0.46)&              0.0069&              1.1374\\
       G37.87-0.40 2&             -0.3993&              37.873&         4.40 (0.01)&      -0.356 (0.019)&        65.13 (0.04)&         2.74 (0.15)&              0.0069&              1.1374\\
       G37.87-0.40 3&             -0.3993&              37.873&         4.40 (0.01)&      -0.324 (0.045)&        72.18 (0.04)&         1.35 (0.14)&              0.0069&              1.1374\\
       G37.87-0.40 4&             -0.3993&              37.873&         4.40 (0.01)&      -0.424 (0.013)&        73.97 (0.13)&         3.01 (0.22)&              0.0069&              1.1374\\
       G37.87-0.40 5&             -0.3993&              37.873&         4.40 (0.01)&      -0.185 (0.012)&        79.98 (0.06)&         1.80 (0.14)&              0.0069&              1.1374\\
       G37.87-0.40 6&             -0.3993&              37.873&         4.40 (0.01)&      -0.114 (0.015)&        91.96 (0.08)&         1.21 (0.18)&              0.0069&              1.1374\\
       G37.87-0.40 7&             -0.3993&              37.873&         4.40 (0.01)&      -0.175 (0.012)&        14.32 (0.14)&         2.94 (0.20)&              0.0069&              1.1374\\
       G37.87-0.40 8&             -0.3993&              37.873&         4.40 (0.01)&      -0.072 (0.022)&        13.16 (0.10)&         0.87 (0.32)&              0.0069&              1.1374\\
       G37.87-0.40 9&             -0.3993&              37.873&         4.40 (0.01)&      -0.137 (0.012)&        20.54 (0.06)&         1.37 (0.14)&              0.0069&              1.1374\\
       G43.89-0.78 0&             -0.7838&             43.8892&         0.66 (0.00)&      -0.181 (0.004)&        54.86 (0.02)&         2.19 (0.06)&              0.0032&              1.1374\\
       G43.89-0.78 1&             -0.7838&             43.8892&         0.66 (0.00)&      -0.020 (0.002)&        50.55 (0.59)&        15.90 (1.20)&              0.0032&              1.1374\\
       G45.07+0.13 0&              0.1323&             45.0711&         0.47 (0.00)&      -0.056 (0.006)&        57.49 (0.10)&         4.24 (0.23)&              0.0035&              1.1374\\
       G45.07+0.13 1&              0.1323&             45.0711&         0.47 (0.00)&      -0.036 (0.006)&        65.44 (0.15)&         4.09 (0.34)&              0.0035&              1.1374\\
       G45.12+0.13 0&              0.1326&             45.1223&         4.28 (0.01)&      -0.188 (0.006)&        55.70 (0.12)&         3.32 (0.24)&              0.0065&              1.1374\\
       G45.12+0.13 1&              0.1326&             45.1223&         4.28 (0.01)&      -0.154 (0.009)&        59.40 (0.13)&         3.11 (0.33)&              0.0065&              1.1374\\
       G45.12+0.13 2&              0.1326&             45.1223&         4.28 (0.01)&      -0.200 (0.010)&        24.86 (0.03)&         1.68 (0.08)&              0.0065&              1.1374\\
       G45.12+0.13 3&              0.1326&             45.1223&         4.28 (0.01)&      -0.027 (0.004)&        65.53 (0.82)&         7.23 (2.03)&              0.0065&              1.1374\\
       G45.45+0.06 0&              0.0593&             45.4548&         4.77 (0.01)&      -1.347 (0.018)&        59.58 (0.02)&         3.18 (0.05)&              0.0063&              1.1374\\
       G45.45+0.06 1&              0.0593&             45.4548&         4.77 (0.01)&      -0.123 (0.040)&        55.34 (0.38)&         3.15 (0.38)&              0.0063&              1.1374\\
       G45.45+0.06 2&              0.0593&             45.4548&         4.77 (0.01)&      -0.056 (0.005)&        25.02 (0.12)&         2.82 (0.28)&              0.0063&              1.1374\\
       G45.47+0.05 0&              0.0455&             45.4655&         0.75 (0.00)&      -0.274 (0.003)&        60.62 (0.03)&         6.59 (0.07)&              0.0039&              1.1374\\
       G45.47+0.05 1&              0.0455&             45.4655&         0.75 (0.00)&      -0.017 (0.004)&        25.55 (0.23)&         2.18 (0.55)&              0.0039&              1.1374\\
       G48.61+0.02 0&              0.0229&             48.6055&         1.01 (0.00)&      -0.067 (0.003)&        18.08 (0.09)&         4.97 (0.22)&              0.0035&              1.1374\\
       G48.61+0.02 1&              0.0229&             48.6055&         1.01 (0.00)&      -0.024 (0.005)&         6.08 (0.13)&         1.20 (0.31)&              0.0035&              1.1374\\
       G48.61+0.02 2&              0.0229&             48.6055&         1.01 (0.00)&      -0.018 (0.003)&        53.73 (0.33)&         4.72 (0.79)&              0.0035&              1.1374\\
       G50.32+0.68 0&              0.6761&             50.3153&         0.24 (0.00)&      -0.011 (0.003)&        26.28 (0.40)&         3.32 (0.94)&              0.0031&              1.1374\\
       G60.88-0.13 0&             -0.1285&             60.8826&         0.66 (0.01)&      -0.093 (0.009)&        22.60 (0.15)&         3.24 (0.35)&              0.0096&              1.1374\\
       G61.48+0.09 0&              0.0893&             61.4769&         6.16 (0.01)&      -0.531 (0.009)&        21.45 (0.02)&         2.81 (0.06)&              0.0084&              1.1374\\
       G69.54-0.98 0&             -0.9759&             69.5398&         0.28 (0.01)&      -0.280 (0.006)&        10.65 (0.05)&         4.55 (0.11)&              0.0076&              1.1374\\
       G70.29+1.60 0&              1.6006&             70.2927&         4.37 (0.13)&      -0.372 (0.008)&       -21.74 (0.07)&         3.92 (0.15)&              0.0108&              1.1374\\
       G70.29+1.60 1&              1.6006&             70.2927&         4.37 (0.13)&      -0.050 (0.007)&       -27.17 (0.58)&         4.86 (1.33)&              0.0108&              1.1374\\
       G70.33+1.59 0&               1.589&             70.3296&         2.21 (0.01)&      -1.201 (0.007)&       -21.24 (0.01)&         3.65 (0.03)&              0.0115&              1.1374\\
   IRAS 20051+3435 0&              0.2088&             32.4662&         0.00 (0.01)&      -0.019 (0.001)&        10.77 (0.07)&         3.60 (0.18)&             0.00071&              2.2747\\
       G41.74+0.10 0&              0.0975&             41.7415&         0.34 (0.00)&      -0.062 (0.004)&        14.60 (0.09)&         2.56 (0.26)&              0.0033&              1.1374\\
       G41.74+0.10 1&              0.0975&             41.7415&         0.34 (0.00)&      -0.020 (0.004)&        10.99 (0.29)&         2.52 (0.71)&              0.0033&              1.1374\\
       G41.74+0.10 2&              0.0975&             41.7415&         0.34 (0.00)&      -0.066 (0.004)&        34.25 (0.05)&         1.63 (0.13)&              0.0033&              1.1374\\
       G41.74+0.10 3&              0.0975&             41.7415&         0.34 (0.00)&      -0.022 (0.005)&        56.61 (0.13)&         1.15 (0.32)&              0.0033&              1.1374\\
       G41.74+0.10 4&              0.0975&             41.7415&         0.34 (0.00)&      -0.043 (0.005)&        17.57 (0.07)&         1.13 (0.18)&              0.0033&              1.1374\\
      IRDC 1923+13 0&             -0.4972&             48.9325&         0.40 (0.00)&      -0.011 (0.001)&        50.20 (0.08)&         1.83 (0.19)&              0.0008&              0.7582\\
      IRDC 1923+13 1&             -0.4972&             48.9325&         0.40 (0.00)&      -0.009 (0.001)&        57.56 (0.09)&         2.57 (0.22)&              0.0008&              0.7582\\
      IRDC 1923+13 2&             -0.4972&             48.9325&         0.40 (0.00)&      -0.005 (0.001)&        47.32 (0.20)&         2.11 (0.51)&              0.0008&              0.7582\\
      IRDC 1916+11 0&             -0.2923&              45.666&         0.00 (0.01)&      -0.005 (0.001)&        25.94 (0.17)&         2.53 (0.41)&             0.00083&              0.7582\\
      IRDC 1916+11 1&             -0.2923&              45.666&         0.00 (0.01)&      -0.013 (0.001)&        55.91 (0.13)&         6.21 (0.34)&             0.00083&              0.7582\\
      IRDC 1916+11 2&             -0.2923&              45.666&         0.00 (0.01)&      -0.003 (0.001)&        48.85 (0.48)&         3.58 (1.13)&             0.00083&              0.7582\\
}{
\tablenotetext{a}{Sources are labeled by the line-of-sight followed by the
number of the component identified, indexed from zero.  The components do not
follow a particular order, but are uniquely identifiable by their velocity,
width, and amplitude.}}
\Table{lccccc}{Measured \formaldehyde\ \twotwo\ line properties}
{\colhead{Source Name}&\colhead{2cm Continuum}&\colhead{Peak\tablenotemark{a}}&\colhead{Center}&\colhead{FWHM }&\colhead{RMS\tablenotemark{b}}\\
\colhead{           }&\colhead{(Jy)}&\colhead{(Jy)}&\colhead{(\kms)}&\colhead{(\kms)}&\colhead{(Jy)}\\ }
{tab:h2comeasured_b}{
       G32.80+0.19 0&         3.68 (0.02)&      -0.519 (0.032)&        15.65 (0.03)&         5.72 (0.08)&              0.0038\\
       G32.80+0.19 1&         3.68 (0.02)&      -0.076 (0.019)&        11.90 (1.18)&         8.17 (0.98)&              0.0038\\
       G32.80+0.19 2&         3.68 (0.02)&      -0.016 (0.001)&        80.47 (0.14)&         4.35 (0.36)&              0.0038\\
       G32.80+0.19 3&         3.68 (0.02)&      -0.065 (0.002)&        84.96 (0.02)&         1.29 (0.05)&              0.0038\\
       G32.80+0.19 4&         3.68 (0.02)&      -0.026 (0.001)&        88.83 (0.06)&         2.31 (0.14)&              0.0038\\
       G33.13-0.09 0&         0.47 (0.02)&      -0.224 (0.003)&        76.17 (0.02)&         3.31 (0.05)&               0.003\\
       G33.13-0.09 1&         0.47 (0.02)&       0.000 (0.000)&         0.00 (0.00)&         0.00 (0.00)&               0.003\\
       G33.13-0.09 2&         0.47 (0.02)&       0.000 (0.000)&         0.00 (0.00)&         0.00 (0.00)&               0.003\\
       G33.13-0.09 3&         0.47 (0.02)&       0.000 (0.000)&         0.00 (0.00)&         0.00 (0.00)&               0.003\\
       G33.92+0.11 0&         0.87 (0.02)&      -0.086 (0.003)&       106.43 (0.03)&         2.17 (0.09)&              0.0032\\
       G33.92+0.11 1&         0.87 (0.02)&      -0.069 (0.002)&       108.83 (0.11)&         6.82 (0.16)&              0.0032\\
       G33.92+0.11 2&         0.87 (0.02)&       0.000 (0.000)&         0.00 (0.00)&         0.00 (0.00)&              0.0032\\
       G34.26+0.15 0&         5.89 (0.02)&      -1.356 (0.006)&        60.99 (0.01)&         3.96 (0.02)&              0.0051\\
       G34.26+0.15 1&         5.89 (0.02)&      -0.046 (0.003)&        27.11 (0.04)&         1.03 (0.09)&              0.0051\\
       G34.26+0.15 2&         5.89 (0.02)&      -0.018 (0.002)&        11.23 (0.16)&         3.19 (0.38)&              0.0051\\
       G34.26+0.15 3&         5.89 (0.02)&      -0.025 (0.004)&        52.82 (0.58)&         6.34 (1.53)&              0.0051\\
       G34.26+0.15 4&         5.89 (0.02)&      -0.018 (0.007)&        47.05 (0.47)&         2.47 (1.15)&              0.0051\\
       G35.20-1.74 0&         5.98 (0.03)&      -0.482 (0.004)&        43.38 (0.02)&         3.71 (0.04)&              0.0055\\
       G35.20-1.74 1&         5.98 (0.03)&      -0.028 (0.005)&        37.91 (0.32)&         3.46 (0.76)&              0.0055\\
       G35.20-1.74 2&         5.98 (0.03)&      -0.056 (0.003)&        14.18 (0.02)&         1.00 (0.05)&              0.0055\\
       G35.20-1.74 3&         5.98 (0.03)&       0.000 (0.000)&         0.00 (0.00)&         0.00 (0.00)&              0.0055\\
       G35.57-0.03 0&         0.32 (0.15)&      -0.075 (0.003)&        52.14 (0.09)&         4.39 (0.21)&              0.0046\\
       G35.57-0.03 1&         0.32 (0.15)&      -0.015 (0.006)&        47.39 (0.25)&         1.31 (0.60)&              0.0046\\
       G35.57-0.03 2&         0.32 (0.15)&       0.000 (0.000)&         0.00 (0.00)&         0.00 (0.00)&              0.0046\\
       G35.57-0.03 3&         0.32 (0.15)&       0.000 (0.000)&         0.00 (0.00)&         0.00 (0.00)&              0.0046\\
       G35.57-0.03 4&         0.32 (0.15)&      -0.024 (0.008)&        29.25 (0.11)&         0.43 (0.15)&              0.0046\\
       G35.58+0.07 0&         0.23 (0.09)&      -0.106 (0.002)&        49.21 (0.06)&         5.00 (0.14)&              0.0031\\
       G35.58+0.07 1&         0.23 (0.09)&       0.000 (0.004)&         0.00 (0.00)&         0.00 (0.00)&              0.0031\\
       G35.58+0.07 2&         0.23 (0.09)&       0.000 (0.004)&         0.00 (0.00)&         0.00 (0.00)&              0.0031\\
       G35.58+0.07 3&         0.23 (0.09)&       0.000 (0.004)&         0.00 (0.00)&         0.00 (0.00)&              0.0031\\
       G37.87-0.40 0&         3.73 (0.02)&      -0.221 (0.003)&        59.99 (0.12)&         8.53 (0.14)&              0.0048\\
       G37.87-0.40 1&         3.73 (0.02)&      -0.045 (0.007)&        54.55 (0.25)&         5.99 (0.34)&              0.0048\\
       G37.87-0.40 2&         3.73 (0.02)&      -0.036 (0.007)&        65.06 (0.11)&         2.57 (0.45)&              0.0048\\
       G37.87-0.40 3&         3.73 (0.02)&      -0.053 (0.003)&        72.44 (0.05)&         1.37 (0.08)&              0.0048\\
       G37.87-0.40 4&         3.73 (0.02)&      -0.047 (0.002)&        74.25 (0.07)&         2.07 (0.18)&              0.0048\\
       G37.87-0.40 5&         3.73 (0.02)&      -0.016 (0.001)&        80.04 (0.03)&         1.28 (0.07)&              0.0048\\
       G37.87-0.40 6&         3.73 (0.02)&      -0.010 (0.002)&        91.99 (0.12)&         1.60 (0.28)&              0.0048\\
       G37.87-0.40 7&         3.73 (0.02)&      -0.026 (0.002)&        14.89 (0.12)&         1.40 (0.20)&              0.0048\\
       G37.87-0.40 8&         3.73 (0.02)&      -0.017 (0.002)&        13.29 (0.19)&         1.52 (0.34)&              0.0048\\
       G37.87-0.40 9&         3.73 (0.02)&      -0.017 (0.001)&        20.52 (0.10)&         3.09 (0.23)&              0.0048\\
       G43.89-0.78 0&         0.53 (0.02)&      -0.059 (0.004)&        54.61 (0.08)&         2.85 (0.23)&               0.003\\
       G43.89-0.78 1&         0.53 (0.02)&      -0.015 (0.002)&        49.59 (0.94)&        14.49 (1.69)&               0.003\\
       G45.07+0.13 0&         0.79 (0.07)&      -0.073 (0.003)&        57.18 (0.08)&         3.45 (0.18)&              0.0029\\
       G45.07+0.13 1&         0.79 (0.07)&      -0.011 (0.003)&        65.67 (0.42)&         3.46 (0.98)&              0.0029\\
       G45.12+0.13 0&         5.20 (0.20)&      -0.086 (0.002)&        56.21 (0.11)&         5.22 (0.21)&              0.0044\\
       G45.12+0.13 1&         5.20 (0.20)&      -0.059 (0.005)&        59.70 (0.06)&         2.42 (0.16)&              0.0044\\
       G45.12+0.13 2&         5.20 (0.20)&      -0.047 (0.002)&        25.14 (0.04)&         1.55 (0.09)&              0.0044\\
       G45.12+0.13 3&         5.20 (0.20)&      -0.021 (0.001)&        64.68 (0.39)&         8.15 (0.87)&              0.0044\\
       G45.45+0.06 0&         3.16 (0.02)&      -0.260 (0.003)&        59.58 (0.01)&         2.06 (0.03)&              0.0043\\
       G45.45+0.06 1&         3.16 (0.02)&      -0.042 (0.002)&        57.90 (0.14)&         9.40 (0.31)&              0.0043\\
       G45.45+0.06 2&         3.16 (0.02)&       0.000 (0.000)&         0.00 (0.00)&         0.00 (0.00)&              0.0043\\
       G45.47+0.05 0&         0.38 (0.02)&      -0.124 (0.003)&        61.67 (0.07)&         5.85 (0.17)&              0.0049\\
       G45.47+0.05 1&         0.38 (0.02)&      -0.000 (0.007)&         0.00 (0.00)&         0.00 (0.00)&              0.0049\\
       G48.61+0.02 0&         0.41 (0.02)&      -0.022 (0.003)&        18.50 (0.25)&         4.39 (0.59)&              0.0033\\
       G48.61+0.02 1&         0.41 (0.02)&      -0.000 (0.000)&         0.00 (0.00)&         0.00 (0.00)&              0.0033\\
       G48.61+0.02 2&         0.41 (0.02)&      -0.005 (0.002)&        52.50 (1.25)&         7.47 (2.94)&              0.0033\\
       G50.32+0.68 0&         0.16 (0.02)&      -0.011 (0.003)&        26.21 (0.44)&         3.10 (1.03)&              0.0036\\
       G60.88-0.13 0&         0.29 (0.02)&      -0.016 (0.003)&        21.63 (0.21)&         2.47 (0.50)&               0.003\\
       G61.48+0.09 0&         3.42 (0.02)&      -0.300 (0.004)&        21.40 (0.02)&         2.39 (0.04)&              0.0037\\
       G69.54-0.98 0&         0.23 (0.02)&      -0.220 (0.002)&         9.97 (0.03)&         5.81 (0.08)&              0.0031\\
       G70.29+1.60 0&         6.21 (0.02)&      -0.159 (0.003)&       -23.52 (0.06)&         5.36 (0.13)&              0.0046\\
       G70.29+1.60 1&         6.21 (0.02)&      -0.000 (0.000)&        -0.00 (0.00)&         0.00 (0.00)&              0.0046\\
       G70.33+1.59 0&         2.68 (0.02)&      -1.081 (0.005)&       -21.17 (0.01)&         2.95 (0.01)&              0.0038\\
   IRAS 20051+3435 0&         0.00 (0.02)&      -0.016 (0.003)&        11.51 (0.37)&         4.14 (0.88)&              0.0032\\
       G41.74+0.10 0&         0.28 (0.02)&      -0.014 (0.002)&        14.36 (0.34)&         3.80 (0.80)&              0.0032\\
       G41.74+0.10 1&         0.28 (0.02)&       0.000 (0.004)&         0.00 (0.00)&         0.00 (0.00)&              0.0032\\
       G41.74+0.10 2&         0.28 (0.02)&       0.000 (0.004)&         0.00 (0.00)&         0.00 (0.00)&              0.0032\\
       G41.74+0.10 3&         0.28 (0.02)&       0.000 (0.004)&         0.00 (0.00)&         0.00 (0.00)&              0.0032\\
       G41.74+0.10 4&         0.28 (0.02)&       0.000 (0.004)&         0.00 (0.00)&         0.00 (0.00)&              0.0032\\
      IRDC 1923+13 0&         0.00 (0.02)&       0.000 (0.000)&         0.00 (0.00)&         0.00 (0.00)&              0.0032\\
      IRDC 1923+13 1&         0.00 (0.02)&       0.000 (0.000)&         0.00 (0.00)&         0.00 (0.00)&              0.0032\\
      IRDC 1923+13 2&         0.00 (0.02)&       0.000 (0.000)&         0.00 (0.00)&         0.00 (0.00)&              0.0032\\
      IRDC 1916+11 0&         0.00 (0.02)&       0.000 (0.000)&         0.00 (0.00)&         0.00 (0.00)&              0.0048\\
      IRDC 1916+11 1&         0.00 (0.02)&       0.000 (0.000)&         0.00 (0.00)&         0.00 (0.00)&              0.0048\\
      IRDC 1916+11 2&         0.00 (0.02)&       0.000 (0.000)&         0.00 (0.00)&         0.00 (0.00)&              0.0048\\
}{
\tablenotetext{a}{ The Upper Limit Flag is 1 when the measurement indicated is
a $3-\sigma$ upper limit on the \twotwo\ line depth when there is a
corresponding \oneone\ line detection. }
\tablenotetext{b}{RMS in 1.011 \kms\ channels.} 
}
\Table{lccccccc}{Distance, BGPS 1.1 mm, and other properties}
{\colhead{Source Name}&\colhead{Distance}&\colhead{Galactocentric}&\colhead{KDA\tablenotemark{a}}&\colhead{$S_{1.1mm}$}&\colhead{Source}&\colhead{\formaldehyde\ }&\colhead{Scenario\tablenotemark{b}}\\
\colhead{           }&\colhead{        }&\colhead{Distance      }&\colhead{Resolution}&\colhead{           }&\colhead{Type  }&\colhead{Spectrum}&\colhead{}\\  
\colhead{           }&\colhead{(kpc)   }&\colhead{         (kpc)}&\colhead{          }&\colhead{(Jy)       }&\colhead{      }&\colhead{Type    }&\colhead{}\\ }
{tab:other}{
       G32.80+0.19 0&                12.9&                 7.4&                 far&                6.94&               UCHII&        red gradient&                 2+3\\
       G32.80+0.19 1&                13.1&                 7.6&                 far&                6.94&               UCHII&            envelope&                 2+3\\
       G32.80+0.19 2&                 9.4&                 5.1&                 far&                6.94&                 GMC&                   -&                 2+3\\
       G32.80+0.19 3&                 9.2&                 5.0&                 far&                6.94&                 GMC&                   -&                 2+3\\
       G32.80+0.19 4&                 9.0&                 4.9&                 far&                6.94&                 GMC&                   -&                 2+3\\
       G33.13-0.09 0&                 9.6&                 5.2&                 far&                2.26&               UCHII&        red gradient&                   2\\
       G33.13-0.09 1&                 9.3&                 5.1&                 far&                2.26&                 GMC&            envelope&                   2\\
       G33.13-0.09 2&                 7.1&                 4.7&             tangent&                2.26&                 GMC&                   -&                   2\\
       G33.13-0.09 3&                 0.9&                 7.6&                near&                2.26&                 GMC&                   -&                   2\\
       G33.92+0.11 0&                 7.0&                 4.6&             tangent&                3.86&               UCHII&        red gradient&                   2\\
       G33.92+0.11 1&                 7.0&                 4.6&             tangent&                3.86&               UCHII&            envelope&                   2\\
       G33.92+0.11 2&                 3.6&                 5.8&                near&                3.86&                 GMC&                   -&                   2\\
       G34.26+0.15 0&                 3.6&                 5.7&                near&               35.69&               UCHII&        red gradient&                   2\\
       G34.26+0.15 1&                 1.9&                 6.9&                near&               35.69&                 GMC&                   -&                   2\\
       G34.26+0.15 2&                 1.0&                 7.6&                near&               35.69&                 GMC&                   -&                   2\\
       G34.26+0.15 3&                 3.6&                 6.0&                near&               35.69&                 GMC&            envelope&                   2\\
       G34.26+0.15 4&                 3.6&                 6.1&                near&               35.69&                 GMC&                   -&                   2\\
       G35.20-1.74 0&                 2.8&                 6.3&                near&                   -&               UCHII&              single&                   4\\
       G35.20-1.74 1&                 2.5&                 6.5&                near&                   -&                 GMC&                   -&                   4\\
       G35.20-1.74 2&                 1.1&                 7.5&                near&                   -&                 GMC&                   -&                   4\\
       G35.20-1.74 3&                 3.2&                 6.1&                near&                   -&                 GMC&                   -&                   4\\
       G35.57-0.03 0&                10.3&                 6.0&                 far&                2.57&               UCHII&              single&                 2+3\\
       G35.57-0.03 1&                10.7&                 6.2&                 far&                2.57&                 GMC&                   -&                 2+3\\
       G35.57-0.03 2&                 3.6&                 5.9&                near&                2.57&                 GMC&                   -&                 2+3\\
       G35.57-0.03 3&                 1.1&                 7.6&                near&                2.57&                 GMC&                   -&                 2+3\\
       G35.57-0.03 4&                 2.0&                 6.8&                near&                2.57&                 GMC&                   -&                 2+3\\
       G35.58+0.07 0&                10.5&                 6.1&                 far&                1.44&               UCHII&       blue gradient&                   2\\
       G35.58+0.07 1&                10.3&                 6.0&                 far&                1.44&               UCHII&                   -&                   2\\
       G35.58+0.07 2&                 3.6&                 5.8&                near&                1.44&                 GMC&                   -&                   2\\
       G35.58+0.07 3&                 1.1&                 7.5&                near&                1.44&                 GMC&                   -&                   2\\
       G37.87-0.40 0&                 9.4&                 5.9&                 far&                4.14&               UCHII&       blue gradient&                   1\\
       G37.87-0.40 1&                 9.8&                 6.1&                 far&                4.14&               UCHII&       blue gradient&                   1\\
       G37.87-0.40 2&                 9.2&                 5.7&                 far&                4.14&               UCHII&       blue gradient&                   1\\
       G37.87-0.40 3&                 8.7&                 5.6&                 far&                4.14&                 GMC&                   -&                   1\\
       G37.87-0.40 4&                 8.6&                 5.5&                 far&                4.14&                 GMC&                   -&                   1\\
       G37.87-0.40 5&                 8.1&                 5.4&                 far&                4.14&                 GMC&                   -&                   1\\
       G37.87-0.40 6&                 6.6&                 5.1&             tangent&                4.14&                 GMC&                   -&                   1\\
       G37.87-0.40 7&                 1.2&                 7.5&                near&                4.14&                 GMC&                   -&                   1\\
       G37.87-0.40 8&                 1.1&                 7.6&                near&                4.14&                 GMC&                   -&                   1\\
       G37.87-0.40 9&                 1.5&                 7.2&                near&                4.14&                 GMC&                   -&                   1\\
       G43.89-0.78 0&                 8.3&                 6.2&                 far&                   -&               UCHII&       blue gradient&                   3\\
       G43.89-0.78 1&                 8.6&                 6.3&                 far&                   -&                 GMC&            envelope&                   3\\
       G45.07+0.13 0&                 7.6&                 6.2&                 far&                4.26&               UCHII&              single&                   2\\
       G45.07+0.13 1&                 6.5&                 6.0&                 far&                4.26&                 GMC&                   -&                   2\\
       G45.12+0.13 0&                 7.4&                 6.2&                 far&                6.78&               UCHII&               other&                   1\\
       G45.12+0.13 1&                 7.4&                 6.1&                 far&                6.78&               UCHII&            envelope&                   1\\
       G45.12+0.13 2&                 1.9&                 7.2&                near&                6.78&                 GMC&                   -&                   1\\
       G45.12+0.13 3&                 7.4&                 6.0&                 far&                6.78&                 GMC&            envelope&                   1\\
       G45.45+0.06 0&                 7.2&                 6.1&                 far&                3.71&               UCHII&       blue gradient&                   2\\
       G45.45+0.06 1&                 7.6&                 6.2&                 far&                3.71&                 GMC&            envelope&                   2\\
       G45.45+0.06 2&                 1.9&                 7.2&                near&                3.71&                 GMC&                   -&                   2\\
       G45.47+0.05 0&                 7.1&                 6.1&                 far&                3.34&               UCHII&        red gradient&               1+2+3\\
       G45.47+0.05 1&                 1.9&                 7.2&                near&                3.34&                 GMC&                   -&               1+2+3\\
       G48.61+0.02 0&                 9.6&                 7.5&                 far&                2.20&               UCHII&        red gradient&                 2+3\\
       G48.61+0.02 1&                 0.7&                 8.0&                near&                2.20&                 GMC&                   -&                 2+3\\
       G48.61+0.02 2&                 6.5&                 6.4&                 far&                2.20&                 GMC&                   -&                 2+3\\
       G50.32+0.68 0&                 2.1&                 7.2&                near&                   -&               UCHII&                   -&                   1\\
       G60.88-0.13 0&                 2.8&                 7.4&                near&                4.90&               UCHII&               limit&                   2\\
       G61.48+0.09 0&                 5.2&                 7.5&                 far&                7.86&               UCHII&              single&                   4\\
       G69.54-0.98 0&                2.57&                 7.9&             tangent&                   -&               UCHII&               thick&                 4+5\\
       G70.29+1.60 0&                 7.3&                 9.1&                 far&                   -&               UCHII&       blue gradient&                   2\\
       G70.29+1.60 1&                 7.8&                 9.3&                 far&                   -&                 GMC&            envelope&                   2\\
       G70.33+1.59 0&                 7.3&                 9.1&                 far&                   -&               UCHII&              single&                 1+2\\
   IRAS 20051+3435 0&                 2.6&                 7.6&             tangent&                   -&                 GMC&               limit&                  -1\\
       G41.74+0.10 0&                11.3&                 7.6&                 far&                0.56&               UCHII&               limit&                  -1\\
       G41.74+0.10 1&                11.6&                 7.7&                 far&                0.56&               UCHII&                   -&                  -1\\
       G41.74+0.10 2&                 2.4&                 6.8&                near&                0.56&                 GMC&                   -&                  -1\\
       G41.74+0.10 3&                 3.8&                 6.1&                near&                0.56&                 GMC&                   -&                  -1\\
       G41.74+0.10 4&                11.2&                 7.4&                 far&                0.56&               UCHII&                   -&                  -1\\
      IRDC 1923+13 0&                 4.2&                 6.5&                near&                   -&                 GMC&               limit&                  -1\\
      IRDC 1923+13 1&                 5.5&                 6.3&             tangent&                   -&                 GMC&                   -&                  -1\\
      IRDC 1923+13 2&                 3.8&                 6.6&                near&                   -&                 GMC&                   -&                  -1\\
      IRDC 1916+11 0&                 2.0&                 7.2&                near&                   -&                 GMC&               limit&                  -1\\
      IRDC 1916+11 1&                 4.2&                 6.2&                near&                   -&                 GMC&                   -&                  -1\\
      IRDC 1916+11 2&                 3.6&                 6.4&                near&                   -&                 GMC&                   -&                  -1\\
}{
\tablenotetext{a}{The Kinematic Distance Ambiguity described in Section \ref{sec:distances}.}
\tablenotetext{b}{Scenario or scenarios most likely to be consistent with the observed spectrum, as described
in Section \ref{sec:scenarios}.  In some cases, the spectrum was consistent with multiple scenarios or some
blend of multiple scenarios.  In others, the source could not be classified, in which case it is marked with 
-1 in this column. }
}
\Table{lcccccccc}{Inferred \formaldehyde\ line properties}
{\colhead{Source Name}&\colhead{$\tau_{1-1}$}&\colhead{$\tau_{1-1}$ (FFC)}&\colhead{$\tau_{2-2}$}&\colhead{$\tau_{2-2}$ (FFC)}&\colhead{2-2 Upper }&\colhead{2cm Area\tablenotemark{a}}&\colhead{6cm Area  \tablenotemark{a}   }&\colhead{FFC Error}\\
\colhead{           }&\colhead{             }&\colhead{                   }&\colhead{             }&\colhead{                   }&\colhead{Limit Flag}&\colhead{\arcsec$^2$                   }&\colhead{\arcsec$^2$                   }&\colhead{}\\ }
{tab:h2coinferred}{
       G32.80+0.19 0&        0.18 (0.055)&         0.2 (0.059)&        0.12 (0.024)&        0.15 (0.031)&                   0&                88.0&               226.2&                 0.1\\
       G32.80+0.19 1&        0.04 (0.013)&       0.043 (0.013)&      0.016 (0.0051)&      0.021 (0.0065)&                   0&                88.0&               226.2&                 0.1\\
       G32.80+0.19 2&      0.027 (0.0089)&      0.029 (0.0095)&    0.0033 (0.00069)&    0.0042 (0.00088)&                   0&                88.0&               226.2&                 0.1\\
       G32.80+0.19 3&        0.11 (0.035)&        0.12 (0.037)&      0.014 (0.0028)&      0.018 (0.0035)&                   0&                88.0&               226.2&                 0.1\\
       G32.80+0.19 4&       0.039 (0.012)&       0.042 (0.013)&     0.0055 (0.0011)&     0.0071 (0.0014)&                   0&                88.0&               226.2&                 0.1\\
       G33.13-0.09 0&          0.34 (0.1)&         0.49 (0.15)&        0.16 (0.032)&         0.63 (0.12)&                   0&                33.5&                33.5&                 0.2\\
       G33.13-0.09 1&       0.035 (0.015)&        0.047 (0.02)&         0 (0.0059)&         0 (0.0031)&                   1&                33.5&                33.5&                 0.2\\
       G33.13-0.09 2&       0.062 (0.022)&       0.084 (0.029)&         0 (0.0059)&         0 (0.0031)&                   1&                33.5&                33.5&                 0.2\\
       G33.13-0.09 3&       0.061 (0.021)&       0.082 (0.028)&         0 (0.0059)&         0 (0.0031)&                   1&                33.5&                33.5&                 0.2\\
       G33.92+0.11 0&       0.084 (0.027)&         0.1 (0.031)&      0.045 (0.0091)&       0.094 (0.018)&                   0&               214.0&               214.0&                 0.2\\
       G33.92+0.11 1&       0.082 (0.026)&       0.098 (0.031)&      0.036 (0.0072)&       0.075 (0.014)&                   0&               214.0&               214.0&                 0.2\\
       G33.92+0.11 2&        0.17 (0.062)&        0.21 (0.074)&         0 (0.0049)&         0 (0.0031)&                   1&               214.0&               214.0&                 0.2\\
       G34.26+0.15 0&         0.38 (0.12)&          0.4 (0.12)&        0.22 (0.043)&        0.26 (0.052)&                   0&                10.9&                10.9&                 0.2\\
       G34.26+0.15 1&      0.028 (0.0089)&      0.029 (0.0092)&     0.0067 (0.0014)&     0.0079 (0.0017)&                   0&                10.9&                10.9&                 0.2\\
       G34.26+0.15 2&      0.017 (0.0059)&       0.018 (0.006)&    0.0026 (0.00059)&     0.0031 (0.0007)&                   0&                10.9&                10.9&                 0.2\\
       G34.26+0.15 3&      0.022 (0.0072)&      0.023 (0.0074)&    0.0036 (0.00092)&     0.0043 (0.0011)&                   0&                10.9&                10.9&                 0.2\\
       G34.26+0.15 4&     0.0082 (0.0036)&     0.0085 (0.0037)&     0.0026 (0.0011)&      0.003 (0.0013)&                   0&                10.9&                10.9&                 0.2\\
       G35.20-1.74 0&        0.21 (0.063)&        0.22 (0.066)&       0.071 (0.014)&       0.084 (0.017)&                   0&                39.5&                39.5&                 0.2\\
       G35.20-1.74 1&      0.028 (0.0085)&      0.029 (0.0088)&     0.0039 (0.0011)&     0.0046 (0.0013)&                   0&                39.5&                39.5&                 0.2\\
       G35.20-1.74 2&       0.063 (0.019)&       0.065 (0.019)&      0.008 (0.0017)&     0.0095 (0.0019)&                   0&                39.5&                39.5&                 0.2\\
       G35.20-1.74 3&     0.0073 (0.0027)&     0.0075 (0.0028)&         0 (0.0023)&         0 (0.0031)&                   1&                39.5&                39.5&                 0.2\\
       G35.57-0.03 0&        0.11 (0.035)&        0.15 (0.049)&       0.056 (0.011)&        0.26 (0.054)&                   0&                 6.7&                 6.7&                 0.1\\
       G35.57-0.03 1&       0.034 (0.018)&       0.046 (0.024)&      0.011 (0.0047)&        0.047 (0.02)&                   0&                 6.7&                 6.7&                 0.1\\
       G35.57-0.03 2&        0.03 (0.017)&       0.042 (0.023)&         0 (0.0099)&          0 (0.019)&                   1&                 6.7&                 6.7&                 0.1\\
       G35.57-0.03 3&        0.05 (0.021)&       0.069 (0.029)&         0 (0.0099)&          0 (0.019)&                   1&                 6.7&                 6.7&                 0.1\\
       G35.57-0.03 4&        0.051 (0.02)&       0.069 (0.028)&      0.017 (0.0065)&       0.077 (0.029)&                   0&                 6.7&                 6.7&                 0.1\\
       G35.58+0.07 0&        0.24 (0.071)&        0.32 (0.097)&       0.085 (0.017)&         0.61 (0.12)&                   0&                 2.1&                 2.1&                 0.2\\
       G35.58+0.07 1&       0.072 (0.029)&       0.096 (0.038)&         0 (0.0072)&          0 (0.019)&                   1&                 2.1&                 2.1&                 0.2\\
       G35.58+0.07 2&       0.037 (0.012)&       0.049 (0.016)&         0 (0.0072)&          0 (0.019)&                   1&                 2.1&                 2.1&                 0.2\\
       G35.58+0.07 3&        0.05 (0.016)&       0.066 (0.021)&         0 (0.0072)&           0 (0.01)&                   1&                 2.1&                 2.1&                 0.2\\
       G37.87-0.40 0&        0.12 (0.037)&        0.13 (0.038)&      0.047 (0.0095)&       0.061 (0.012)&                   0&                27.5&               170.9&                 0.2\\
       G37.87-0.40 1&      0.028 (0.0089)&      0.029 (0.0092)&     0.0095 (0.0024)&      0.012 (0.0031)&                   0&                27.5&               170.9&                 0.2\\
       G37.87-0.40 2&       0.081 (0.025)&       0.084 (0.026)&     0.0075 (0.0021)&     0.0096 (0.0027)&                   0&                27.5&               170.9&                 0.2\\
       G37.87-0.40 3&       0.074 (0.024)&       0.076 (0.025)&      0.011 (0.0023)&       0.014 (0.003)&                   0&                27.5&               170.9&                 0.2\\
       G37.87-0.40 4&       0.097 (0.029)&          0.1 (0.03)&      0.0098 (0.002)&      0.013 (0.0026)&                   0&                27.5&               170.9&                 0.2\\
       G37.87-0.40 5&       0.041 (0.013)&       0.043 (0.013)&    0.0033 (0.00068)&    0.0043 (0.00087)&                   0&                27.5&               170.9&                 0.2\\
       G37.87-0.40 6&      0.025 (0.0083)&      0.026 (0.0086)&    0.0021 (0.00052)&    0.0026 (0.00066)&                   0&                27.5&               170.9&                 0.2\\
       G37.87-0.40 7&       0.039 (0.012)&        0.04 (0.012)&     0.0054 (0.0012)&     0.0069 (0.0015)&                   0&                27.5&               170.9&                 0.2\\
       G37.87-0.40 8&      0.016 (0.0069)&      0.016 (0.0071)&    0.0035 (0.00081)&      0.0046 (0.001)&                   0&                27.5&               170.9&                 0.2\\
       G37.87-0.40 9&       0.03 (0.0095)&      0.031 (0.0098)&    0.0035 (0.00073)&    0.0045 (0.00094)&                   0&                27.5&               170.9&                 0.2\\
       G43.89-0.78 0&        0.25 (0.074)&        0.32 (0.096)&      0.037 (0.0078)&        0.12 (0.024)&                   0&                13.5&                13.5&                 0.1\\
       G43.89-0.78 1&      0.025 (0.0077)&      0.031 (0.0097)&     0.0097 (0.0022)&      0.029 (0.0067)&                   0&                13.5&                13.5&                 0.1\\
       G45.07+0.13 0&       0.092 (0.029)&         0.13 (0.04)&       0.04 (0.0081)&        0.096 (0.02)&                   0&                 2.5&                 2.5&                 0.2\\
       G45.07+0.13 1&        0.058 (0.02)&        0.08 (0.027)&     0.0061 (0.0019)&      0.014 (0.0045)&                   0&                 2.5&                 2.5&                 0.2\\
       G45.12+0.13 0&       0.043 (0.013)&       0.045 (0.013)&      0.014 (0.0028)&      0.017 (0.0033)&                   0&                15.4&               516.6&                 0.2\\
       G45.12+0.13 1&       0.035 (0.011)&       0.036 (0.011)&      0.0095 (0.002)&      0.011 (0.0025)&                   0&                15.4&               516.6&                 0.2\\
       G45.12+0.13 2&       0.046 (0.014)&       0.048 (0.014)&     0.0075 (0.0015)&      0.009 (0.0019)&                   0&                15.4&               516.6&                 0.2\\
       G45.12+0.13 3&       0.006 (0.002)&     0.0062 (0.0021)&    0.0033 (0.00068)&     0.004 (0.00082)&                   0&                15.4&               516.6&                 0.2\\
       G45.45+0.06 0&        0.32 (0.096)&        0.32 (0.095)&       0.063 (0.013)&       0.069 (0.012)&                   0&              1963.0&              1963.0&                 0.2\\
       G45.45+0.06 1&       0.025 (0.011)&       0.026 (0.011)&       0.01 (0.0021)&      0.011 (0.0019)&                   0&              1963.0&              1963.0&                 0.2\\
       G45.45+0.06 2&      0.011 (0.0036)&      0.012 (0.0035)&         0 (0.0031)&           0 (0.01)&                   1&              1963.0&              1963.0&                 0.2\\
       G45.47+0.05 0&         0.35 (0.11)&         0.45 (0.14)&       0.089 (0.018)&        0.39 (0.079)&                   0&                 3.0&                 3.0&                 0.2\\
       G45.47+0.05 1&      0.018 (0.0068)&      0.023 (0.0084)&           0 (0.01)&           0 (0.01)&                   1&                 3.0&                 3.0&                 0.2\\
       G48.61+0.02 0&       0.058 (0.018)&       0.068 (0.021)&      0.015 (0.0034)&       0.053 (0.012)&                   0&                25.5&                25.5&                 0.2\\
       G48.61+0.02 1&       0.02 (0.0075)&      0.023 (0.0088)&         0 (0.0067)&         0 (0.0026)&                   1&                25.5&                25.5&                 0.2\\
       G48.61+0.02 2&      0.016 (0.0052)&      0.018 (0.0061)&     0.0033 (0.0013)&      0.012 (0.0046)&                   0&                25.5&                25.5&                 0.2\\
       G50.32+0.68 0&        0.027 (0.01)&       0.045 (0.017)&     0.0089 (0.0031)&       0.058 (0.019)&                   0&               108.0&               108.0&                 0.2\\
       G60.88-0.13 0&        0.12 (0.037)&        0.14 (0.043)&      0.011 (0.0031)&      0.031 (0.0071)&                   0&               615.0&               615.0&                 0.2\\
       G61.48+0.09 0&       0.088 (0.026)&        0.09 (0.027)&       0.069 (0.014)&       0.088 (0.017)&                   0&               355.0&               355.0&                 0.2\\
       G69.54-0.98 0&         0.98 (0.29)&           5.7 (1.7)&        0.18 (0.037)&          2.9 (0.57)&                   0&                 0.5&                 0.5&                 0.2\\
       G70.29+1.60 0&       0.086 (0.026)&       0.089 (0.027)&      0.022 (0.0044)&      0.026 (0.0052)&                   0&                52.8&                52.8&                 0.1\\
       G70.29+1.60 1&      0.011 (0.0037)&      0.012 (0.0038)&         0 (0.0019)&         0 (0.0026)&                   1&                52.8&                52.8&                 0.1\\
       G70.33+1.59 0&          0.7 (0.21)&         0.78 (0.24)&        0.34 (0.068)&          0.52 (0.1)&                   0&                16.4&                16.4&                 0.1\\
   IRAS 20051+3435 0&        0.12 (0.036)&        0.13 (0.014)&      0.015 (0.0041)&      0.016 (0.0034)&                   0&             2747.75&             2747.75&                 0.0\\
       G41.74+0.10 0&         0.13 (0.04)&          0.2 (0.06)&       0.01 (0.0027)&       0.045 (0.012)&                   0&                75.2&                75.2&                 0.2\\
       G41.74+0.10 1&        0.04 (0.014)&        0.06 (0.021)&         0 (0.0071)&         0 (0.0026)&                   1&                75.2&                75.2&                 0.2\\
       G41.74+0.10 2&        0.14 (0.043)&        0.21 (0.065)&         0 (0.0071)&         0 (0.0026)&                   1&                75.2&                75.2&                 0.2\\
       G41.74+0.10 3&       0.045 (0.017)&       0.067 (0.025)&         0 (0.0071)&         0 (0.0026)&                   1&                75.2&                75.2&                 0.2\\
       G41.74+0.10 4&       0.089 (0.029)&        0.13 (0.043)&         0 (0.0071)&         0 (0.0028)&                   1&                75.2&                75.2&                 0.2\\
      IRDC 1923+13 0&       0.02 (0.0062)&       0.02 (0.0047)&          0 (0.009)&         0 (0.0028)&                   1&             2747.75&             2747.75&                 0.0\\
      IRDC 1923+13 1&      0.016 (0.0051)&      0.017 (0.0038)&          0 (0.009)&         0 (0.0028)&                   1&             2747.75&             2747.75&                 0.0\\
      IRDC 1923+13 2&     0.0081 (0.0028)&     0.0083 (0.0023)&          0 (0.009)&         0 (0.0028)&                   1&             2747.75&             2747.75&                 0.0\\
      IRDC 1916+11 0&       0.033 (0.011)&      0.036 (0.0062)&          0 (0.014)&          0 (0.042)&                   1&             2747.75&             2747.75&                 0.0\\
      IRDC 1916+11 1&       0.082 (0.025)&      0.089 (0.0095)&          0 (0.014)&          0 (0.042)&                   1&             2747.75&             2747.75&                 0.0\\
      IRDC 1916+11 2&      0.017 (0.0064)&      0.018 (0.0046)&          0 (0.014)&          0 (0.042)&                   1&             2747.75&             2747.75&                 0.0\\
}{\tablenotetext{a}{The beam area is 2747.75\arcsec$^2$, which is used when the CMB is the only background continuum illumination}}
\Table{lccccccc}{Derived physical properties from \formaldehyde\ }
{\colhead{Source Name}&\colhead{N(\formaldehyde)\tablenotemark{a}}&\colhead{N(\formaldehyde) (FFC)\tablenotemark{b}}&\colhead{n(\hh) \tablenotemark{a} }&\colhead{n(\hh) (FFC)\tablenotemark{b}}&\colhead{X$_{\formaldehyde}$\tablenotemark{a}}&\colhead{X$_{\formaldehyde}$ (FFC)\tablenotemark{b}}&\colhead{Flag\tablenotemark{c}}\\
\colhead{           }&\colhead{(\persc)        }&\colhead{(\persc)              }&\colhead{(\percc)}&\colhead{(\percc)    }&\colhead{                    }&\colhead{                          }&\colhead{                      }\\ }
{tab:h2coderived}{
       G32.80+0.19 0&$12.79^{+0.11}_{-0.16}$&$\mathbf{12.94^{+0.16}_{-0.24}}$&$5.10^{+0.25}_{-0.26}$&$\mathbf{5.21^{+0.27}_{-0.29}}$&$-10.79^{+0.15}_{-0.20}$&$\mathbf{-10.75^{+0.15}_{-0.18}}$&                   2\\
       G32.80+0.19 1&$12.05^{+0.12}_{-0.11}$&$\mathbf{12.14^{+0.13}_{-0.13}}$&$4.96^{+0.22}_{-0.28}$&$\mathbf{5.05^{+0.21}_{-0.28}}$&$-11.39^{+0.20}_{-0.23}$&$\mathbf{-11.39^{+0.17}_{-0.20}}$&                   2\\
       G32.80+0.19 2&$11.66^{+0.10}_{-0.10}$&$\mathbf{11.71^{+0.10}_{-0.10}}$&$4.16^{+0.39}_{-0.38}$&$\mathbf{4.33^{+0.31}_{-0.32}}$&$-10.97^{+0.44}_{-0.46}$&$\mathbf{-11.10^{+0.37}_{-0.37}}$&                   2\\
       G32.80+0.19 3&$12.18^{+0.10}_{-0.09}$&$\mathbf{12.23^{+0.09}_{-0.09}}$&$4.07^{+0.38}_{-0.39}$&$\mathbf{4.23^{+0.32}_{-0.32}}$&$-10.37^{+0.44}_{-0.45}$&$\mathbf{-10.48^{+0.36}_{-0.38}}$&                   2\\
       G32.80+0.19 4&$11.82^{+0.10}_{-0.09}$&$\mathbf{11.87^{+0.10}_{-0.09}}$&$4.30^{+0.31}_{-0.32}$&$\mathbf{4.44^{+0.26}_{-0.29}}$&$-10.97^{+0.37}_{-0.37}$&$\mathbf{-11.05^{+0.31}_{-0.32}}$&                   2\\
       G33.13-0.09 0&$>12.80                 $&$\mathbf{>13.56}       $&$>4.54                 $&$\mathbf{>5.10}       $&$>-10.62                 $&$\mathbf{>-11.70}       $&                   8\\
       G33.13-0.09 1&$<11.96                 $&$\mathbf{<11.90}       $&$<4.50                 $&$\mathbf{<3.91}       $&$<-8.44                 $&$\mathbf{<-8.45}       $&                   6\\
       G33.13-0.09 2&$\mathbf{<12.20}       $&$<0.00                 $&$\mathbf{<4.29}       $&$<0.00                 $&$\mathbf{<-8.29}       $&$<0.00                 $&                   5\\
       G33.13-0.09 3&$\mathbf{<12.20}       $&$<0.00                 $&$\mathbf{<4.32}       $&$<0.00                 $&$\mathbf{<-8.29}       $&$<0.00                 $&                   5\\
       G33.92+0.11 0&$>12.35                 $&$\mathbf{>12.64}       $&$>4.86                 $&$\mathbf{>5.16}       $&$>-11.29                 $&$\mathbf{>-12.30}       $&                   8\\
       G33.92+0.11 1&$12.34^{+0.07}_{-0.08}$&$\mathbf{12.65^{+0.11}_{-0.17}}$&$4.97^{+0.22}_{-0.23}$&$\mathbf{5.26^{+0.22}_{-0.24}}$&$-11.11^{+0.19}_{-0.22}$&$\mathbf{-11.09^{+0.13}_{-0.16}}$&                   2\\
       G33.92+0.11 2&                   -&                   -&                   -&                   -&                   -&                   -&                   9\\
       G34.26+0.15 0&$13.01^{+0.10}_{-0.17}$&$\mathbf{13.13^{+0.15}_{-0.23}}$&$4.91^{+0.28}_{-0.29}$&$\mathbf{5.01^{+0.31}_{-0.32}}$&$-10.38^{+0.18}_{-0.23}$&$\mathbf{-10.36^{+0.17}_{-0.23}}$&                   2\\
       G34.26+0.15 1&$11.79^{+0.09}_{-0.08}$&$\mathbf{11.83^{+0.09}_{-0.08}}$&$4.67^{+0.23}_{-0.25}$&$\mathbf{4.75^{+0.21}_{-0.24}}$&$-11.36^{+0.26}_{-0.27}$&$\mathbf{-11.40^{+0.23}_{-0.25}}$&                   2\\
       G34.26+0.15 2&$11.53^{+0.10}_{-0.10}$&$\mathbf{11.56^{+0.10}_{-0.10}}$&$4.38^{+0.30}_{-0.33}$&$\mathbf{4.48^{+0.28}_{-0.30}}$&$-11.33^{+0.36}_{-0.37}$&$\mathbf{-11.40^{+0.32}_{-0.34}}$&                   2\\
       G34.26+0.15 3&$11.63^{+0.11}_{-0.10}$&$\mathbf{11.66^{+0.10}_{-0.10}}$&$4.43^{+0.29}_{-0.32}$&$\mathbf{4.53^{+0.26}_{-0.30}}$&$-11.28^{+0.34}_{-0.35}$&$\mathbf{-11.34^{+0.31}_{-0.32}}$&                   2\\
       G34.26+0.15 4&$11.40^{+0.17}_{-0.14}$&$\mathbf{11.45^{+0.17}_{-0.16}}$&$4.87^{+0.31}_{-0.43}$&$\mathbf{4.94^{+0.30}_{-0.42}}$&$-11.95^{+0.30}_{-0.35}$&$\mathbf{-11.98^{+0.27}_{-0.34}}$&                   2\\
       G35.20-1.74 0&$12.60^{+0.08}_{-0.07}$&$\mathbf{12.65^{+0.08}_{-0.07}}$&$4.72^{+0.25}_{-0.25}$&$\mathbf{4.79^{+0.25}_{-0.26}}$&$-10.61^{+0.23}_{-0.27}$&$\mathbf{-10.62^{+0.22}_{-0.26}}$&                   2\\
       G35.20-1.74 1&$11.69^{+0.11}_{-0.10}$&$\mathbf{11.72^{+0.11}_{-0.10}}$&$4.30^{+0.33}_{-0.37}$&$\mathbf{4.41^{+0.30}_{-0.33}}$&$-11.09^{+0.39}_{-0.39}$&$\mathbf{-11.16^{+0.34}_{-0.36}}$&                   2\\
       G35.20-1.74 2&$11.97^{+0.09}_{-0.09}$&$\mathbf{12.00^{+0.09}_{-0.09}}$&$4.20^{+0.34}_{-0.35}$&$\mathbf{4.31^{+0.30}_{-0.30}}$&$-10.70^{+0.39}_{-0.41}$&$\mathbf{-10.79^{+0.34}_{-0.36}}$&                   2\\
       G35.20-1.74 3&$<11.41                 $&$\mathbf{<11.44}       $&$<4.89                 $&$\mathbf{<5.02}       $&$<-9.24                 $&$\mathbf{<-9.30}       $&                   6\\
       G35.57-0.03 0&$>12.42                 $&$\mathbf{>13.38}       $&$>4.82                 $&$\mathbf{>5.61}       $&$>-11.20                 $&$\mathbf{>-12.02}       $&                   8\\
       G35.57-0.03 1&$>11.72                 $&$\mathbf{>12.25}       $&$>4.51                 $&$\mathbf{>5.13}       $&$>-11.80                 $&$\mathbf{>-12.72}       $&                   8\\
       G35.57-0.03 2&$<11.98                 $&$\mathbf{<12.12}       $&$<4.96                 $&$\mathbf{<5.12}       $&$<-8.71                 $&$\mathbf{<-8.72}       $&                   6\\
       G35.57-0.03 3&$<12.09                 $&$\mathbf{<12.23}       $&$<4.60                 $&$\mathbf{<4.78}       $&$<-8.38                 $&$\mathbf{<-8.37}       $&                   6\\
       G35.57-0.03 4&$>11.93                 $&$\mathbf{>12.47}       $&$>4.58                 $&$\mathbf{>5.20}       $&$>-11.55                 $&$\mathbf{>-12.48}       $&                   8\\
       G35.58+0.07 0&$>12.58                 $&$\mathbf{>14.06}       $&$>4.50                 $&$\mathbf{>5.48}       $&$>-10.79                 $&$\mathbf{>-11.71}       $&                   8\\
       G35.58+0.07 1&$<12.19                 $&$\mathbf{<12.32}       $&$<4.08                 $&$\mathbf{<4.55}       $&$<-8.07                 $&$\mathbf{<-8.14}       $&                   6\\
       G35.58+0.07 2&$<11.96                 $&$\mathbf{<12.10}       $&$<4.57                 $&$\mathbf{<4.95}       $&$<-8.53                 $&$\mathbf{<-8.63}       $&                   6\\
       G35.58+0.07 3&$<12.06                 $&$\mathbf{<12.17}       $&$<4.35                 $&$\mathbf{<4.38}       $&$<-8.32                 $&$\mathbf{<-8.24}       $&                   6\\
       G37.87-0.40 0&$12.44^{+0.07}_{-0.07}$&$\mathbf{12.53^{+0.07}_{-0.09}}$&$4.86^{+0.22}_{-0.23}$&$\mathbf{4.98^{+0.21}_{-0.24}}$&$-10.90^{+0.20}_{-0.24}$&$\mathbf{-10.92^{+0.18}_{-0.21}}$&                   2\\
       G37.87-0.40 1&$11.87^{+0.10}_{-0.09}$&$\mathbf{11.95^{+0.10}_{-0.09}}$&$4.89^{+0.22}_{-0.26}$&$\mathbf{5.00^{+0.21}_{-0.24}}$&$-11.50^{+0.21}_{-0.24}$&$\mathbf{-11.53^{+0.18}_{-0.22}}$&                   2\\
       G37.87-0.40 2&$12.10^{+0.18}_{-0.26}$&$\mathbf{12.09^{+0.14}_{-0.28}}$&$3.16^{+1.15}_{-1.20}$&$\mathbf{3.79^{+1.78}_{-0.71}}$&$-9.54^{+1.32}_{-1.41}$&$\mathbf{-10.18^{+0.77}_{-2.06}}$&                   4\\
       G37.87-0.40 3&$12.05^{+0.10}_{-0.10}$&$\mathbf{12.10^{+0.10}_{-0.09}}$&$4.33^{+0.32}_{-0.33}$&$\mathbf{4.49^{+0.28}_{-0.29}}$&$-10.76^{+0.37}_{-0.39}$&$\mathbf{-10.87^{+0.31}_{-0.34}}$&                   2\\
       G37.87-0.40 4&$12.17^{+0.17}_{-0.25}$&$\mathbf{12.14^{+0.10}_{-0.09}}$&$3.36^{+1.35}_{-0.98}$&$\mathbf{4.13^{+0.35}_{-0.36}}$&$-9.67^{+1.12}_{-1.60}$&$\mathbf{-10.46^{+0.41}_{-0.42}}$&                   2\\
       G37.87-0.40 5&$11.85^{+0.18}_{-0.21}$&$\mathbf{11.88^{+0.18}_{-0.25}}$&$3.05^{+1.04}_{-1.17}$&$\mathbf{3.34^{+1.33}_{-1.03}}$&$-9.68^{+1.32}_{-1.25}$&$\mathbf{-9.94^{+1.17}_{-1.58}}$&                   4\\
       G37.87-0.40 6&$11.67^{+0.19}_{-0.23}$&$\mathbf{11.70^{+0.19}_{-0.26}}$&$3.08^{+1.07}_{-1.21}$&$\mathbf{3.33^{+1.32}_{-1.13}}$&$-9.89^{+1.36}_{-1.30}$&$\mathbf{-10.11^{+1.26}_{-1.58}}$&                   4\\
       G37.87-0.40 7&$11.81^{+0.10}_{-0.09}$&$\mathbf{11.86^{+0.10}_{-0.09}}$&$4.28^{+0.31}_{-0.33}$&$\mathbf{4.45^{+0.27}_{-0.29}}$&$-10.95^{+0.37}_{-0.38}$&$\mathbf{-11.07^{+0.32}_{-0.32}}$&                   2\\
       G37.87-0.40 8&$11.56^{+0.11}_{-0.11}$&$\mathbf{11.63^{+0.11}_{-0.10}}$&$4.67^{+0.30}_{-0.33}$&$\mathbf{4.80^{+0.29}_{-0.31}}$&$-11.59^{+0.34}_{-0.38}$&$\mathbf{-11.65^{+0.30}_{-0.34}}$&                   2\\
       G37.87-0.40 9&$11.72^{+0.13}_{-0.29}$&$\mathbf{11.74^{+0.10}_{-0.09}}$&$3.98^{+1.97}_{-0.52}$&$\mathbf{4.32^{+0.31}_{-0.32}}$&$-10.74^{+0.60}_{-2.26}$&$\mathbf{-11.06^{+0.36}_{-0.37}}$&                   2\\
       G43.89-0.78 0&$12.49^{+0.10}_{-0.09}$&$\mathbf{12.76^{+0.08}_{-0.07}}$&$4.18^{+0.34}_{-0.33}$&$\mathbf{4.68^{+0.28}_{-0.28}}$&$-10.17^{+0.37}_{-0.40}$&$\mathbf{-10.40^{+0.24}_{-0.30}}$&                   2\\
       G43.89-0.78 1&$\mathbf{11.87^{+0.09}_{-0.08}}$&$12.80^{+0.61}_{-1.00}$&$\mathbf{4.95^{+0.20}_{-0.23}}$&$6.16^{+0.96}_{-1.84}$&$\mathbf{-11.56^{+0.19}_{-0.22}}$&$-11.84^{+0.88}_{-0.42}$&                   1\\
       G45.07+0.13 0&$12.38^{+0.08}_{-0.08}$&$\mathbf{12.75^{+0.13}_{-0.20}}$&$4.96^{+0.22}_{-0.24}$&$\mathbf{5.25^{+0.25}_{-0.27}}$&$-11.06^{+0.19}_{-0.22}$&$\mathbf{-10.97^{+0.15}_{-0.18}}$&                   2\\
       G45.07+0.13 1&                   -&                   -&                   -&                   -&                   -&                   -&                   9\\
       G45.12+0.13 0&$12.02^{+0.08}_{-0.07}$&$\mathbf{12.07^{+0.08}_{-0.07}}$&$4.83^{+0.21}_{-0.21}$&$\mathbf{4.92^{+0.19}_{-0.21}}$&$-11.30^{+0.20}_{-0.23}$&$\mathbf{-11.32^{+0.19}_{-0.21}}$&                   2\\
       G45.12+0.13 1&$11.90^{+0.09}_{-0.08}$&$\mathbf{11.95^{+0.08}_{-0.08}}$&$4.74^{+0.22}_{-0.23}$&$\mathbf{4.83^{+0.21}_{-0.23}}$&$-11.32^{+0.23}_{-0.25}$&$\mathbf{-11.36^{+0.21}_{-0.23}}$&                   2\\
       G45.12+0.13 2&$11.90^{+0.10}_{-0.08}$&$\mathbf{11.93^{+0.09}_{-0.09}}$&$4.41^{+0.26}_{-0.28}$&$\mathbf{4.52^{+0.24}_{-0.26}}$&$-11.00^{+0.32}_{-0.32}$&$\mathbf{-11.06^{+0.29}_{-0.30}}$&                   2\\
       G45.12+0.13 3&$11.48^{+0.08}_{-0.09}$&$\mathbf{11.55^{+0.08}_{-0.12}}$&$5.15^{+0.19}_{-0.21}$&$\mathbf{5.23^{+0.19}_{-0.22}}$&$-12.15^{+0.16}_{-0.18}$&$\mathbf{-12.16^{+0.14}_{-0.17}}$&                   2\\
       G45.45+0.06 0&$12.62^{+0.08}_{-0.08}$&$\mathbf{12.64^{+0.07}_{-0.08}}$&$4.33^{+0.29}_{-0.31}$&$\mathbf{4.37^{+0.28}_{-0.28}}$&$-10.19^{+0.33}_{-0.35}$&$\mathbf{-10.21^{+0.30}_{-0.33}}$&                   2\\
       G45.45+0.06 1&$11.89^{+0.08}_{-0.08}$&$\mathbf{11.92^{+0.07}_{-0.07}}$&$5.00^{+0.26}_{-0.28}$&$\mathbf{5.04^{+0.26}_{-0.27}}$&$-11.59^{+0.25}_{-0.29}$&$\mathbf{-11.60^{+0.23}_{-0.28}}$&                   2\\
       G45.45+0.06 2&$<11.55                 $&$\mathbf{<11.66}       $&$<4.77                 $&$\mathbf{<5.39}       $&$<-9.04                 $&$\mathbf{<-9.43}       $&                   6\\
       G45.47+0.05 0&$12.71^{+0.09}_{-0.07}$&$\mathbf{13.48^{+0.32}_{-0.50}}$&$4.46^{+0.28}_{-0.28}$&$\mathbf{5.21^{+0.40}_{-0.34}}$&$-10.23^{+0.28}_{-0.31}$&$\mathbf{-10.21^{+0.21}_{-0.19}}$&                   2\\
       G45.47+0.05 1&$\mathbf{<11.91}       $&$<11.92                 $&$\mathbf{<5.34}       $&$<5.23                 $&$\mathbf{<-8.65}       $&$<-8.57                 $&                   5\\
       G48.61+0.02 0&$12.06^{+0.09}_{-0.09}$&$\mathbf{12.53^{+0.13}_{-0.17}}$&$4.68^{+0.23}_{-0.25}$&$\mathbf{5.29^{+0.22}_{-0.24}}$&$-11.10^{+0.25}_{-0.26}$&$\mathbf{-11.25^{+0.13}_{-0.16}}$&                   2\\
       G48.61+0.02 1&$\mathbf{<11.88}       $&$<11.91                 $&$\mathbf{<5.09}       $&$<4.47                 $&$\mathbf{<-8.61}       $&$<-8.58                 $&                   5\\
       G48.61+0.02 2&$11.54^{+0.14}_{-0.13}$&$\mathbf{11.94^{+0.18}_{-0.23}}$&$4.60^{+0.28}_{-0.39}$&$\mathbf{5.22^{+0.23}_{-0.36}}$&$-11.54^{+0.33}_{-0.33}$&$\mathbf{-11.76^{+0.14}_{-0.19}}$&                   2\\
       G50.32+0.68 0&$>11.71                 $&$\mathbf{>12.41}       $&$>4.61                 $&$\mathbf{>5.31}       $&$>-11.77                 $&$\mathbf{>-12.63}       $&                   8\\
       G60.88-0.13 0&$12.24^{+0.18}_{-0.25}$&$\mathbf{12.35^{+0.09}_{-0.09}}$&$3.20^{+1.19}_{-1.16}$&$\mathbf{4.51^{+0.27}_{-0.28}}$&$-9.44^{+1.29}_{-1.43}$&$\mathbf{-10.64^{+0.28}_{-0.31}}$&                   2\\
       G61.48+0.09 0&$>12.51                 $&$\mathbf{>12.62}       $&$>5.07                 $&$\mathbf{>5.19}       $&$>-11.27                 $&$\mathbf{>-12.33}       $&                   8\\
       G69.54-0.98 0&                   -&                   -&                   -&                   -&                   -&                   -&                  11\\
       G70.29+1.60 0&$12.21^{+0.09}_{-0.08}$&$\mathbf{12.25^{+0.08}_{-0.08}}$&$4.67^{+0.23}_{-0.23}$&$\mathbf{4.74^{+0.23}_{-0.24}}$&$-10.94^{+0.24}_{-0.26}$&$\mathbf{-10.97^{+0.23}_{-0.26}}$&                   2\\
       G70.29+1.60 1&$<11.53                 $&$\mathbf{<11.55}       $&$<4.50                 $&$\mathbf{<4.67}       $&$<-8.92                 $&$\mathbf{<-8.98}       $&                   6\\
       G70.33+1.59 0&$13.16^{+0.09}_{-0.14}$&$\mathbf{13.41^{+0.19}_{-0.35}}$&$4.64^{+0.34}_{-0.32}$&$\mathbf{4.83^{+0.39}_{-0.37}}$&$-9.96^{+0.22}_{-0.31}$&$\mathbf{-9.90^{+0.21}_{-0.26}}$&                   2\\
   IRAS 20051+3435 0&$\mathbf{12.20^{+0.11}_{-0.10}}$&$12.23^{+0.04}_{-0.05}$&$\mathbf{4.12^{+0.39}_{-0.41}}$&$4.11^{+0.21}_{-0.23}$&$\mathbf{-10.40^{+0.45}_{-0.46}}$&$-10.35^{+0.22}_{-0.22}$&                   3\\
       G41.74+0.10 0&$12.25^{+0.17}_{-0.23}$&$\mathbf{12.48^{+0.10}_{-0.09}}$&$2.99^{+0.99}_{-1.18}$&$\mathbf{4.50^{+0.28}_{-0.31}}$&$-9.23^{+1.31}_{-1.22}$&$\mathbf{-10.50^{+0.29}_{-0.32}}$&                   2\\
       G41.74+0.10 1&$\mathbf{<12.12}       $&$<0.00                 $&$\mathbf{<4.72}       $&$<0.00                 $&$\mathbf{<-8.37}       $&$<0.00                 $&                   5\\
       G41.74+0.10 2&$\mathbf{<12.18}       $&$<0.00                 $&$\mathbf{<3.21}       $&$<0.00                 $&$\mathbf{<-8.91}       $&$<0.00                 $&                   5\\
       G41.74+0.10 3&$\mathbf{<12.17}       $&$<0.00                 $&$\mathbf{<4.70}       $&$<0.00                 $&$\mathbf{<-8.32}       $&$<0.00                 $&                   5\\
       G41.74+0.10 4&$\mathbf{<12.27}       $&$<0.00                 $&$\mathbf{<4.11}       $&$<0.00                 $&$\mathbf{<-8.22}       $&$<0.00                 $&                   5\\
      IRDC 1923+13 0&$\mathbf{<11.86}       $&$<11.84                 $&$\mathbf{<5.19}       $&$<4.47                 $&$\mathbf{<-8.63}       $&$<-8.65                 $&                   5\\
      IRDC 1923+13 1&$\mathbf{<11.86}       $&$<11.77                 $&$\mathbf{<5.30}       $&$<4.58                 $&$\mathbf{<-8.70}       $&$<-8.72                 $&                   5\\
      IRDC 1923+13 2&$\mathbf{<13.29}       $&$<11.53                 $&$\mathbf{<8.00}       $&$<5.00                 $&$\mathbf{<-8.95}       $&$<-8.96                 $&                   5\\
      IRDC 1916+11 0&$\mathbf{<12.05}       $&$<12.57                 $&$\mathbf{<5.16}       $&$<5.62                 $&$\mathbf{<-8.44}       $&$<-8.46                 $&                   5\\
      IRDC 1916+11 1&$\mathbf{<12.36}       $&$<12.40                 $&$\mathbf{<4.61}       $&$<5.03                 $&$\mathbf{<-8.13}       $&$<-8.16                 $&                   5\\
      IRDC 1916+11 2&$\mathbf{<12.09}       $&$<13.77                 $&$\mathbf{<5.55}       $&$<8.00                 $&$\mathbf{<-8.68}       $&$<-8.68                 $&                   5\\
}{
\tablenotetext{a}{The values used in this paper are shown in boldface.
Uncorrected values are listed in this column.  The filling-factor corrected
values are shown for comparison in the next column even though they were not used for analysis.}
\tablenotetext{b}{The values used in this paper are shown in boldface.
Filling-factor corrected values are listed in this column.  The uncorrected
values are shown for comparison in the previous column even though they not used for analysis.}
\tablenotetext{c}{Flags:\begin{enumerate}
  \item  No filling factor correction (no FFC) is the most reliable.                                             
  \item  Filling factor correction (FFC) is the most reliable                                                    
  \item  There is an ambiguity between low density / high abundance and low abundance / high density (no FFC)    
  \item  There is an ambiguity between low density / high abundance and low abundance / high density (FFC)       
  \item  Upper Limit (No FFC)                                                                                    
  \item  Upper Limit (FFC)                                                                                       
  \item  Lower Limit (No FFC)                                                                                    
  \item  Lower Limit (FFC)                                                                                       
  \item  Unreliable estimate because of continuum / filling factor uncertainty.                                  
  \item  No limit (S/N)                                                                                         
  \item  Optically Thick                                                                                        
\end{enumerate}
}}

\subsection{Green Bank Telescope}
We observed the \formaldehyde\ \twotwo\ line at 2 cm (14.488789 GHz) with the
Green Bank Telescope (GBT)\footnote{ The National Radio Astronomy Observatory operates
the GBT and VLA and is a facility of the National Science Foundation operated
under cooperative agreement by Associated Universities, Inc.  } dual-beam
Ku-band receiver as part of project GBT09C-049.  The GBT dual-beam Ku-band
receiver was used for 4 hours on January 18th, 2010 in beam-switched nodding
mode.  System temperatures ranged from 27 to 38 K in the \formaldehyde\
band centered on the  \twotwo\ line. A bandwidth of 12.5 MHz
(258.8 \kms) and channel width of 3.052 kHz (0.063 \kms) were used with 9-level
sampling, with receiver temperature $\approx21$K.   Three additional tunings
were acquired simultaneously, centered between the H and He 75$\alpha$,
76$\alpha$, and 77$\alpha$ radio recombination lines (RRLs) with the same channel widths and
bandwidths as above at 14.1315, 14.6930, and 15.2846 GHz.  Each source was
observed for 150 seconds in each receiver for a total on-source integration
time of 300 seconds.  Each observation in the pair was independently inspected
to search for emission/absorption in the off position, which was 5.5\arcmin\
away in azimuth.  When absorption was detected in one of the off positions,
that on/off pair was discarded if one of the detected lines was affected, but
otherwise was noted and ignored.  Pointing and focus observations on the
calibrator source 1822-0938 were taken at the start of and two hours into the
observations.  

The gain was assumed to be 1.91 K/Jy based on previous calibration observations
on point sources in Ku-band; our flux density measurements will therefore be
overestimates for extended sources.  The aperture efficiency was
$\eta_{A}=0.671$, and the main beam efficiency $\eta_{MB}=1.32\eta_{A}=0.886$,
so our main-beam corrected measurements could overestimate extended source flux
densities by at most $13\%$ (ignoring atmospheric absorption).
The data were calibrated using the normal {\sc getnod} procedure in GBTIDL 
\footnote{ GBTIDL (http://gbtidl.nrao.edu/) is the data reduction package
produced by NRAO and written in the IDL language for the reduction of GBT data.
The National Radio Astronomy Observatory is a facility of the National Science
Foundation operated under cooperative agreement by Associated Universities,
Inc.
} ,
which assumes an atmospheric opacity at zenith $\tau=0.014$ at 14.488 GHz. 

We assume primary beam $\theta_{FWHM}=51.1\arcsec$ 
per the GBT
observers manual.  We assume a conservative 10\% error in the beam area
$\Omega=7.8\times10^{-8}$ sr, which governs the flux density received from the
CMB over the observed area.  Beam size error should be dominated by small errors
in focus.  By utilizing the 305 m Arecibo telescope at 6 cm and the 100 m GBT
at 2 cm, we acquired beam-matched (FWHM$\sim 50\arcsec$) observations of the
\formaldehyde\ \oneone\ and \twotwo\ lines.

\subsubsection{GBT Data Reduction}
In the 24 lines of sight, 75 independent components were identified from the
\oneone\ spectra.  These were fit with gaussians using GBTIDL's {\sc fitgauss}
routine.  Out of these 75 components, 51 had corresponding \twotwo\ detections.
The fitted gaussian spectral lines are listed by line of sight in Table
\ref{tab:h2comeasured_a}.  The gaussian fits may not be representative of the
true spectral line profile; complex spectral line profiles are discussed in Section
\ref{sec:lineprofiles}.

The 2 cm continua were measured by fitting a first-order baseline in each
reduced nodded pair excluding the line and the bandpass edges.  Figure
\ref{fig:specexample} shows the flat baselines achieved in the observations,
though the RRL spectrum shows an example of the artifacts seen at the edges of
the bandpass.  The continuum error listed in the table is the RMS of only the
data included in the baseline fit after the baseline was subtracted from the
spectrum; the systematic error from flux calibration uncertainty is 20\% and
dominant.  

\subsection{Arecibo} 
The Arecibo 4.829660 GHz \formaldehyde\ \oneone\ observations used in this
project were previously presented in \citet{Araya2002} and \citet{Araya2004}
and were kindly provided in reduced form by E. Araya.  All observations were performed
using standard on/off position switching and 5 minute integration times in both the 
on and off positions, resulting in off positions 1.25 degrees away from the pointing center.
We assume a 30\% error
in the continuum \citep[based on measured gains in the range 2.0-2.5
as reported in][]{Araya2002} and an effective diameter of 227m ($\theta_{FWHM}
= 56\arcsec$, $\Omega=9.0\times10^{-8}$ sr ) with
10\% uncertainty\footnote{\url{http://www.naic.edu/$\sim$phil/sysperf/misc/hpbw\_vs\_lambda\_2004.html}}.

The Arecibo spectral lines were re-fit for this paper by converting the Arecibo data
from CLASS\footnote{CLASS is part of the GILDAS software developed by IRAM.} to
GBTIDL's {\sc SDFITS} format\footnote{Code for the CLASS-GBTIDL conversion is available from
\url{http://code.google.com/p/casaradio/wiki/class\_to\_gbt}}
and using GBTIDL's {\sc fitgauss} routine.  The 6 cm continua were taken
directly from \citet{Araya2002} Table 3.

\OneColFigure{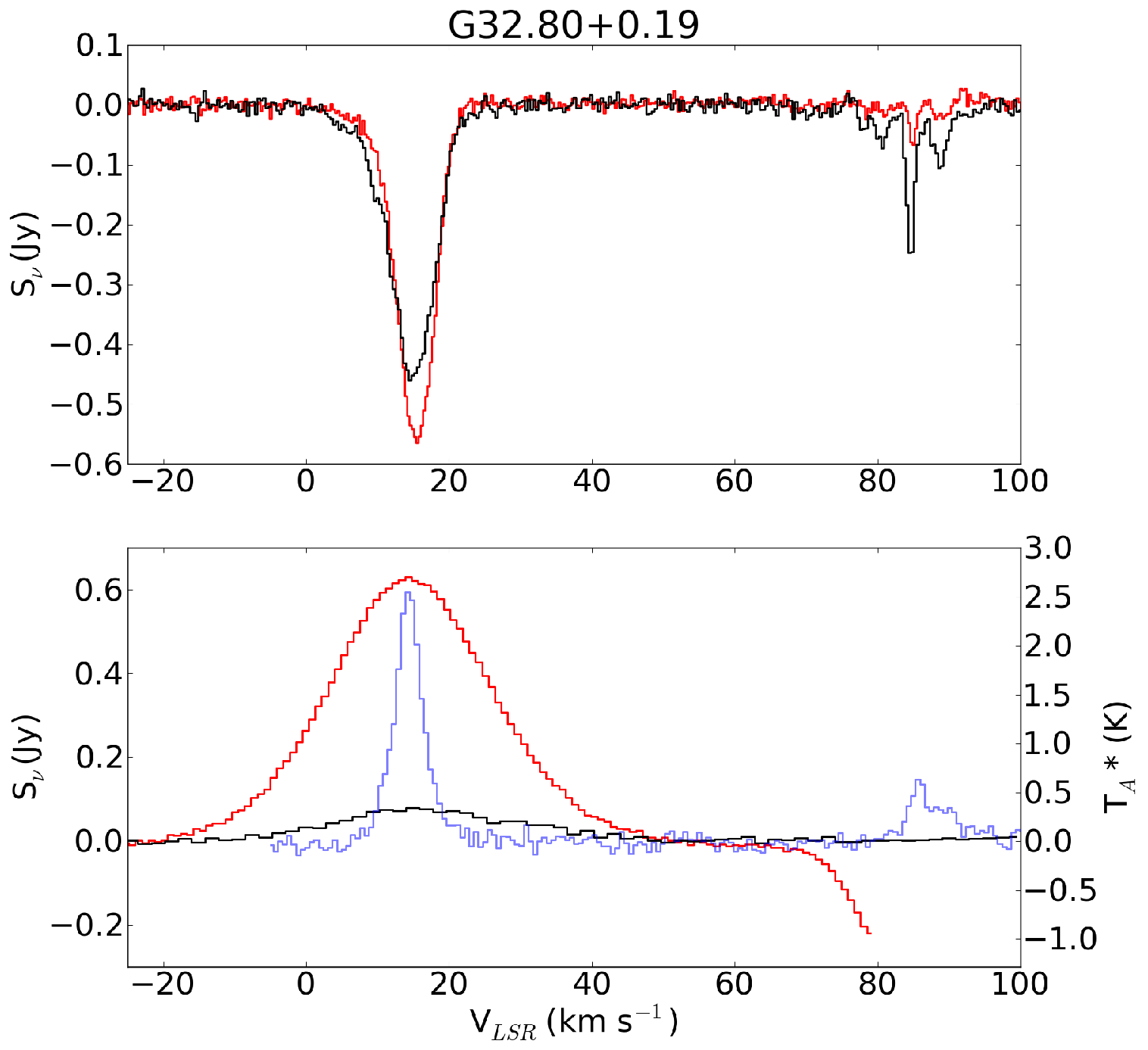}
{ {\it Top:} The GBT \twotwo\ (red) and Arecibo \oneone\ (black) spectra of G32.80+0.19.  
{\it Bottom:} The GBT H75$\alpha$ (red) and Arecibo H110$\alpha$ (black) spectra with the GRS
\thirteenco\ spectrum (light blue) overlaid.   The left axis is for the RRLs and the right
axis is for the \thirteenco.  The C and He RRLs are not displayed.}
{fig:specexample}
{0.30}{0}

\subsection{Other Archival Data}

\subsubsection{Very Large Array}
We acquired VLA archival images from the Multi-Array Galactic Plane Imaging
Survey (MAGPIS) 6 cm Epoch 3 data set \citep{Becker2005} and the NRAO VLA
Archive Survey (NVAS)\footnote{The NVAS is run by Lourant Sjouwerman at the
NRAO.  It has not yet been published.}.  MAGPIS has a resolution of
$\sim4\arcsec$ and sensitivity $\sigma\sim 2.5 $mJy/bm.  The NVAS has variable
resolution and sensitivity since it is based on VLA archival data.  The VLA
data was used to estimate source sizes and interferometer-to-single-dish flux
ratios.

\subsubsection{ Bolocam 1.1 mm }
We extract 1.1 mm dust continuum fluxes from the Bolocam Galactic Plane Survey
(BGPS) v1.0 release summing over a 25\arcsec\ radius aperture after subtracting
the median in a 50-200\arcsec\ annulus to remove background
contributions. The aperture size is selected to match the 1.1 mm data to the 
2 and 6 cm data.  We assume a uniform 50\% systematic error in BGPS fluxes from
combined uncertainties in the calibration and background subtraction.
\citet{Aguirre2011} contains a complete discussion of the uncertainties in the
BGPS. 
We apply the
\citet{Aguirre2011} recommended flux correction of 1.5 and aperture correction
for a 25\arcsec\ aperture of 1.21.  Additionally, data from the Bolocam catalog
\citep{Rosolowsky2010} was used with the flux correction and an aperture correction
of 1.46 for 20\arcsec\ apertures.

\subsubsection{Boston University / Five College Radio Observatory Galactic Ring Survey}
The BU FCRAO GRS \citep{Jackson2006} is a survey of the Galactic plane in the
\thirteenco\ 1-0 line with $\sim 46\arcsec$ resolution.  We extracted spectra in 25\arcsec\ radius
apertures from the publicly available data for comparison with the \formaldehyde\ spectra.

\subsubsection{GLIMPSE}
The Galactic Legacy Infrared Mid-Plane Survey Extraordinaire
\citep[GLIMPSE]{Benjamin2003} maps were used to examine the morphology of the
objects in our survey in order to determine whether an IRDC was present.

\section{Models and Error Estimation}
\label{sec:models}
A grid of large velocity gradient (LVG) models was run using both the RADEX
\citep{VanDerTak2007} code and a proprietary code by \citet{henkel1980} with a 
gradient of 1 \kmspc.  The
models from the two codes were consistent to within $\sim10\%$ in predicted
optical depth and $T_{ex}$.  Both utilized collision
rates from \citet{Green1991} extracted from the LAMDA
database\footnote{\url{http://www.strw.leidenuniv.nl/$\sim$moldata/}} and multiplied by
the recommended factor of 1.6 to account for collisions with \hh\ being more
efficient than He.  The expected accuracy is $\sim30\%$.
\citet{Zeiger2010} demonstrated that the errors in collision rates lead to
systematic errors $\lesssim50\%$  (0.3 dex) in the measured quantities
(N(\formaldehyde), n$(\hh)$).  When measuring density and column, we used the
RADEX models because of their extensively tested code and documentation.  All of the
models used a kinetic temperature of 40 K and covered a range of 500 densities
$\times$ 500 columns logarithmically sampled over $10^1 < n(\hh) < 10^7$ \percc\
and $10^{11} < N(\ortho) < 10^{16} $ \persc.  The assumption $T_K=40$ K is
reasonable in UC\ion{H}{2} regions, which should be warmer than IRDCs and other
cold molecular clouds.  Dust temperatures measured towards UC\ion{H}{2}
regions are around 40 K \citep{Rivera2010}.  In the foreground clouds, this
assumption is less well supported, but as long as the temperatures are higher
than $\sim20$ K, the models change little with temperature (Figure
\ref{fig:modeltau}).

Because of a collisional selection effect, above its critical density
\citep[$n_{cr}(\formaldehyde\ \oneone)\approx 8\ \percc$, \ensuremath{n_{cr}(\formaldehyde\ 
\twotwo) \approx 76\ \percc}, ][]{Mangum2008}
\formaldehyde\ preferentially overpopulates lower states of the K-doublet
\citep[$\Delta$J =0, $\Delta K_a=0$, $\Delta K_c = \pm1$,][]{henkel1980}.  These
spectral lines are cooled to excitation temperatures lower than the CMB and can
therefore be seen in absorption against it.  The \oneone/\twotwo\
absorption line ratio is sensitive to the density of \hh\ at densities
$\gtrsim10^{3.5}\ $ \percc, allowing measurements of the density to within
$\sim$0.3 dex with little sensitivity to gas kinetic temperature
\citep{Mangum2008}.  When density is `measured' with critical density based
tracers such as CO, CS, HCN, or HCO$^+$, the estimate can be off by as much as
2 orders of magnitude because of radiative trapping effects.  Similarly,
measurements of density assuming spherical symmetry can be very far from the
local values.

The collision rates of \formaldehyde\ with \hh\ have been re-derived with a
claimed accuracy of 10\% by \citet{Troscompt2009}.  \citet{Troscompt2009b}
showed that collisions with para-\hh\ are more efficient at cooling
\formaldehyde\ into absorption against the CMB than He or ortho-\hh\ and that
\formaldehyde\ absorption is therefore sensitive to the Ortho/Para ratio of
\hh.   These improved rates are not used in this paper since they are only
computed over a more limited range of temperatures, but may be used in future
works.

\OneColFigure{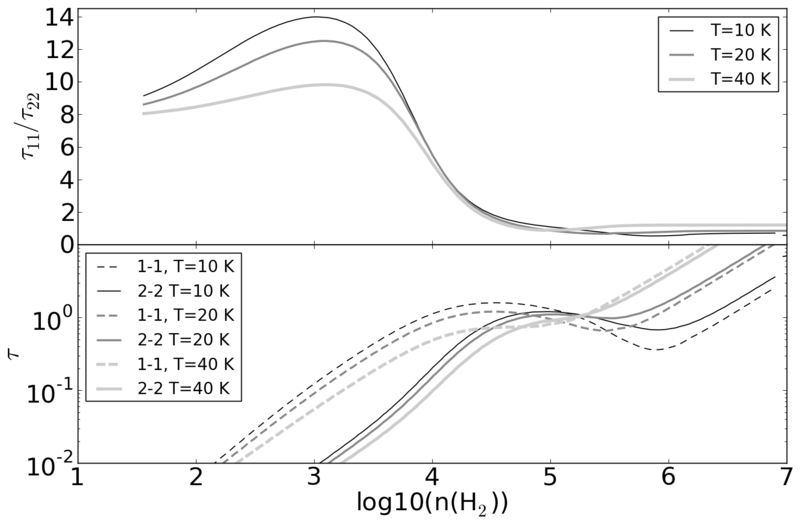}
{The predicted optical depth ratio ({\it top}) and optical depth ({\it bottom}) vs. volume density
assuming a fixed abundance  $X_{\ortho}=10^{-9}$ \perkmspc\ shows that the
dependence of the derived density on temperature is weak.  At lower abundances,
these curves shift to the right, providing sensitivity to moderately higher
densities.  Our 5-$\sigma$ detection limit is generally around $\tau\sim0.01$.
}
{fig:modeltau}
{0.30}{0}

\subsection{Turbulence}
\label{sec:turbulence}
Molecular gas is often observed to have spectral line widths consistent with
supersonic turbulence \citep{Kainulainen2009} and therefore a lognormal density
distribution \citep{Kritsuk2007}.  Our LVG models assume constant density per
velocity bin, so the resulting models should be smoothed by the probability
distribution function (PDF) of the density.  For clouds with a narrow density
distribution (logarithmic standard deviation of the density $\sigma_s \equiv
\sigma_{ln(\rho)/ln(\bar{\rho})}\lesssim0.5$)\footnote{We use $\rho$ to
indicate number density in this section in order to be consistent with the
cited literature.  Because the widths are relative to a mean density, the
scaling between mass and number density is unimportant.}, the effect of
smoothing is smaller than other systematic errors, but for more turbulent
clouds the density PDF width can exceed an order of magnitude \citep[e.g.,
][]{Federrath2010} and will substantially change the derived density.  Because
the Mach numbers of the turbulence in the observed clouds are unconstrained, we
cannot correct for this added uncertainty.  The change in measured density is
$|\Delta\log(\rho)|<0.25$ for $\sigma_s \le 0.5$, with a slight bias towards
higher densities at lower optical depth ratio $\tau_{\oneone}/\tau_{\twotwo}$
(Figure \ref{fig:turbtaurat}).  However, for $\sigma_s=1.5$, the bias exceeds
an order of magnitude at some densities.

Additionally, we consider the effects of ``gravoturbulence'', in which a
high-density tail inconsistent with a lognormal distribution is observed.
\citet{Kainulainen2009} report column density distributions derived from 2MASS
extinction measurements that can be used as a proxy for the density
distribution for a wide variety of clouds.  Non-star-forming clouds retain a
lognormal distribution and are consistent with the analysis presented above.
However, evolved star-forming regions develop a high-column density tail.  For
evolved (actively star-forming) regions like Ophiucus, Orion, and Perseus,
the high-column density tail is substantial, and \formaldehyde\ density measurements
will be highly biased towards the highest density gas.  More quiescent regions
like the Pipe and Coalsack nebulae are consistent with a lognormal column
distribution to a degree that the high-column density tail would not affect
\formaldehyde\ density measurements significantly.

\OneColFigure{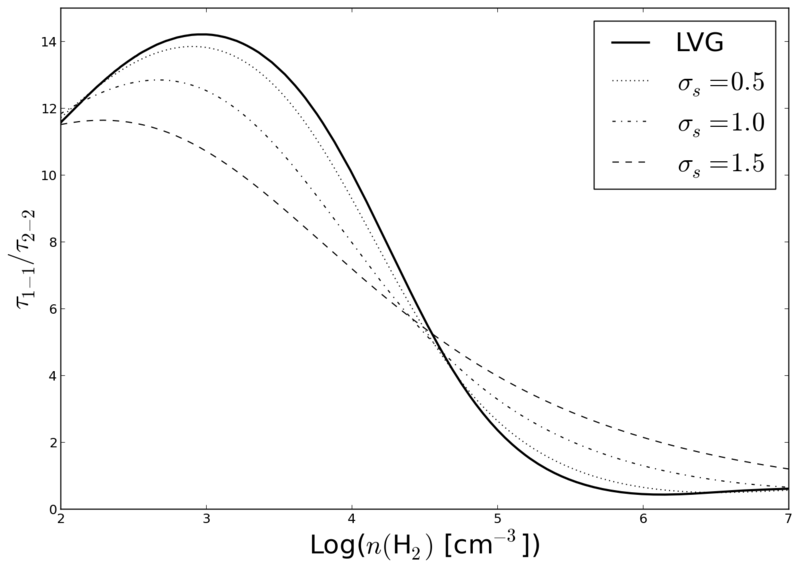} 
{
The optical depth ratio as a function of density for turbulent density
distributions with widths specified in the legend.  The optical depth ratio
varies more slowly with density than in the pure LVG model (the solid line is
the same as the black 10 K line in Figure \ref{fig:modeltau}a).}
{fig:turbtaurat}{0.30}{0}

To demonstrate the effects of turbulent distributions, we calculate the optical
depth ratio as a function of the mean density for three turbulent widths in
Figure \ref{fig:turbtaurat}.  We compare the density that would be inferred
from the spectral line ratio assuming no turbulence (just LVG) to the `correct'
density including turbulent effects in Figure \ref{fig:turbcorr}.  We have also
compared the LVG and turbulent densities to ``gravoturbulent'' density
distributions, in which a power law tail of high-density gas begins at about
$10^{-2}$ times the peak density \citep[e.g., ][]{Klessen2000,KimCho2011}, but
because the density distributions in these simulations are relatively narrow,
the effects of the high-density tail on the measured density are negligible
except for the most turbulent cases.

Figure \ref{fig:turbtaurat} is meant to demonstrate the effects of turbulence,
but it is \emph{not} used to derive densities, since the true density
distribution in observed clouds is unknown.  However, future measurements of
the density distribution can be used to apply the `correction' shown in Figure
\ref{fig:turbcorr}.

\OneColFigure{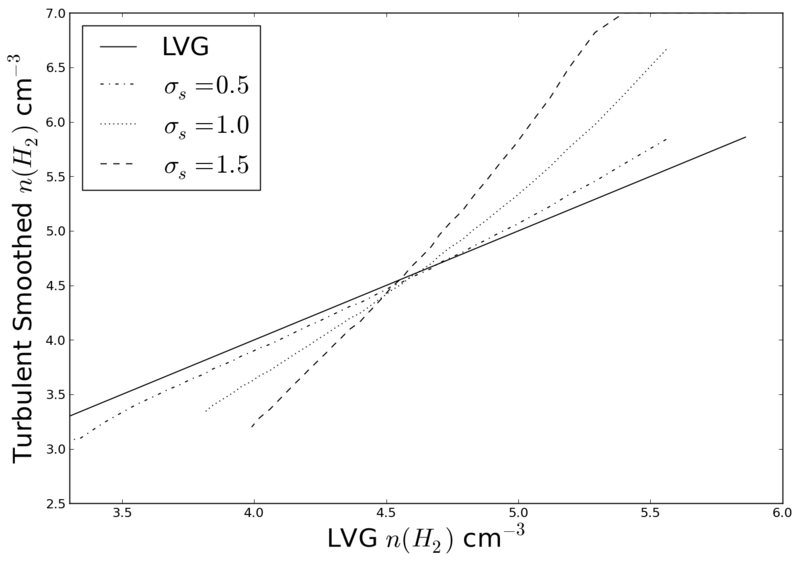}
{ The mean density from a lognormal density distribution plotted against the
density derived assuming a single density per region (i.e., the directly
LVG-derived density).   At low densities, the wider turbulent
distributions  are heavily biased towards ``observing'' higher densities than
the true mean density.  The distributions cut off at the low end where the
optical depth ratio becomes a double-valued function of density; at these low
densities, no detections are expected at our survey's sensitivity.  The cutoff
at the high end is where the optical depth ratio becomes constant.  
}
{fig:turbcorr}{0.30}{0}

\section{Analysis}
\label{sec:analysis}

\subsection{Measuring Line Optical Depth}
\label{sec:linedepth}
In order to measure physical properties of an absorbing source, measurements
must be obtained of the optical depths of both the \oneone\ and \twotwo\ lines.
These measurements are presented in Tables \ref{tab:h2comeasured_a} and \ref{tab:h2comeasured_a}.
Once an optical depth with errors is determined, the spectral line depths can be matched
to large velocity gradient (LVG) models to determine column and spatial
density.  The spectral line optical depth depends both on the nadir flux density of the
absorption line and the strength of the illuminating background continuum
source.  If the background is the CMB, the `filling factor' of the molecular
cloud is simply its size relative to the beam size.  If there is a continuum
source in addition to the CMB, the size of the continuum source and the
intervening molecular cloud both affect the absorption depth.  Throughout this
paper, we use the term `filling factor' to refer to the fraction of the beam area filled
by the absorbing molecular cloud and `covering factor' to refer to the fraction
of the background continuum source that is covered by the intervening molecular
material.

The VLA archival images were used to estimate the size of the illuminating
background source.  When images at both wavelengths were available, we
separately determined the 2 cm and 6 cm source sizes.  The source size
determination is imprecise because we select a single source size for
non-uniform surface brightness profiles, and in many cases the VLA observation
did not recover the full flux density seen in single-dish measurements.
\citet{Araya2002} estimated the interferometer-to- single-dish flux ratio at
6 cm in this sample and found that the interferometer observations recovered
anywhere from 3\% to 100\% of the single-dish flux.  We repeat these
measurements at 2 cm and find that the typical recovery fraction is higher,
$\sim40\%$ to $100\%$, although sources for which only VLA upper limits could
be measured have recovery fractions $<1\%$.

\OneColFigure{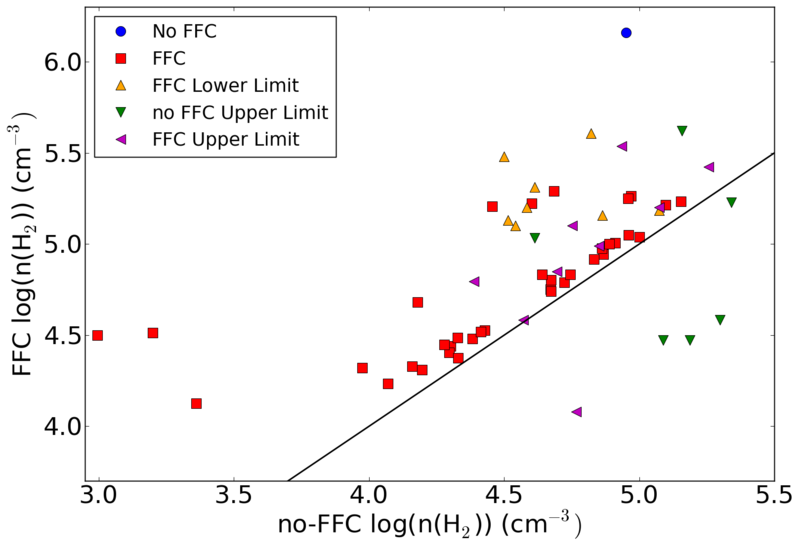} 
{The filling factor corrected (FFC) density vs. the derived density with no
filling factor correction.  While there are some cases where the correction
results in an order of magnitude or more increase in the density, most points
show a small correction.  The black line is the one-one line.  Red squares
show where the filling factor corrected point was used, while blue circles show
where the uncorrected point was used.  Magenta left-pointing triangles are
limits where the filling factor correction was used, green downward triangles
are limits where the uncorrected points were used, and orange upward triangles
are lower limits where the filling-factor correction was used.}
{fig:ffc}
{0.30}{0}

The optical depth measurements were ``filling factor corrected'' by assuming
the CMB only contributed flux density over the same area as the \ion{H}{2}
region (i.e., the foreground cloud covers the exact same patch of sky as the
\uchii\ region).  When the \ion{H}{2} region is small (e.g., 10\% of the beam
area or less), the contribution of the CMB to the continuum is negligible, but
in cases of more diffuse \ion{H}{2} regions, the CMB contribution is
significant, particularly at 2 cm.  The inferred optical depths and source areas
are presented in Table \ref{tab:h2coinferred}.  Both ``filling factor corrected'' and
uncorrected densities are presented in Table \ref{tab:h2coderived}.  The effect
of the filling factor correction (FFC) on density measurements is shown in
Figure \ref{fig:ffc}.  In a few cases, no volume density-column density
parameter space in the models (Section \ref{sec:models}) was consistent with
the spectral line ratio after filling factor correction: in these cases, the
filling factor correction was not used.  Similarly, no filling factor
correction was applied to sources without detected continuum.  These exceptions
are noted in Table \ref{tab:h2coderived} in the ``Flag'' column.

The above definitions are summarized briefly in the following equations:
\begin{eqnarray*}
  S_{\nu,obs} &=& S_{\nu,cont} (1-CF e^{-\tau_{\nu}}) - S_{\nu,CMB} (FF e^{-\tau_{\nu}}) \\
  FF &=& \Omega_{cloud} / \Omega_{beam} \\ 
  CF &=& \Omega_{cloud} / \Omega_{continuum} \\ 
\end{eqnarray*}
in which CF is the ``covering factor'', FF is the ``filling factor'', and there
is no positive contribution from the CMB because it is assumed to be removed by
position-switching.

The systematic uncertainties in the continuum and the filling factor result in
similar errors in the optical depth measurement, and together dominate the
total error budget for our measurements.  A 30\% error in the \oneone\ and 20\%
error in the \twotwo\ continuum levels were assumed because of flux calibration
uncertainty characteristic of the instruments. An additional 10\% error in the beam area, which sets the maximum
coupling to the CMB (assuming a beam-filling source), was included to account 
for focus error.
A 20\% statistical error in the cloud filling factor was assumed for the majority of the
survey, but it was decreased to 10\% when the ratio of continuum to CMB flux was $>0.5$ and the source
size was small, indicating that the VLA-measured source is indeed the dominant
continuum component in the beam.  The statistical error does not account for systematic
errors in the geometric assumptions.  Note that changes to the filling factor
should have a minimal effect on the derived density unless the source sizes at
2 cm and 6 cm differ substantially, while changes in the filling factor will
always have a large effect on the derived column density (Figure
\ref{fig:ffcdependence}).

\OneColFigure{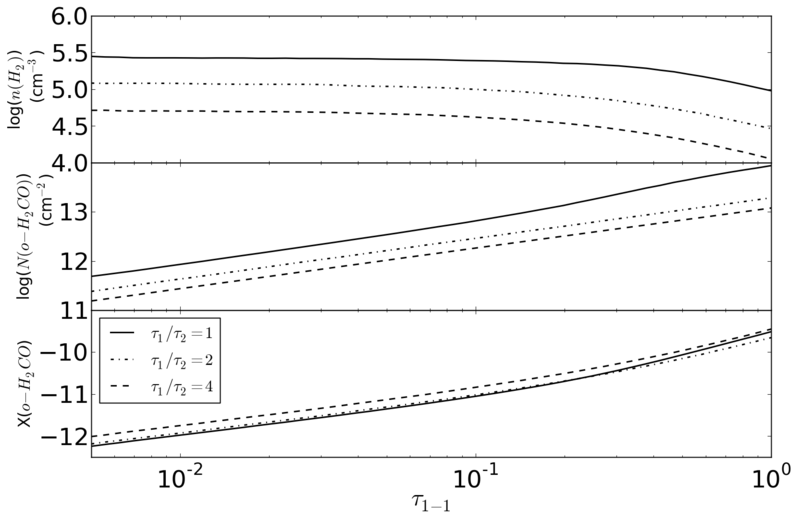}
{The dependence of derived parameters on the filling factor, assuming an
optical depth ratio $\tau_{\oneone}/\tau_{\twotwo} =$1 (solid), 2 (dash-dot),
or 4 (dashed).  
The X-axis is the ``real'' optical depth, $\tau_{1-1}(real) = \tau_{1-1}(observed) / FF$.
Assuming the same filling factor correction is applied to both
the \oneone\ and \twotwo\ lines, filling factor correction will only move the
measurements along the X-axis of these plots.  A decrease in the filling factor
requires an increase in the true optical depth to maintain a constant apparent
$\tau(observed)$, which in turn drives up the derived abundance and column density while
leaving the volume density unchanged (except at high optical depths,
$\tau\gtrsim0.2$).
}
{fig:ffcdependence}{0.30}{0}

Measurements of volume and column density were taken by averaging over the regions of
LVG model parameter space consistent with both spectral line optical depth measurements
to within $1\sigma$.  The ``$1\sigma$'' (68\% confidence; errors are
non-gaussian) error bars on the derived parameters ($N,n,X$) were taken to be
the extrema of these regions.  An example of this fitting process is shown in
Figure \ref{fig:fitexample}.  A second example demonstrating a lower-limit on the density
(instead of a direct measurement) is shown in Figure \ref{fig:fitexample2}.  This method is not as robust as $\chi^2$ fitting,
but because there are no free fit parameters, a statistically meaningful
$\chi^2_\nu$ cannot be computed.

In some cases, the ratio of the spectral line optical depths was consistent with low
density ($n\lesssim 100$ \percc) and high abundances ($X(\ortho)>10^{-8}$ \perkmspc), but
these were ruled out based on the prior assumption that extremely high
\formaldehyde\ abundances should not be observed at very low densities, since
it is formed at higher densities and destroyed by hard UV at low columns
\citep[see discussion in][]{Troscompt2009b}.

\OneColFigure{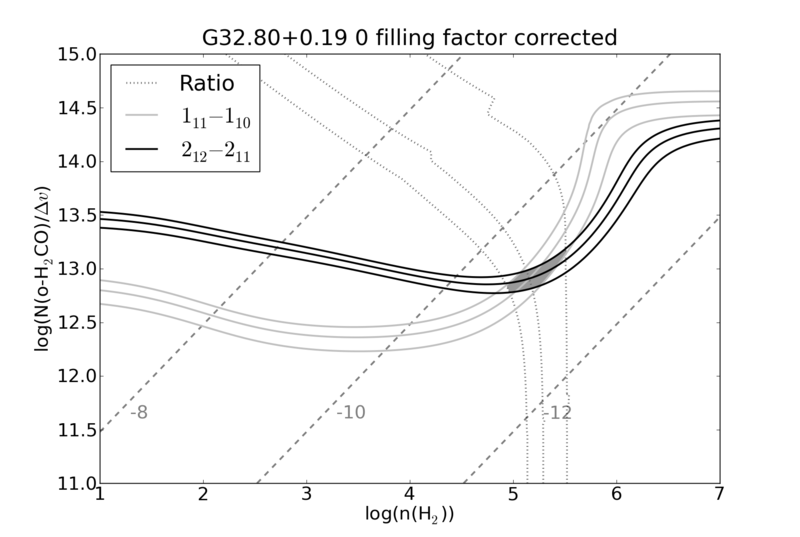} 
{An example of the column density - density parameter space available given
measured \oneone\ and \twotwo\ optical depths.  The dashed lines show
abundances $\log_{10}(X(\ortho))$ \perkmspc.  The contours show the
regions allowed by the measurements of optical depth (\oneone: black, \twotwo: grey,
ratio: dotted);
the middle curve is the measured value, while the pair of curves around it are
$\pm 1\sigma$ including systematic error.  The shaded region shows the allowed
parameter space from which the physical parameters are derived. }
{fig:fitexample}
{0.30}{0}

\OneColFigure{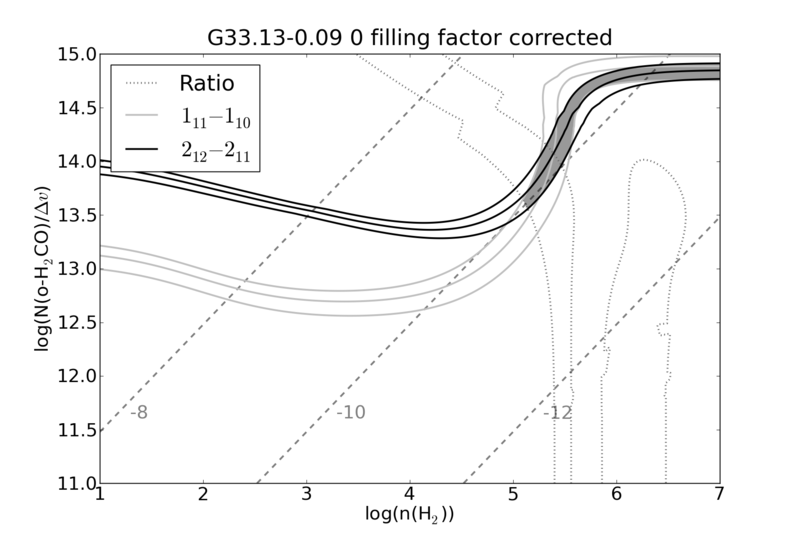}
{  Same description as Figure
\ref{fig:fitexample} but for the strongest component in G33.13-0.09.  It was
only possible to measure lower limits on the volume and column density for this
line; it is therefore assigned flag 8 in Table \ref{tab:h2coderived}.
}
{fig:fitexample2}
{0.30}{0}

\subsection{Systematic Errors: Absorption Geometry}
There are potential systematic errors associated with geometric assumptions,
i.e. the filling factor.  There are four geometric configurations possible;
these are outlined in Table \ref{tab:systematics}.  The ``small source''
geometry (3 and 4) is technically impossible given that the CMB is always
present in these observations, but it is equivalent to the scenario in which
the small illuminating compact source (\uchii) is much brighter than the CMB in
the beam.  The second column shows the effects of applying the `true' filling
factor correction for errors 2 and 4.  For error type 3, the optical depth will
only be overestimated if the absorber is ``corrected'' to be smaller than the
background source (i.e., if a correction is applied when none should have
been).

Figure \ref{fig:ffcdependence} shows the effects of incorrect geometric
assumptions.  Type 1 and 3 errors - i.e. filling factor overcorrections -  will
result in measurements of column and abundance that are \emph{greater} than
the real values, while type 2 and 4 errors will result in column and abundance
measurements that are \emph{lower} than the real values.

Additionally, it is possible that an observation will include a beam-filling,
low-density source that will contribute negligibly in \twotwo\ line absorption
but substantially in \oneone\ absorption over most of the beam area.  This type
of error will result in an underestimate of the volume density.

Since these errors are failures of assumptions, they cannot be quantified, but
Figure \ref{fig:ffc} shows the effects of correcting for these errors to the
extent possible with the available data.

\Table{lcc}{\formaldehyde\ Geometric Systematic Errors}
{\colhead{Real Geometry} & \colhead{Assumed filling factor $= 1$} & \colhead{Assumed filling factor $< 1$} \\}
{tab:systematics}
{
1. Beam-filling source, beam-filling absorber & \tablenotemark{a}$\tau_M=\tablenotemark{b}\tau_R$ & $\tau_M> \tau_R$ \\
2. Beam-filling source, small absorber        & $\tau_M<\tau_R$ & $\tau_M= \tau_R$ \\
3. Small source, beam-filling absorber        & $\tau_M=\tau_R$ & $\tau_M>=\tau_R$ \\
4. Small source, smaller absorber             & $\tau_M<\tau_R$ & $\tau_M= \tau_R$ \\
}
{
\tablenotetext{a}{$\tau_M = $ measured optical depth}
\tablenotetext{b}{$\tau_R = $ real optical depth}
}

\subsection{RRLs}
Radio recombination lines are used to measure the velocity of the \uchii\ regions.
The recombination lines 75-77$\alpha$ were independently fitted with gaussians
because the signal-to-noise in each spectrum with a detection was high. Out of our 24
spectra, there were 21 H detections, 13 He detections, and 12 C detections;
Table \ref{tab:rrls76} shows the fitted parameters using the 76$\alpha$ lines
(75$\alpha$ and 77$\alpha$ were also measured but are not reported for
brevity).  For some of the analysis in later sections, we additionally use the
deeper and more careful RRL study by \citet{Roshi2005}, who observed 17 of our
sample in the 89-92$\alpha$ lines.  We attempted to measure carbon RRLs in the
\citet{Araya2002} spectra, who only measured hydrogen RRLs.  We detected one
carbon line in G61.48 and tentatively ($\sim2\sigma$) detected another three in
G32.80, G34.26, and G45.45; we report the low-significance detections in these
sources because of corresponding detections of C75-77$\alpha$.

We compare the central velocities of the H and C $\alpha$ lines to the
velocities of the \formaldehyde\ absorption lines on a case-by-case basis in
Figures \ref{fig:g33pt13spectrum}-\ref{fig:g61.48+0.09spectrum}.  The spectral line
profiles are used to fit the observations into the models discussed in detail
in Sections \ref{sec:lineprofiles} and \ref{sec:scenarios}.

\Table{lccccccccc}{Measured RRL 76 properties}
{
 & \multicolumn{3}{c}{H}&\multicolumn{3}{c}{He}&\multicolumn{3}{c}{C}\\
\cline{3-3} \cline{6-6} \cline{9-9} \\
\colhead{Source}&\colhead{Peak}&\colhead{Center}&\colhead{FWHM}&\colhead{Peak}&\colhead{Center}&\colhead{FWHM}&\colhead{Peak}&\colhead{Center}&\colhead{FWHM}\\
\colhead{Name}&\colhead{H76$\alpha$\tablenotemark{a}}&\colhead{H76$\alpha$}&\colhead{H76$\alpha$}&\colhead{He76$\alpha$}&\colhead{He76$\alpha$}&\colhead{He76$\alpha$}&\colhead{C76$\alpha$}&\colhead{C76$\alpha$}&\colhead{C76$\alpha$}\\
\colhead{           }&\colhead{(Jy)}&\colhead{(\kms)}&\colhead{(\kms)}&\colhead{(Jy)}&\colhead{(\kms)}&\colhead{(\kms)}&\colhead{(Jy)}&\colhead{(\kms)}&\colhead{(\kms)}\\ }
{tab:rrls76}{
         G32.80+0.19&               0.622&               15.69&               12.09&               0.066&               16.49&                9.25&               0.015&               15.40&                8.27\\
                    &             (0.001)&              (0.03)&              (0.03)&             (0.002)&              (0.36)&              (0.38)&             (0.002)&              (1.45)&              (1.64)\\
         G33.13-0.09&               0.067&               73.49&               14.10&                   -&                   -&                   -&                   -&                   -&                   -\\
                    &             (0.001)&              (0.17)&              (0.17)&                    &                    &                    &                    &                    &                    \\
         G33.92+0.11&               0.157&              101.86&               12.16&               0.013&               99.07&               13.60&                   -&                   -&                   -\\
                    &             (0.001)&              (0.07)&              (0.07)&             (0.001)&              (0.87)&              (0.87)&                    &                    &                    \\
         G34.26+0.15&               0.367&               54.68&               10.43&               0.034&               51.98&                6.54&               0.026&               59.54&                5.66\\
                    &             (0.004)&              (0.06)&              (0.09)&             (0.002)&              (0.46)&              (0.49)&             (0.002)&              (0.55)&              (0.56)\\
                    &               0.251&               37.46&               22.76&                   -&                   -&                   -&                   -&                   -&                   -\\
                    &             (0.003)&              (0.29)&              (0.12)&                    &                    &                    &                    &                    &                    \\
         G35.20-1.74&               1.016&               47.94&               10.70&               0.105&               48.26&                8.27&               0.045&               44.18&                4.05\\
                    &             (0.002)&              (0.02)&              (0.02)&             (0.002)&              (0.21)&              (0.21)&             (0.003)&              (0.33)&              (0.33)\\
         G35.57-0.03&               0.036&               52.38&               13.71&                   -&                   -&                   -&                   -&                   -&                   -\\
                    &             (0.001)&              (0.41)&              (0.41)&                    &                    &                    &                    &                    &                    \\
         G35.58+0.07&               0.044&               46.68&               10.55&               0.007&               43.15&                6.30&                   -&                   -&                   -\\
                    &             (0.001)&              (0.20)&              (0.20)&             (0.001)&              (0.94)&              (0.94)&                    &                    &                    \\
         G37.87-0.40&               0.446&               59.99&               15.47&               0.042&               60.16&               11.88&               0.018&               59.27&                7.93\\
                    &             (0.001)&              (0.08)&              (0.07)&             (0.001)&              (0.55)&              (0.55)&             (0.001)&              (0.98)&              (0.89)\\
                    &               0.049&               26.21&               10.49&                   -&                   -&                   -&                   -&                   -&                   -\\
                    &             (0.002)&              (0.52)&              (0.42)&                    &                    &                    &                    &                    &                    \\
         G41.74+0.10&               0.038&               11.46&               13.89&                   -&                   -&                   -&                   -&                   -&                   -\\
                    &             (0.001)&              (0.29)&              (0.29)&                    &                    &                    &                    &                    &                    \\
         G43.89-0.78&               0.103&               54.98&               10.83&               0.010&               54.18&                7.72&               0.007&               54.08&                0.82\\
                    &             (0.001)&              (0.08)&              (0.08)&             (0.001)&              (0.68)&              (0.68)&             (0.002)&              (0.34)&              (0.30)\\
         G45.07+0.13&               0.041&               58.22&               10.05&                   -&                   -&                   -&                   -&                   -&                   -\\
                    &             (0.004)&              (0.41)&              (0.64)&                    &                    &                    &                    &                    &                    \\
                    &               0.043&               41.57&               20.01&                   -&                   -&                   -&                   -&                   -&                   -\\
                    &             (0.003)&              (1.53)&              (0.59)&                    &                    &                    &                    &                    &                    \\
         G45.12+0.13&               0.461&               58.70&               17.42&               0.039&               59.85&               10.70&               0.023&               59.58&               12.37\\
                    &             (0.002)&              (0.08)&              (0.08)&             (0.005)&              (2.50)&              (1.62)&             (0.003)&              (4.92)&              (3.76)\\
         G45.45+0.06&               0.493&               55.38&               11.80&               0.050&               56.41&                8.02&               0.014&               63.40&               10.42\\
                    &             (0.001)&              (0.03)&              (0.03)&             (0.004)&              (0.93)&              (0.56)&             (0.002)&              (4.57)&              (3.05)\\
         G45.47+0.05&               0.040&               64.01&               14.51&                   -&                   -&                   -&                   -&                   -&                   -\\
                    &             (0.001)&              (0.41)&              (0.41)&                    &                    &                    &                    &                    &                    \\
         G48.61+0.02&               0.076&               16.77&               10.53&               0.007&               16.33&                7.73&               0.006&               19.08&                4.71\\
                    &             (0.001)&              (0.14)&              (0.14)&             (0.001)&              (1.30)&              (1.39)&             (0.001)&              (1.26)&              (1.29)\\
         G50.32+0.68&               0.034&               26.94&               10.27&                   -&                   -&                   -&                   -&                   -&                   -\\
                    &             (0.001)&              (0.27)&              (0.27)&                    &                    &                    &                    &                    &                    \\
         G60.88-0.13&               0.067&               18.30&                9.08&                   -&                   -&                   -&               0.023&               21.77&                2.54\\
                    &             (0.001)&              (0.12)&              (0.12)&                    &                    &                    &             (0.002)&              (0.19)&              (0.19)\\
         G69.54-0.98&               0.017&                3.69&               16.24&                   -&                   -&                   -&                   -&                   -&                   -\\
                    &             (0.001)&              (0.64)&              (0.64)&                    &                    &                    &                    &                    &                    \\
         G70.33+1.59&               0.343&              -19.18&               12.59&               0.032&              -20.14&               10.12&               0.025&              -21.67&                3.28\\
                    &             (0.001)&              (0.05)&              (0.05)&             (0.001)&              (0.44)&              (0.46)&             (0.002)&              (0.33)&              (0.33)\\
         G70.29+1.60&               0.545&              -26.97&               17.82&               0.042&              -26.32&               14.53&               0.032&              -24.78&                4.71\\
                    &             (0.001)&              (0.12)&              (0.09)&             (0.001)&              (0.64)&              (0.68)&             (0.002)&              (0.36)&              (0.42)\\
                    &               0.066&              -64.41&               13.12&                   -&                   -&                   -&                   -&                   -&                   -\\
                    &             (0.002)&              (0.70)&              (0.50)&                    &                    &                    &                    &                    &                    \\
         G61.48+0.09&               0.566&               25.96&               11.16&               0.046&               28.80&                7.86&               0.059&               21.27&                2.48\\
                    &             (0.001)&              (0.02)&              (0.02)&             (0.001)&              (0.24)&              (0.24)&             (0.002)&              (0.11)&              (0.11)\\
}{\tablenotetext{a}{Some H lines were fit with two gaussian components, in which case the second fit component is on the second line below.  Errors (1$\sigma$) are indicated by the numbers in parentheses on the line below the measurement.}}

\section{Results}
\label{sec:results}
\subsection{Derived Properties}
The average properties of the spectral line components associated with the \uchii\ regions
and the other spectral lines representing molecular clouds are
shown in Table \ref{tab:properties}.  The table includes the mean and median
only of spectral lines with both \oneone\ and \twotwo\ detections that yielded
measurements of density; upper and lower limits are not included.  The full
results are presented in Table \ref{tab:h2coderived}.

\Table{cccccccc}
{Inferred properties}
{
          &  \uchii &&& Other Lines (GMC)&&& \\
\hline
Parameter & Median\tablenotemark{a} & Mean \tablenotemark{a} & RMS \tablenotemark{a} & Median \tablenotemark{b} & 
Mean \tablenotemark{b} & RMS \tablenotemark{b} & KS PTE \\}
{tab:properties}
{
log(\hh~Density) (\percc)          &       4.95 &       4.91 &       0.27 &       4.49 &       4.61 &       0.32 &      0.022 \\
log(\ortho~Column) (\persc)         &      12.59 &      12.59 &       0.44 &      11.86 &      11.83 &       0.20 &    6\ee{-6} \\
$X(\ortho)$                         &     -10.84 &     -10.80 &       0.46 &     -11.16 &     -11.26 &       0.45 &      0.028 \\
} {
\tablenotetext{a}{Spectral line components associated with \uchii\ regions}
\tablenotetext{b}{Other spectral lines (associated with line-of-sight molecular clouds)}
}

There is statistical evidence that the deepest spectral line components have higher
\formaldehyde\ column and/or abundance than the other (GMC) components (Table
\ref{tab:properties}).  It is unlikely that this difference could be caused by
underestimates of the optical depths in the GMC components (type 2 and 4
errors, see Table \ref{tab:systematics}) because the filling factor correction
should tend to cancel out these errors.  However, it is possible that, in those
cases where the \ion{H}{2} emission and the CMB emission in the beam are the
same order of magnitude, type 1 errors have occurred: the \ion{H}{2} region
absorber is much larger than the \ion{H}{2} region and a significant fraction
of the spectral line depth comes from absorption against the CMB; this error
should have little effect on the derived density (see Figure
\ref{fig:ffcdependence}) but may lead to overestimates of the derived column
density.  

Each identified Gaussian component was associated with an UC\ion{H}{2} region
if it was within 5 \kms\ of the RRL peak, since RRLs are assumed to be
generated in the UC\ion{H}{2} regions.  Any spectral lines blended with the
\uchii\ \formaldehyde\ lines were also associated with the \uchii\ region.
Other velocity components, including those without corresponding RRL
detections, were assumed to be from GMCs along the line of sight or part of the
larger cloud not directly associated with the UC\ion{H}{2} region; 29
components were associated with UC\ion{H}{2} regions and 46 were associated
with unrelated line-of-sight GMCs (Table \ref{tab:other}).

The density difference between the two populations is significant by a
Kolmogorov-Smirnov (KS) test with $\sim2\%$ probability of being drawn from the
same distribution (the `probability to exceed' or PTE in Table
\ref{tab:properties}). This result is in contradiction to the results of
\citet{Wadiak1988}, who found no significant density difference between ``warm
clouds'' and ``cold clouds'' selected and observed in the same manner (though
with larger beams).  The difference is likely because the larger beam sizes in
their study and a failure to include the continuum contribution of the CMB
(which is more substantial in a larger beam, especially at 2 cm), resulting in
a type 3 error and an underestimate of density for their ``warm clouds'' in
particular.

The measured \hh\ densities do not display any trend with heliocentric distance
over the range 2-14 kpc, contrasing with mm-continuum surveys of star forming
regions that tend to measure lower densities at greater distances
\citep{Reid2010}.  The lack of correlation in Figure \ref{fig:densvsdist}
demonstrates the strength of the \formaldehyde\ densitometry method: the
properties of star-forming gas can be explored throughout the galaxy with
distance bias largely removed.   Similarly, no trend with
Galactocentric distance was readily apparent.

\OneColFigure{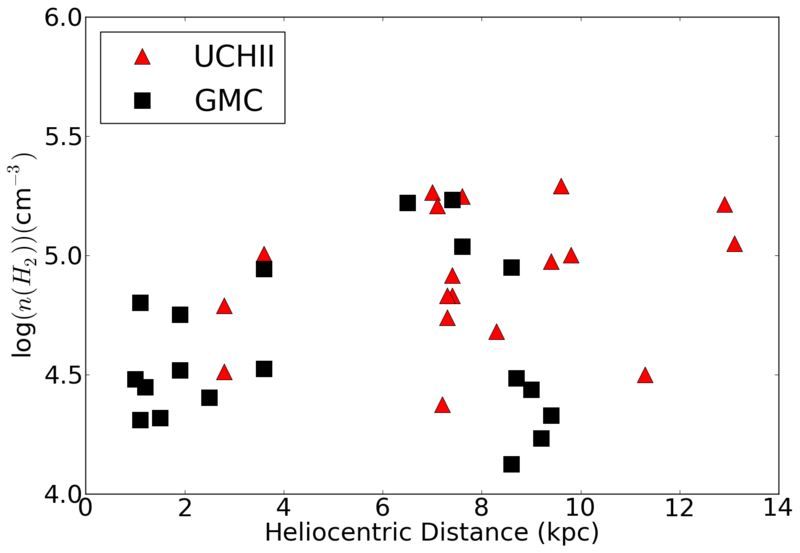}
{Derived density plotted against kinematic distance.  No trend is obvious, demonstrating
that the \formaldehyde\ densitometer is not biased by source distance.
Black squares represent GMCs along the line of sight; red triangles represent
UC\ion{H}{2} regions.}
{fig:densvsdist}{0.30}{0}

Densities were measured within a range $10^4\ \percc\ \lesssim n(\hh)
\lesssim10^6$ \percc\ due to sensitivity cutoffs at low densities and
thermalization of the spectral line ratio (ratio $\rightarrow$ 1) at high densities (see 
Section \ref{sec:strengthsweaknesses} for a discussion of the limitations of the
densitometer).  On the high density end, a lower limit on the density remains
interesting, as densities $n(\hh)\gtrsim10^6$ \percc\ are close to those of low-mass
protostellar cores and are a strong indication of runaway gravitational
collapse, since such high densities are rarely observed in non-star-forming
regions.  On the low density end, it should be possible to detect the \twotwo\
transition with sensitivity improvements $\sim2-10\times$, a consideration that
will govern the allocated time-on-source for future \twotwo\ observations.

\subsection{Free-free Contribution to 1.1 mm Flux Density Measurements}
It is expected that all young star-forming regions should be dust-rich and
therefore bright at 1.1 mm.  We therefore compare the BGPS 1.1 mm, GBT 2 cm,
and Arecibo 6 cm continuum measurements for sources covered by the BGPS in
Figure \ref{fig:MassVsCm}.  For a flat-spectrum \citep[$\alpha \approx -0.1$,
$\tau_{ff}<<1$;][]{rohlfs} free-free continuum source, the 2 cm flux density
should be $1.34\times$ the 1.1 mm flux density.  For an optically thick source,
$S_{1.1 mm} = 330~S_{2 cm}$.

The objects targeted in our survey include 9 of the 13 brightest ($S_{1.1 mm,40\arcsec}>1.5$
Jy) sources in the range $32<\ell<48$, and 11 of 26 with $S_{1.1mm,
40\arcsec}>1.0$ Jy. We use flux density measurements from the 40\arcsec\ apertures in the BGPS catalog
because they are most appropriate for determining peak brightness of point-like
sources \citep{Rosolowsky2010}.  Out of the sample within the BGPS survey area,
6 of 15 sources have free-free fractions of at least 30\%, but potentially much higher if 
the free-free emission is not optically thin.  Since the sample was selected from
well-known \uchii\ regions, these (rather incomplete) statistics are a warning
that most of the brightest 1.1 mm emission sources in the BGPS are likely to be
active \uchii\ regions and therefore may include a significant contribution
from free-free emission to their measured flux densities (Figure
\ref{fig:freefreefraction}).  The same warning applies to other mm-wavelength galactic
plane surveys, though the contamination should be less severe at shorter wavelengths.

\FigureTwo{f10}{f11}
{Bolocam 1.1 millimeter flux density versus the cm continuum flux density at 2
cm (left) and 6 cm (right).  The BGPS 1.1 mm flux density is moderately
correlated with both cm continuum measurements; the legend shows the regression
parameter.  The expectation for optically-thin free-free
emission ( $\alpha = -0.1$, dotted) and for intermediate spectral index emission
($\alpha > 0$, dashed) are shown to illustrate that some sources have
significant free-free contributions at 1.1 mm (the optically thick case is not
shown for either 2 or 6 cm because it does not fit on the plot).
The legend shows the predicted flux densities for a given spectral index
$\alpha$, the regression parameter $r$, and its likelihood $p$.  The brighter
sources are likely to be less optically thick in the free-free continuum than
the faint sources. }
{fig:MassVsCm}{1}

\OneColFigure{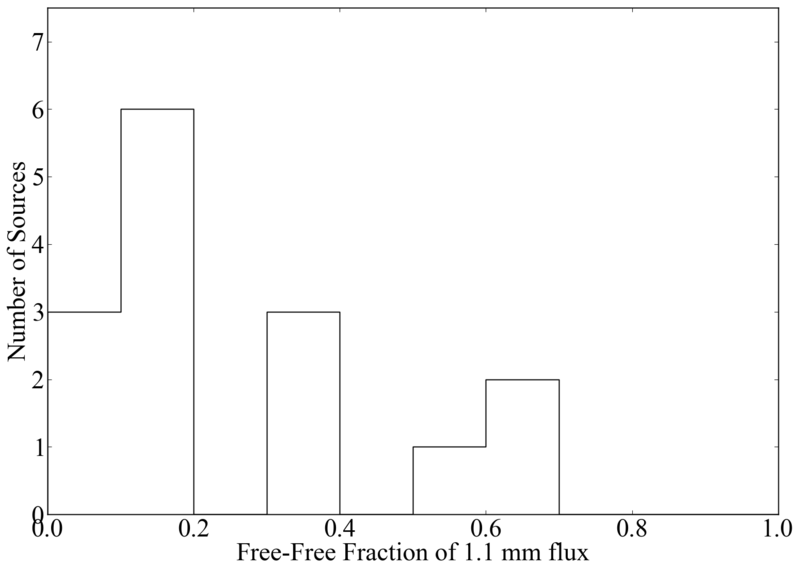}
{The distribution of free-free contributions to the 1.1 mm flux density
assuming the \uchii\ region is optically thin at 2 cm, $f_{ff} = (S_{2
cm}/1.34)/S_{1.1 mm}$.  While 9 sources are either dust-dominated or optically
thick at 2 cm, 6 sources have free-free contributrions of 30\% or greater.  The
other sources in the sample were missing 1.1 mm flux density measurements
because they are outside the BGPS survey area.
}
{fig:freefreefraction}{0.30}{0}

In order to evaluate the impact of this conclusion on the BGPS, we examine the
flux distribution of 6 cm continuum sources from the MAGPIS survey compared to
the BPGS in the same area, $5<\ell<42$ and $|b|<0.42$, which is the full range
of the MAGPIS survey excluding the galactic center, where the BGPS catalog
follows a different flux distribution \citep{Bally2010}.  

In Figure
\ref{fig:contfluxdistr}, we plot histograms of the MAGPIS 6 cm flux density and
the BGPS 40\arcsec\ aperture flux density along with the best-fit power-law
distribution line \footnote{The power law was fit using the python translation
of the \citet{Clauset2009} power-law fitter provided at
\url{http://code.google.com/p/agpy/wiki/PowerLaw}.  The fitter computes the
maximum likelihood value of the power-law $\alpha$ and the cutoff of the
distribution, below which a power law is no longer valid either because of
incompleteness or a change in the underlying distribution.}.  Since the 6 cm
power-law distribution is
shallower than the 1.1 mm distribution, the 6 cm sources can dominate at high
flux densities, although the power-law fit for the 6 cm sources significantly
overpredicts the highest-flux bins and therefore the power-law is not an acceptable fit
above $S_{6 cm}>1$ Jy \footnote{We have tested the consistency of the two data
sets with a low-cutoff power-law distribution by the Monte-Carlo process
described in \citet{Clauset2009}.  The BGPS 40\arcsec\ aperture flux densities
are consistent with a power-law distribution at the $p=0.64$ level, while the
MAGPIS 6 cm fluxes are inconsistent, with $p<0.001$ (where p measures the
probability that the data are drawn from a low-cutoff power-law distribution) }.
The dashed line in Figure \ref{fig:contfluxdistr} shows the best-fit power-law
distribution of the MAGPIS flux densities scaled down by
0.67, which is the expected decrement for an optically thin free-free source 
from 6 cm to 1.1 mm
(spectral index $\alpha=-0.1$).  

Figure \ref{fig:contfluxdistr}b shows the
ratio of the BGPS to the MAGPIS best-fit power-law distribution, indicating
that the free-free contamination fraction is only large ($\sim10\%$) at values
much greater than the valid range of the 6 cm power law fit, which overpredicts the
number of sources at $S_{6 cm} \approx 1 $ Jy.  However, if any
of these sources are \emph{not} optically thin at 6 cm, this fraction could be
much larger.  Additionally, these numbers only describe the sources in which
\emph{all} of the 1.1 mm flux is free-free emission; the implication remains
that a large number of 1.1 mm sources have a substantial (if not dominant)
free-free contribution.  

Finally, we emphasize that unless a large fraction of 6 cm sources are
optically thick in free-free continuum, the lower flux-density BGPS
dust-continuum sample should be negligibly contaminated by free-free emission
sources, but the brightest BGPS sources may have a significant free-free
contribution.

\FigureTwo{f13}{f14}
{{\it Left:} Histograms of BGPS 1.1 mm 40\arcsec\ aperture flux densities (red)
and the MAGPIS 6 cm flux densities (black), and their respective best-fit
power-law distributions ($\alpha(1.1 mm)=2.41\pm0.03$, $\alpha(6
cm)=1.72\pm0.03$).  The dashed black line shows the MAGPIS best-fit power-law
scaled down to the expected flux density at 1.1 mm assuming all sources are
optically thin.  Both distributions appear to be reasonably well-fit by
power-laws above a cutoff (presumably set by completeness), although the power-law
significantly over-predicts the number of sources with $S_{6 cm}>1$Jy.  The
histograms are binned by 0.1 dex, and while the best-fit $\alpha$ and $x_{min}$
values are independent of the binning scheme, the normalization is not.
{\it Right:} The ratio of the number of MAGPIS 6 cm sources to BGPS 1.1 mm
sources as a function of flux density for the best-fit power laws.  Only 10 1.1
mm sources are detected above 5 Jy (in 40\arcsec\ apertures), so even the
brightest detected 1.1 mm sources are not purely free-free, but they probably 
have a substantial free-free component.}
{fig:contfluxdistr}{1}

\subsection{Distances}
\label{sec:distances}
We measure a kinematic distance to each source using the \citet{Reid2009}
rotation curve.  We resolved the Kinematic Distance Ambiguity (KDA) towards
each line of sight using a variety of methods described below.  The method
in \citet{Sewilo2004} allows a resolution in favor of the far
distance for \uchii\ regions with an intervening molecular absorption line at
more positive velocities in the first Galactic quadrant.  Associations with
infrared dark clouds (IRDCs) can resolve the KDA in favor of the near distance.
We compare our KDA resolutions to \citet{Anderson2009}, with whom we agree on
all common sources except for G33.13-0.09, which we place at the far distance
based on the \citet{Sewilo2004} method.  The derived distances are listed in
Table
\ref{tab:other}.

\subsubsection{Size Estimates}
\label{sec:sizecomp}

We estimate the source size using two methods.  First, we use the VLA
measurements of \uchii\ region sizes.  As stated in Section
\ref{sec:linedepth}, the VLA size measurements are very uncertain and are
simplifications of an evidently complicated geometry.  We estimate a spherical
radius $r=\sqrt{area/\pi}$.  Second, we assume the gas traced by \formaldehyde\
and the BGPS 1.1 mm images are the same and get a `size scale'
$r=2 N_{mm}(\hh)/n(\hh)$ where $n(\hh)$ is derived from the \formaldehyde\ line
ratio.

The sizes derived from the two methods are plotted against each other in Figure
\ref{fig:sizecomp}.  The sizes estimated from the two different methods are not
well correlated and disagree by around an order of magnitude in most sources.
The disagreement could be because of poor VLA-based size estimates, substantial
1.1 mm emission from low-density gas, or incorrect dust temperature or opacity
estimates.  While additional line-of-sight GMCs could in principle contribute
to the $N/n$ size estimate, the disagreement for sources even without
associated GMCs prevents this hypothesis from fully explaining the
disagreement.  Therefore, any quantities derived from the size - i.e.
mass, which depends on $r^3$ - are even less constrained.  We therefore do not
derive any quantities dependent on the intrinsic source size.

\OneColFigure{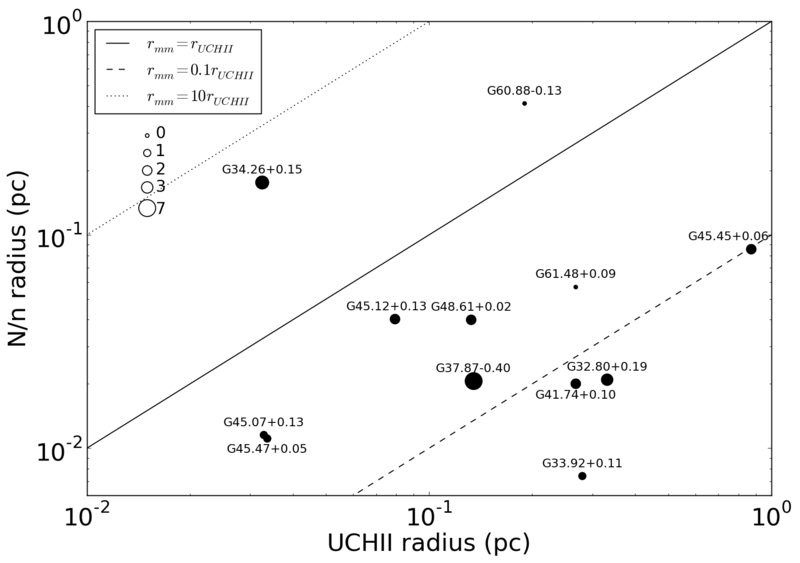}
{A plot of the two derived sizes discussed in Section \ref{sec:sizecomp}.  The
two size estimates are at best very weakly correlated.  Because of the
substantial disagreement between the two methods, we choose not to explore any
parameters with a strong dependence on the size.  The plotted point size
indicates the number of associated line-of-sight GMCs, which in principle could
lead to an overestimate of the $N/n$ size because of additional mass included in
the 1.1 mm continuum measurement.}
{fig:sizecomp}{0.30}{0}

\section{Discussion}
\label{sec:discussion}
\subsection{Comparison to extragalactic observations}
\label{sec:exgal}

We compare our measured column and volume densities to a selection of starburst
galaxies from \citet{Mangum2008} in Figure \ref{fig:exgalcolden}.  All of the
extragalactic observations have much lower column densities per \kms\ than we
measure in the main lines of most \uchii\ regions, but similar volume densities.
This discrepancy can be easily explained by a difference in the area filling factor
of molecular clouds in observations of galaxies and \uchii\ regions.  In a
galaxy, the total area filling factor of molecular clouds per \kms\ (which is
similar but not identical to the volume filling factor) is likely
to be $<1$, even in extreme starbursts; although the galaxy may appear to be
uniformly filled with molecular gas in projection, at any given velocity it is
likely to have significant gaps of ionized or neutral atomic gas.  In contrast,
an \uchii\ region should be completely embedded in a molecular cloud that is
much larger than the free-free emitting continuum region, so the covering
factor of molecular gas should be $\sim1$.
 
It is therefore interesting to note that Arp 220, possibly the most extreme nearby
example of a starburst galaxy, has nearly the same column per channel as the
low end of the \uchii\ regions, suggesting that it is analagous to a scaled-up
\uchii\ region to within a factor of a few; the measured density in Arp 220 is
consistent with only the highest-density \uchii\ regions.  M82, on the other
hand, has a bright continuum background analagous to an \uchii\ region, but a
correspondingly low filling factor, implying that it consists of many 
compact but bright sources with a total filling factor 0.001-0.1.
Alternatively, the density and column measurements are consistent with M82
being dominated by quiescent GMCs, but that is unlikely given the starburst
nature of the galaxy. 

The gravitational lens source B0218+357 is a different scenario.  Its low
density is consistent with that of a non-star-forming GMC, while its column per
\kms\ is comparable to the Galactic sample.  This source is therefore likely to
be a sightline through a `normal' quiescent molecular cloud in its host galaxy,
similar to the narrow beam of an \uchii\ region through the Galactic disk.
\citet{Zeiger2010} note that there is a range of covering factors cited in the
literature, which can affect the measured density and column, but should not
affect the conclusion that the B0128+357 cloud's density is not consistent with
that of massive-star forming regions.  The low-density gas is detected partly
because the \citet{Zeiger2010} data are 3.5$\times$ more sensitive than ours 
with a background continuum source of similar brightness.

\OneColFigure{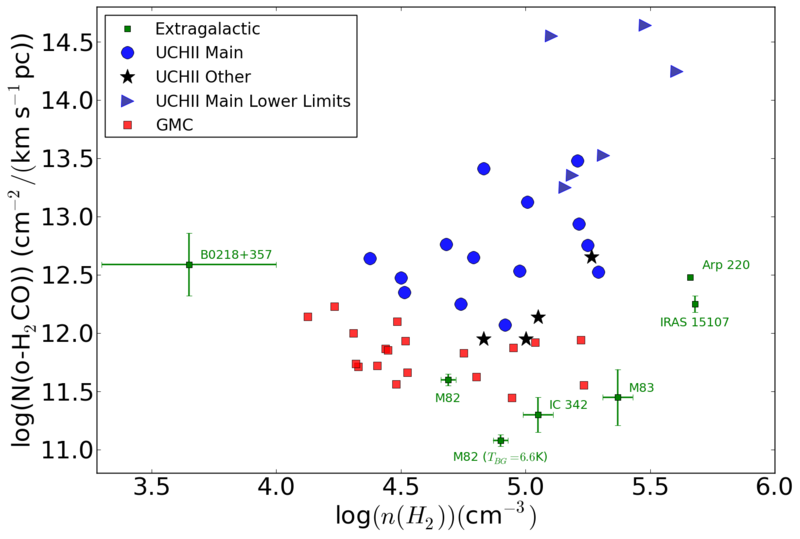}
{Comparison of the UC\ion{H}{2} sample (blue circles are measurements, blue
triangles are lower limits on volume density with poorly constrained column
densities), the GMC sample (red squares), secondary lines associated with
\uchii\ regions (black stars) and the extragalactic sample of
\citet{Mangum2008} (green squares).  The errorbars on the
Galactic data points are excluded for clarity.  The observed galaxies have
similar densities to the Galactic \uchii\ sample, but significantly lower
column densities, suggesting that the molecular gas in these galaxies has a
filling factor $<<1$.  The lack of direct density measurements of UCHII regions
at high densities is due to the presence of a dominant background source; in Arp 220 a
direct measurement of density was possible because \formaldehyde\ was seen in
emission.}
{fig:exgalcolden}
{0.30}{0}

\subsection{Line Profiles}
\label{sec:lineprofiles}
Despite the many systematics discussed above that can affect \formaldehyde\
absorption measurements with a compact illumination source, it is possible to
directly compare the properties of gas along a given line of sight without most
of these hindering factors.  Since most of our spectra have kinematically
resolved spectral line profiles, it is possible to make many density measurements at
different velocities towards each source.  An example of this type of analysis
is shown in Figure \ref{fig:g3280densspec}.  An example demonstrating the need for
this type of analysis is shown in Figure \ref{fig:G70compare}, in which two lines
well-fit by gaussian profiles nonetheless display a density gradient because the
line centers are significantly offset; the figure also demonstrates that the offset
cannot be accounted for by any instrumental effects.

Of our sample, 18 of the 24 observed lines-of-sight had high enough signal-to-noise spectra 
(S/N$\gtrsim5$ in at least four adjacent 0.4 \kms\ channels in both lines) to
measure the density in many velocity bins.  Of these, 12 have different peak
velocities in the \oneone\ and \twotwo\ lines, indicating density gradients in
the molecular gas with velocity.  Figure \ref{fig:g3280densspec}b is an example
density-velocity plot.

We have classified each high S/N spectrum as {\it gradient}, {\it envelope}, or
{\it single} based on spectral line morphology.  The {\it gradient} classification was used
for gaussian or nearly gaussian lines in which the \oneone\ and \twotwo\ line
centers were offset, indicating a gradient in the density with velocity; the
color listed in the table indicates the direction of \emph{increasing} density.  The {\it
envelope} classification was used for flat profiles on the wings of deeper
gaussian lines.  The {\it single} classification was used for lines where the
\oneone\ and \twotwo\ velocities matched.  Low S/N spectra were not classified.
Classifications are given in Table \ref{tab:other}.

Of the 12 sources with density gradients, 6 show an increased density towards
the red and 5 towards the blue.  One source, G45.12+0.13, shows a slight
increase towards the red over a broad (8 \kms) velocity range, but a sharp
increase towards the blue over only 1 \kms\ and is therefore classified as {\it
other}.

Figures \ref{fig:g33pt13spectrum}-\ref{fig:g61.48+0.09spectrum} show the `main
line' (associated with the \uchii\ region) profile and the associated density,
column, and abundance velocity profiles.  The density, column, and abundance
measured for each main line via the gaussian fit technique are shown
overplotted on the profiles with blue squares.  In all cases, the gaussian fit
measurement of density is consistent with the individual channels nearby and
the gaussian fit measurements of column and abundance are consistent with the
peak column and abundance.  The consistency of adjacent velocity bins confirms
the validity of associating gaussian components in observations of whole
galaxies \citep[e.g., ][]{Mangum2008} or kpc-scale regions, since on these scales
the \oneone\ and \twotwo\ lines should be blended by kinematics to have the
same shape.

\Figure{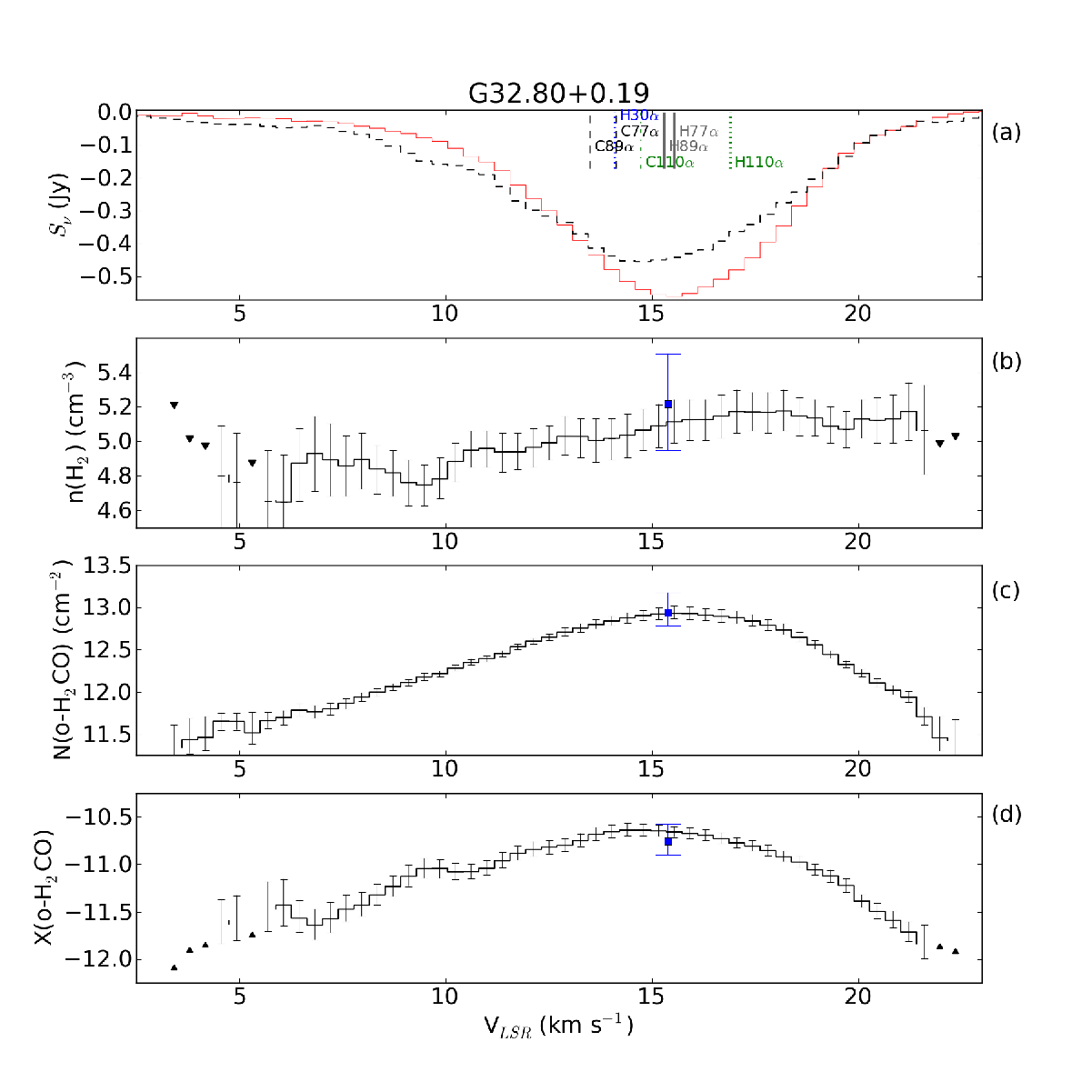}
{Plot of the derived parameters per velocity bin for the main line of
G32.80+0.19; the full spectrum is shown in Figure \ref{fig:specexample}.  The
density peak around 16 \kms\ is slightly redshifted of the H and C RRL velocity
centers, although the C RRLs are blueshifted of the H
RRLs, indicating that the PDR has been accelerated towards us along the line of
sight.  The blueshifted emission tail is suggestive of an outflow.  This source
cannot therefore be easily classified under any of the scenarios in Section
\ref{sec:scenarios}, but is consistent with components of scenarios 2 and 3. \\
{\it a.} The spectra of G32.80+0.19.  The GBT \twotwo\ spectrum (red solid) has
been smoothed to a resolution of 0.38 \kms\ to match the Arecibo (black dashed)
spectral resolution.   Labeled vertical bars indicate the measured velocity centers
of H and C RRLs from this work, \citet{Roshi2005}, and \citet{Churchwell2010}. \\
{\it b.} The measured densities in each spectral bin.  The Y-scale is in log$_{10}$
units. Error bars include a 10\% systematic uncertainty in the continuum and
therefore errors in adjacent channels are not independent.  Limits are
indicated by triangles.  Bins with no information above the 1-$\sigma$ noise
cutoff are left blank.  The increase of density towards higher velocities led us
to classify this source as a {\it red gradient} in Table \ref{tab:other}. \\
{\it c.} The measured column densities per spectral bin.  Because these column
densities are derived from a large velocity gradient code, they are in
\perkmspc\ units.\\
{\it d.} The measured abundances per spectral bin.  The column and abundance are 
somewhat degenerate, but it is possible in some cases to place tight constraints
on the total \ortho\ column while only placing upper limits on abundance
and density.  The abundance must also be interpreted \perkmspc. \\
In plots {\it b} through {\it d}, the blue square with error bars
represents the measured value from Table \ref{tab:h2coderived} using gaussian
fits to the lines.}
{fig:g3280densspec}
{0.5}{0}

\OneColFigure{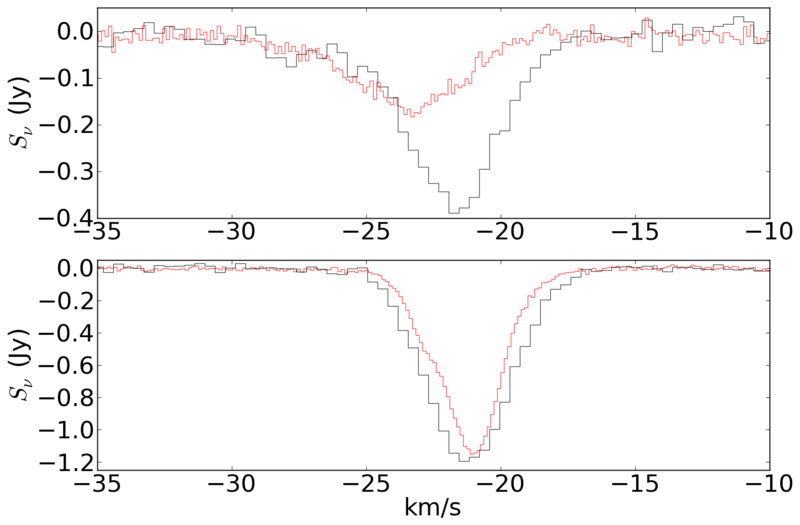}
{Comparison of G70.29+1.60 (top) and G70.33+1.59 (bottom) spectra as observed
by Arecibo (black) and GBT (red/grey).  Note that in G70.29+1.60, the \twotwo\
line is shifted towards the blue of the \oneone\ line, while in G70.33+1.59 the
line centers match well.}
{fig:G70compare}
{0.3}{0}

\subsection{Comparison of RRLs and \formaldehyde\ lines}
\label{sec:scenarios}
We compare the density spectra with the fitted RRL centroids and attempt to
interpret these spectra in the context of various simple models of \ion{H}{2}
region interaction with molecular clouds.  The simple models described below
may actually be short-lived but recurring stages in the normal life cycle of a
collapsing clump that is forming massive ($M\gtrsim10 \msun$) stars
\citep{Peters2010}.

We consider five simple models of embedded \uchii\ regions.  For each scenario,
we include a brief description of the model and an analysis of the
observational consequences in terms of C and H RRL velocities and
velocity-density structure.  We assume that the carbon RRLs are only detected
if seen in the foreground of a bright source.  This assumption is based on
predictions that C RRLs will be amplified by an order of magnitude even in the
presence of a weak background \citep{Natta1994}.   It is backed by a strong
correlation between the continuum and the C RRL intensity  \citep{Roshi2005}.
We also assume that lower-frequency RRLs will have a stronger stimulated
emission component than higher-frequency RRLs \citep{Lockman1978}.  All
\formaldehyde\ absorption is assumed to be against the \uchii\ region in this
section.  The scenario that describes a given spectrum is listed in the figure
caption for each spectrum and in Table \ref{tab:other}.

\OneColFigure{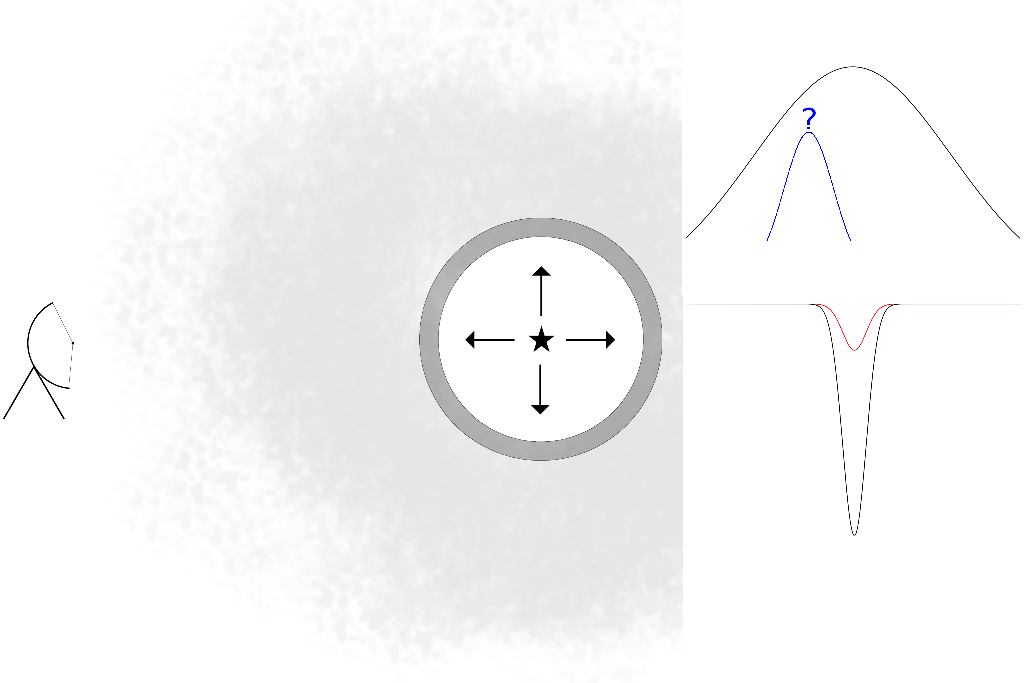}
{Scenario 1: An \uchii\ region forms and begins expanding spherically
in a uniform density gas cloud.  A cartoon of the geometry seen by the observer
is shown on the left side of the figure, with arrows indicating expansion and
darkness of the gray shading indicating relative density.  The white region
around the central star is the ionized \uchii\ region.  On the right side, a
cartoon of the relative velocity and width of the RRLs and \formaldehyde\ lines
is shown.  The relative heights of the \formaldehyde\ lines is representative
of the observed density;  black is \oneone\ and red is \twotwo.  The narrow
emission line with a ? above it indicates a possible blueshifted carbon RRL;
its height has no physical meaning.  In this scenario, the hydrogen
recombination and \formaldehyde\ lines should occur at the same velocity, and
the \formaldehyde\ lines should show relatively low-density (high
\oneone/\twotwo\ ratio) and modest spectral line widths.  A blueshifted carbon
RRL may form, but is not guaranteed.}
{fig:scenario1}
{0.15}{0}

\OneColFigure{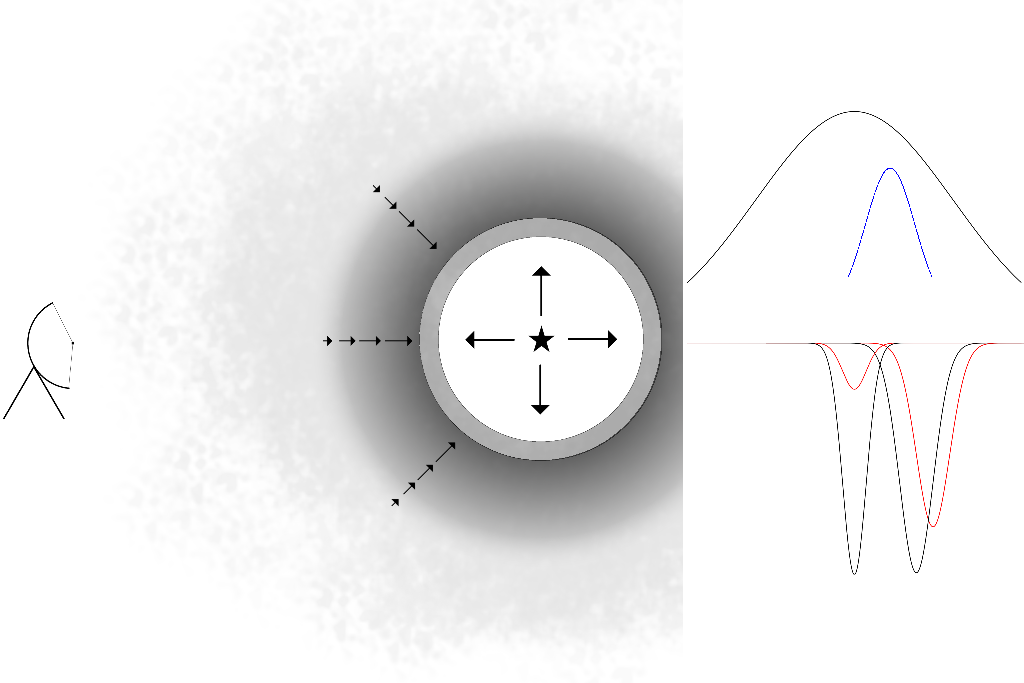}
{Scenario 2: An \uchii\ region forms from a gravitationally unstable cloud
undergoing inside-out collapse.  See Figure \ref{fig:scenario1} for a complete
description of the figure.  The highest density should correspond to the
highest-velocity infall, so the \twotwo\ line peak should be redshifted of the
\oneone\ line peak.  The hydrogen recombination line may align with a
low-density cloud but should be blueshifted of the infalling gas.  The carbon
RRL should be redshifted from the hydrogen RRL and blueshifted from the
\formaldehyde\ line.}
{fig:scenario2}
{0.15}{0}

\OneColFigure{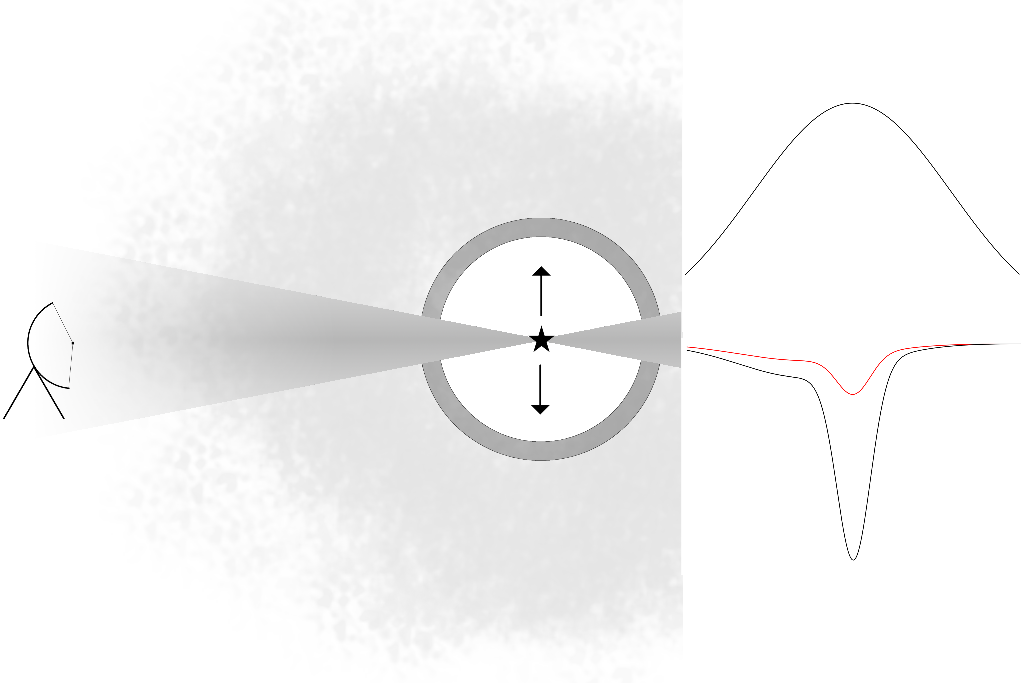}
{Scenario 3: An \uchii\ region expanding in a uniform medium ejects a bipolar
outflow.  Presumably the bipolar outflow comes from a disk-accreting source.
See Figure \ref{fig:scenario1} for a complete
description of the figure.
The outflow (indicated by the cones emitting from the central source) should
have lower column density but could have high or low volume density.  It will be 
observed as high-velocity blueshifted absorption in a line wing.  Carbon
recombination line emitting regions may be destroyed by the outflowing
material.  As in the simple scenario 1, the hydrogen recombination line should
be at the same velocity as the molecular cloud.}
{fig:scenario3}
{0.15}{0}

\OneColFigure{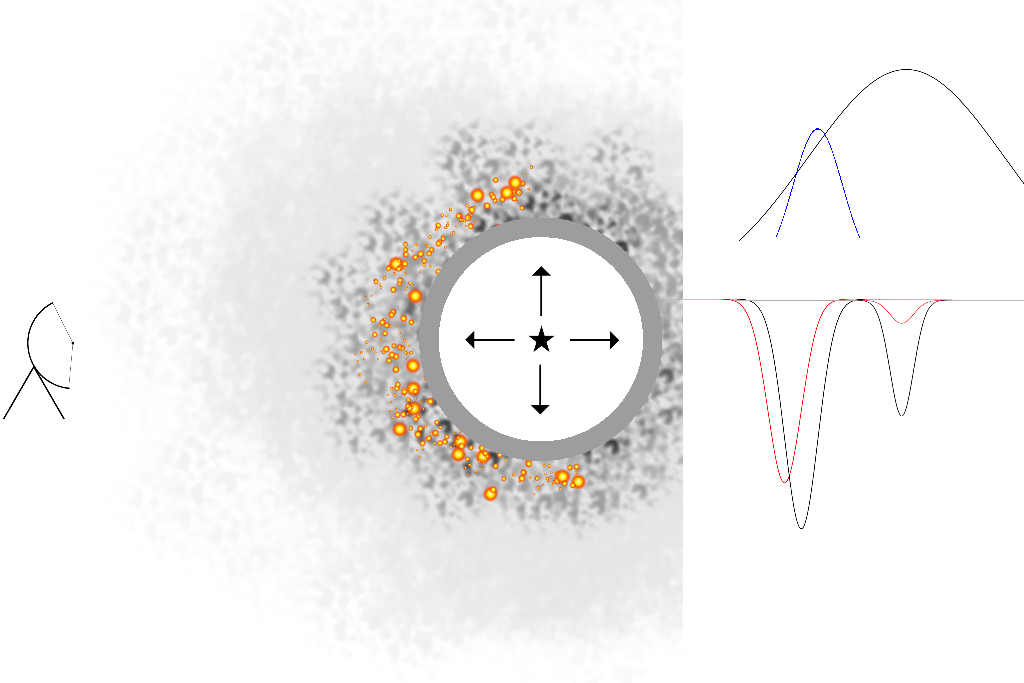}
{Scenario 4: An \uchii\ region expanding in a uniform medium sweeps up and
accelerates material that undergoes triggered star formation.  Because the
highest-density material is the swept up material, it should be the most
blueshifted.  See Figure \ref{fig:scenario1} for a complete description of
the figure.  The orange and yellow circles are meant to indicate triggered star
formation.}
{fig:scenario4}
{0.15}{0}

\OneColFigure{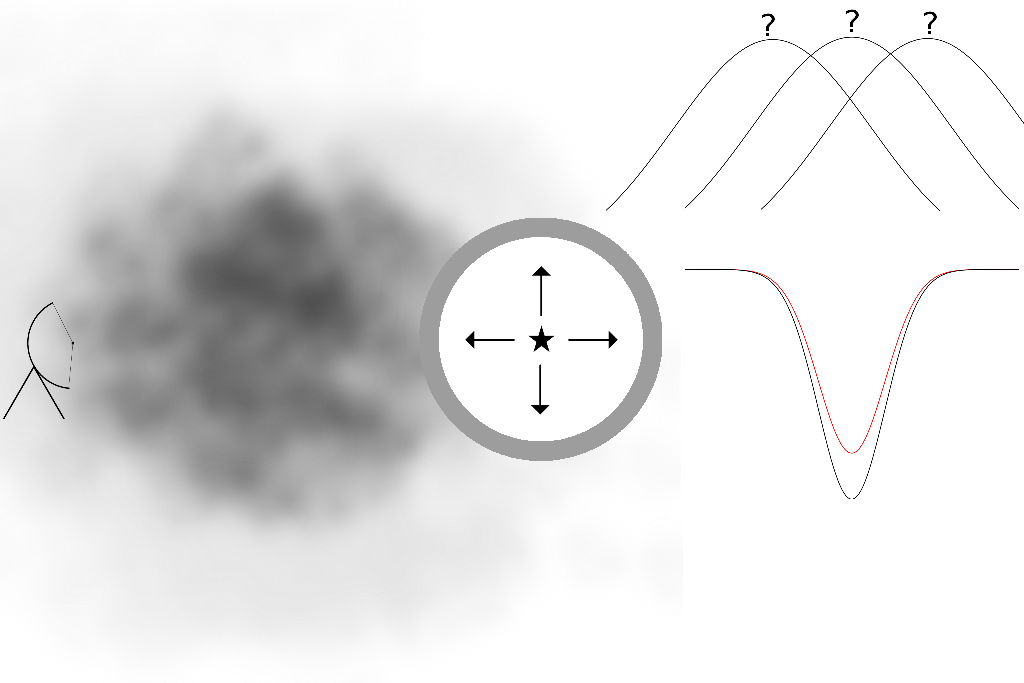}
{Scenario 5: An \uchii\ region is seen behind a high-density, turbulent gas
cloud.  The turbulence drives large spectral line widths, while the high density makes
the \oneone\ and \twotwo\ line depths very close.   The RRL velocity could in
principle be at any velocity relative to the foreground turbulent cloud.  See
Figure \ref{fig:scenario1} for a complete description of the figure.  In this case,
the ?'s indicate an uncertain velocity for the hydrogen RRLs; a carbon RRL is not
expected because the \ion{H}{2} region is not necessarily interacting with the
molecular gas.}
{fig:scenario5}
{0.15}{0}

\begin{itemize}
    \item SCENARIO 1: STATIC
        \\* In a uniform medium with no bulk motions (i.e., no collapse), a massive star
      ignites and generates an expanding \ion{H}{2} region.  Figure \ref{fig:scenario1}.

  \begin{enumerate}
    \item Lower frequency RRLs are blueshifted from higher-frequency RRLs
      because of an increased stimulated emission component \citep{Lockman1978}
    \item A carbon RRL should be seen at the same velocity as or blueshifted from
      the hydrogen RRL line center. 
    \item Molecular gas closest to the \ion{H}{2} region should have the
      highest density because of compression by the expanding \ion{H}{2}
      region.  It will be at a similar velocity or blueshifted from the H RRLs.
\end{enumerate}

  \item SCENARIO 2: COLLAPSE
      \\* A massive star ignites while spherically accreting from a
    molecular cloud undergoing bulk (inside-out) collapse.  Figure \ref{fig:scenario2}.

  \begin{enumerate}
    \item The \formaldehyde-measured density should peak at the velocities
      most redshifted relative to the hydrogen RRLs.  Inside-out collapse
      dictates that the highest densities should be infalling at the highest
      speeds.
    \item The C RRL velocity should be between the \formaldehyde\ and H RRL
      velocity since the PDR will be decelerated by radiation and gas pressure
      from the \ion{H}{2} region
    \item Since the accreting star should be at approximately the rest
      velocity of the cloud, there should be little to no gas blueshifted from
      the RRL velocity
\end{enumerate}

  \item SCENARIO 3: OUTFLOW
      \\* An accreting massive star generates a massive outflow with
    a significant component along the line of sight. Figure \ref{fig:scenario3}.

  \begin{enumerate}
    \item Substantial low-column, low-abundance \perkms\ gas should be observed
      at velocities blue of the RRL velocities.  Densities can range from low
      to high.  Covering factors may be low.
    \item No carbon RRL is expected from the outflow, though if the flow is
      accelerated by ionization pressure a C RRL should be observed blueshifted of
      the H RRL velocity.
\end{enumerate}

\item SCENARIO 4: SWEEPING
    \\* An expanding \ion{H}{2} region pushes on a low-density
  envelope, possibly triggering a new stage of star formation as in the ``collect and
  collapse'' scenario.  This scenario is similar to \#1 but with either a
  higher-density envelope or with more gas swept up (i.e., \#4 may represent
  a more evolved region).  Figure \ref{fig:scenario4}.

  \begin{enumerate}
    \item The hydrogen RRLs should be red of the dense gas and the carbon
      RRLs.  The expanding \ion{H}{2} region should accelerate the dense gas
      blue along the line of sight.
    \item A low-density envelope should persist at the same velocity as the
      \ion{H}{2} region
\end{enumerate}

\item SCENARIO 5: FOREGROUND CLUMP
    \\* A high density, highly turbulent or high mass and rotating clump of gas
  is in front of the \uchii\ region or surrounds it.  This physical situation may exist
  in all of the above and provides alternate explanations for any spectral line wings.  Figure \ref{fig:scenario5}.
  \begin{enumerate}
    \item Moderate density gas from a molecular cloud will result in high
      column but moderate density at the center velocity
    \item Wide wings of high density gas will exist both blue and redshifted of
      the highest-column point
  \end{enumerate}

\end{itemize}

\subsection{The Filling Factor of Molecular Clouds}
\label{sec:gmcdensity}
We have measured the density in 19 line-of-sight molecular clouds in addition
to the 18 measurements of densities around \uchii\ regions (we only include
measurements, not limits, in these counts).  The measured density from the
\formaldehyde\ line ratio can be compared to other measures of density, e.g.
the mean molecular cloud density measured by \citet{Roman-Duval2010} from the
BU-FCRAO GRS.  It is clear from Figure \ref{fig:denshist} that the average
density in GMCs is typically $\sim2-3$ orders of magnitude lower than densities
measured in our sample of line-of-sight GMCs.

\citet{Roman-Duval2010} point out that the mean densities they measure are
significantly below the critical density of \thirteenco,
$n_{cr}=2.7\ee{3}$ \percc, indicating that they do not resolve the high-density
clumps that make up the GMCs.  Our data indicate that a typical GMC consists of
$n\sim3\ee{4}$ \percc\ gas (the median of our GMC subsample excluding upper
limits), substantially higher than the critical density of \thirteenco.  Taking
the ratio of the median density in the \citet{Roman-Duval2010} catalog to that
in our sample, we derive a volume filling factor of $5\ee{-3}$ of dense gas in
molecular clouds.

\OneColFigure{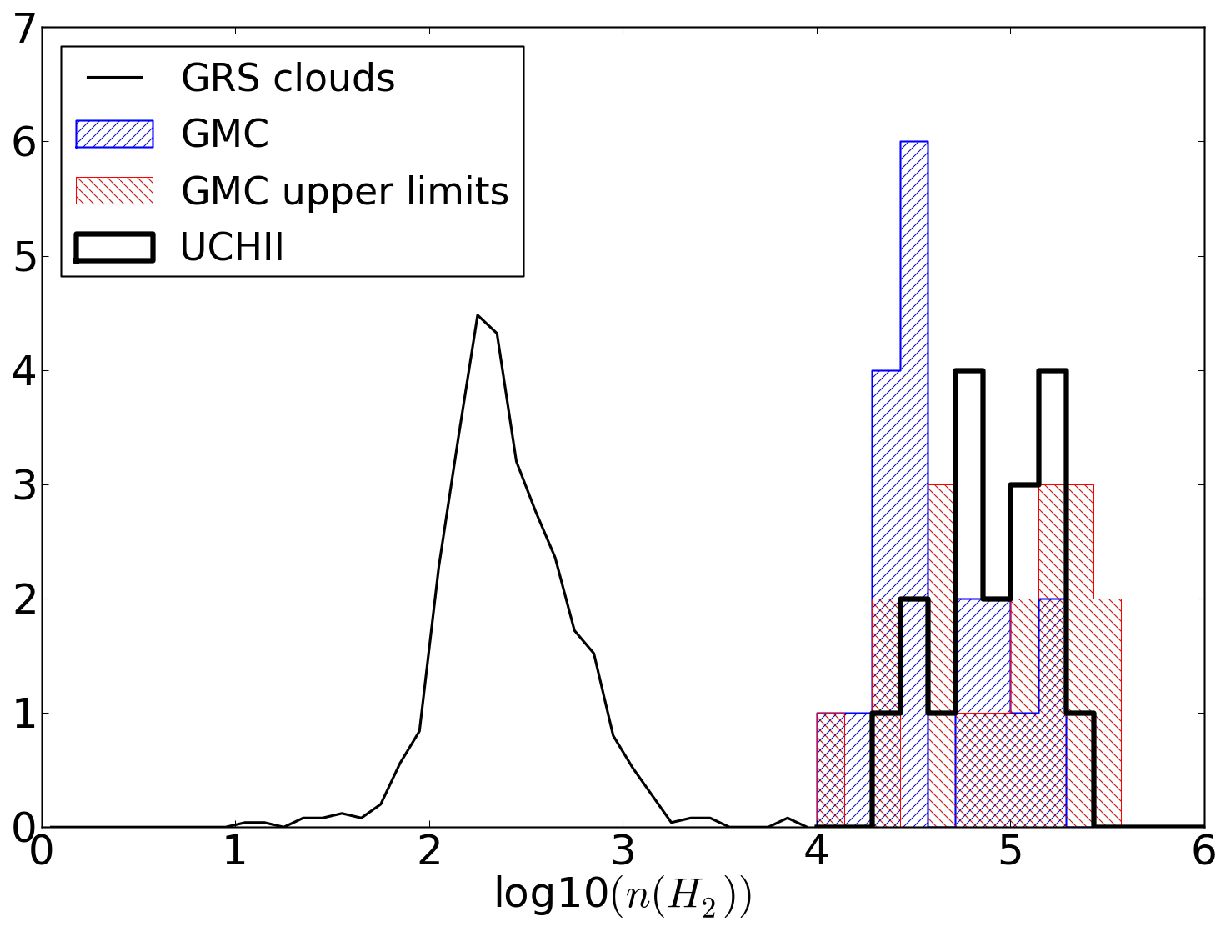}
{Histograms of the GMC and \uchii\ subsamples from our data plotted along with
the GMC-averaged densities from the \thirteenco\ \citet{Roman-Duval2010} GRS
measurements arbitrarily scaled to fit on this plot.  The measured densities in
\uchii\ regions are significantly (by a KS test) higher than densities in GMCs.
The \formaldehyde-measured densities in GMCs are 2-4 orders of magnitude higher
than volume-averaged densities of GMCs from the GRS, suggesting that GMCs
consist of very low volume-filling factor ($\sim5\ee{-3}$) high-density
($n(\hh)\sim3\ee{4}$ \percc) clumps. In Section \ref{sec:gmcdensity}, we argue
that the observed difference is most likely not a selection effect imposed by
the different gas tracers.  The
GMC upper limits shown are $3-\sigma$ upper limits, and all are consistent with
the measured GMC densities.
}
{fig:denshist}
{0.30}{0}

We measured an additional 20 upper limits towards GMCs, all of which are
consistent with high densities ($n(\hh)>10^4$ \percc), but could represent a sample
of lower density ($n(\hh)\sim10^3$ \percc) gas, in which case our `measurement' of
the cloud volume filling factor is biased to be too low.  In order to test for
this bias, we need to acquire more sensitive observations of the upper-limit
systems.  However, we continue analysis below based on the assumption that the
cloud filling factor measurement is realistic, i.e. assuming that the density upper
limit measurements have densities consistent with the other observed GMCs.

Can a medium with supersonic turbulence produce the same density measurements
without having to invoke high-density clumping?  Below about
$n(\hh)\approx10^5$ \percc, measurements of density in a turbulent medium are biased
towards higher densities, i.e. the densities we report may be overestimates for
GMCs since they have a median density $n(\hh)=10^{4.49}$ \percc.  For turbulent
density PDFs with logarithmic widths $\sigma_{ln(\rho)/ln(\bar{\rho})} \lesssim
1.5$, the overestimate is no more than 0.4 dex, and therefore can only bring
the filling factor up by a factor $<3$.  As discussed in Section
\ref{sec:turbulence} and Figure \ref{fig:turbcorr}, a high-density tail could
create a larger discrepancy ($\sim 0.5$ dex).  However, at the measured
densities, these are extreme upper limits on the `turbulent correction', and
therefore (gravo)turbulence alone cannot account for the measured densities.

What clumping properties are required to reproduce the observed density?  As
long as the clumps are all optically thin in the \formaldehyde\ absorption
lines, the spectral line optical depths and ratio are independent of clumping.  However,
a large number of low-density ($n(\hh)\approx10^{3.5}$ \percc) clumps optically
thick in the \oneone\ line and thin in the \twotwo\ line would appear to have a
higher density.  This phenomenon could only occur at densities
$\lesssim10^{4.5}$ \percc, where the \oneone\ absorption line is much stronger
than the \twotwo\ line, and column densities $10^{14} \persc \gtrsim
N(\ortho) \gtrsim 10^{13.5} $ \persc\ per clump (at higher columns, both
lines are optically thick; at lower columns, both lines are optically thin).
Assuming a typical \formaldehyde\ abundance (ortho+para) $X_{\formaldehyde}=10^{-9}$, the
required spherical clump radius would be $\sim0.3 $ pc, which would be
Jeans-unstable at the assumed density and temperature (40 K) and is therefore
unlikely to persist for long time periods \footnote{However, the lifetime of
such clumps in a turbulent medium in which small-scall turbulence supports the
clumps against collapse is unconstrained.}. We therefore regard a collection
of optically thick clumps in the \formaldehyde\ \oneone\ line to be 
unlikely; clumps optically thick in both lines are even less likely following
the same line of reasoning.

The combination of the observed large spatial scales (and therefore low
volume-averaged density) of GMCs and the high densities measured along
essentially arbitrary sightlines through these GMCs suggests that GMCs are not
consistent with a purely turbulent medium with a lognormal density
distribution.  The observations also require a more substantial high-density
tail than typically seen in gravoturbulent simulations, i.e. they require a
clumpier medium.

Alternatively, it is possible that \formaldehyde\ is chemically enriched in
high-density pockets within a turbulent medium, which would imply that
\formaldehyde\ observations probe different gas than CO.  No such mechanism has
been proposed on theoretical grounds, and the timescales for enhancement would
have to be very short \citep[intermittent density enhancement occurs on
timescales much shorter than the dynamical timescale; ][]{Kritsuk2007}, so we
regard this possibility as unlikely but include it for completeness.

Another alternative is that the \thirteenco\ systematically underestimates the
mass or overestimates the volume of the cloud, resulting in an underestimate of
the cloud density.  Sub-thermal excitation of \thirteenco\ in the low-density
parts of the cloud can lead to an underestimate of the mass \citep[][Section
9.3]{Roman-Duval2010}.  Since the cloud sizes were derived using an assumed
spherical symmetry, but molecular gas is typically observed in filamentary
structures, the densities in \citet{Roman-Duval2010} are likely to be lower
limits on the mean density in the molecular gas.  While both of these factors
bring the \formaldehyde\ and \thirteenco\ densities into closer agreement, it
is difficult to quantify these effects.

\subsection{Strengths and Weaknesses of the \formaldehyde\ K-doublet Densitometer}
\label{sec:strengthsweaknesses}
The dynamic range of a spectral line as a tracer of a physical quantity is an
important consideration when designing an experiment.  We have demonstrated
only a modest dynamic range in density measurements using the \oneone\ and
\twotwo\ lines in absorption against bright background sources: above
$n(\hh)\approx10^{5.6}$ \percc, we are only able to set lower limits on the
density because the spectral line ratio asymptotes to 1 , and below $n(\hh)\approx10^4$
\percc, the \twotwo\ line optical depth drops to very low levels (Figure
\ref{fig:modeltau}). 

The lower limits on density at $n(\hh)\gtrsim10^{5.6}$ \percc\ can only be modestly
improved upon by using higher K-doublet transitions (e.g.  \threethree) when
observing \formaldehyde\ in absorption against bright continuum sources.
However, when observing anomolous absorption against the CMB, an additional
density diagnostic is the transition from absorption to emission at higher
densities, which expands the sensitivity of the \oneone/\twotwo\ pair to
$n(\hh)\approx10^{6.5}$ \percc.

The low-density end can only be probed by more sensitive observations of
the \twotwo\ line.  Because the \formaldehyde\ line depths become negligible
below $n(\hh)\approx10^3$ \percc, the densitometer is not a useful probe below these
densities.  However, at such low densities, even within a molecular cloud,
it is questionable whether any molecular probes are reliable, as even CO will be
underabundant in these environments \citep[e.g., ][]{Glover2010}.

As noted in \citet{Mangum2008} and \citet{Zeiger2010}, the K-doublet
densitometer has the advantage that line detection only depends on the
brightness of the background source and the gas density.  It can therefore be
used nearly independent of distance when observing clouds against the CMB or
bright illuminating background sources.  The \citet{Zeiger2010} measurements
are more sensitive than any presented in our study because of longer
integration times and the selection of a bright illuminating background source
despite their target being at a distance z=0.68.  Following this line of
reasoning, any bright synchrotron or free-free source can be used to
sensitively probe the density of a line-of-sight molecular cloud in the Galaxy.
The observation will have angular resolution limited only by the size of the
background source as long as it is much brighter than the CMB in the beam.

\section{Conclusions}
\label{sec:conclusions}
We have presented a pilot study to measure molecular gas densities in clouds
along 24 lines of sight in the \formaldehyde\ \oneone\ and \twotwo\ transitions
primarily toward \uchii\ regions .  We have shown that the \formaldehyde\
densitometer is robust within reasonable ranges of turbulent density
distributions, most cloud geometries, and different cloud clumping properties.
We have presented the methodology and discussed the errors intrinsic to the
\formaldehyde\ densitometer.

Gas volume densities were measured toward 14 of the 24 sources using the
best-fit gaussian profiles; density limits were measured for the remaning 10.
In 18 sources, it was possible to estimate the density in each 0.4 \kms-wide
channel centered on the main line.  Of these, 12 showed some sign of a density
gradient with velocity, 5 appeared to have a single-valued density (i.e. only a
single spectral line component well-fit by a gaussian), and one, G69.54-0.98,
had a spectral line optical depth that was beyond our ability to model.  

Velocity-density gradients have been used to fit 18 sources with simple
models of \uchii\ regions embedded in molecular clouds.  We have found some  
examples consistent with inside-out collapse onto the \uchii\ region,
 \uchii\ region expansion, and bulk outflow.  \formaldehyde\
absorption provides a unique probe of the physical conditions around \uchii\
regions because it is only seen in absorption against the continuum background
(for sources much brighter than the CMB), giving different constraints than mm
and sub-mm spectral lines that are seen both in front of and behind the \uchii\ region.

The measurements of serendipitously detected line-of-sight GMCs revealed
densities $\sim200$ times higher than volume-average densities measured using
\thirteenco.  The high density measured suggests that GMCs consist of many
sparsely distributed high-density clumps and have density distributions
inconsistent with the lognormal distribution predicted by supersonic turbulent
models.  The implied density distribution is also more skewed to high densities
than predicted by typical gravoturbulent simulations.  Alternatively, the 
\thirteenco-based mean densities may be lower than the mean densities within
the molecular gas either because they underestimate the mass or overestimate
the volume of GMCs.

The density measurements show that UC\ion{H}{2}s are associated with
high-density ($n(\hh)>10^{4.5}$ \percc) gas, and UC\ion{H}{2}s are associated with
higher column and volume densities than other line-of-sight molecular clouds,
in contradiction with previous results \citep{Wadiak1988}.  

The 6 cm, 2 cm, and 1.1 mm flux density measurements are strongly correlated
and in most objects in our sample the 1.1 mm flux density has a substantial ($>30\%$) contribution 
free-free continuum emission.  This result implies that the
brightest sources detected in the BGPS have significant free-free emission.
A comparison to the 6 cm MAGPIS survey suggests that the sample of 1.1 mm
sources below about 3 Jy is not significantly contaminated by pure free-free
emission sources.

Comparison of the density measurements in our sample to the starburst sample of
\citet{Mangum2008} suggest that the molecular gas volume filling factor in most of
these galaxies is small ($\sim 0.01$), but in Arp 220 it is quite high
($\gtrsim 0.1$).  The physical properties measured by \formaldehyde\ in Arp 220
are similar to those in \uchii\ regions.  Although velocity-density gradients
are observed in our sample, we argue that kinematic spectral line blending
should uphold the assumption of a single spectral line profile in galaxies as
robust for radiative transfer purposes.

\section{Acknowledgements}
We thank Jim Braatz for assistance with data acquisition and processing,
Esteban Araya for providing us with reduced data, and our referee Jeff Mangum
for a helpful and timely review.  This work was supported by the National
Science Foundation through NSF grant AST-0708403 to John Bally and AST-0707713
to Jeremy Darling.  This research has made use of the SIMBAD database, operated
at CDS, Strasbourg, France.  This research made use of pyspeckit, an
open-source spectroscopic toolkit hosted at \url{http://pyspeckit.bitbucket.org}.

{\it Facilities:} \facility{GBT}, \facility{Arecibo}, \facility{VLA},
\facility{FCRAO}, \facility{CSO}

\Figure{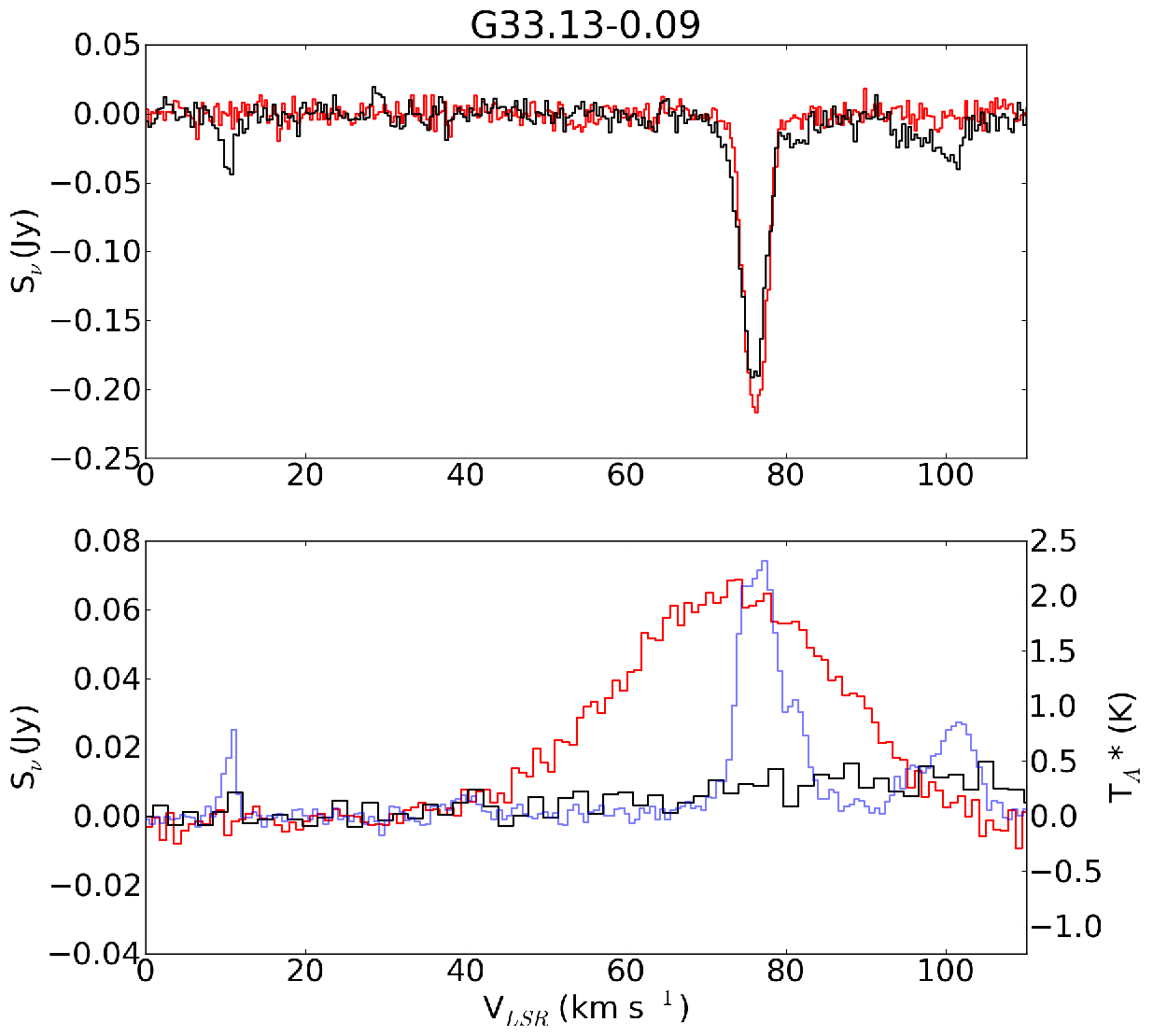}
{ {\it Top:}  The GBT \twotwo\ and Arecibo \oneone\ spectra of G33.13-0.09.
{\it Bottom:} The GBT H75$\alpha$ and Arecibo H110$\alpha$ spectra with the GRS
\thirteenco\ spectrum overlaid.   The left axis is for the RRLs and the right
axis is for the \thirteenco.  }
{fig:g33pt13spectrum}
{0.5}{0}

\OneColFigure{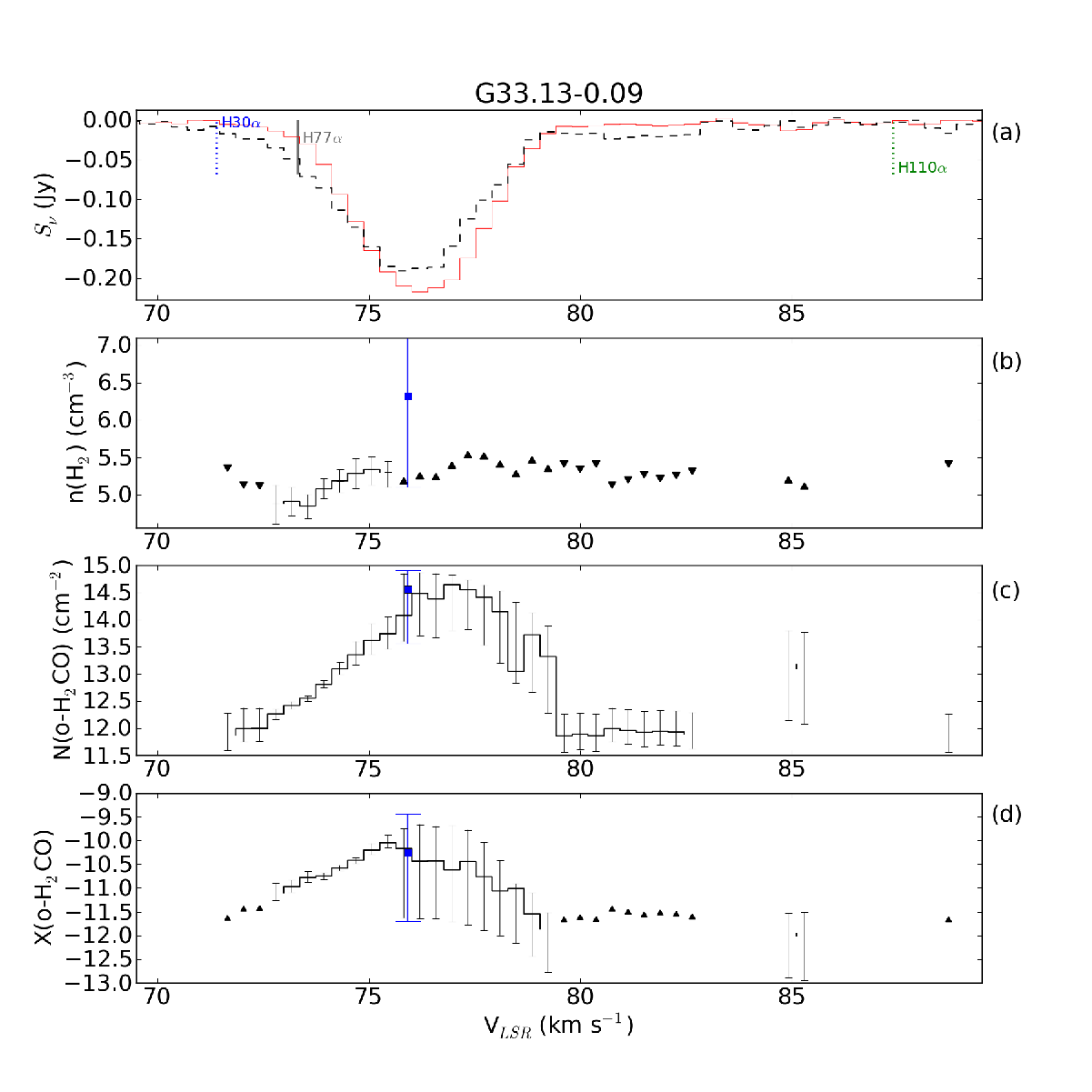}
{  Same panel description as Figure \ref{fig:g3280densspec}.
The recombination lines in this source show enormously different velocities.
H110$\alpha$ is $14.6 \pm 2.9 \kms$ redshifted from H75-77$\alpha$, which is a
significant fraction of the total line width ($\sigma_{110}=41.6\pm6.35,
\sigma_{75}=31.8\pm2.2$).  \citet{Churchwell2010} measure H30$\alpha$ slightly
bluer still, $v_{30} = 71.4 \pm 1.7 \kms, \sigma_{30} = 27.8 \pm 3.2 \kms$.
This discrepancy could be explained by a significant gradient in the density of the
ionized gas with velocity, i.e. low-density gas blowing out from the back side of the
\uchii\ region.
The tail of redshifted gas from 80-85 \kms\ is either rapidly-infalling, low-column
gas (Scenario 2) or unassociated line-of-sight cloud material (Scenario 5).  The main line has a number of
points in which only a lower limit on density could be measured, but because of
the rising lower limits out to $\sim 78 \kms$, the spectrum is classified as a
red gradient.
}  
{fig:g33.13-0.09spectrum}
{0.5}{0}

\FigureTwo{f27}{f28}
{{\it Left:}  Same description as Figure \ref{fig:specexample}. 
{\it Right:}  Panels described in Figure \ref{fig:g3280densspec}.
The red gradient and carbon recombination line at a velocity between the
hydrogen lines and the \formaldehyde\ density peak are consistent with the
inside-out collapse model, Scenario 2.  The larger systematic error bars
included for the main line resulted in a lower limit on density being measured.
In this spectrum, a gaussian line fit is not representative of the
line profile.
}
{fig:g33.92+0.11spectrum}
{1}

\FigureTwo{f29}{f30}
{{\it Left:}   Same description as Figure \ref{fig:specexample}.
 {\it Right:}  Panels described in Figure \ref{fig:g3280densspec}.
The red velocity gradient in this source is very steep and widely separated (by
$6-15 \kms$) from the RRL velocities, indicating that the dense gas is rapidly
approaching the \uchii\ region.  The line profile is exemplary of Scenario 2,
inside-out collapse, since the dense gas is redshifted of the recombination
lines and the C RRL is at an intermediate velocity.
The derived density at the peak is lower than previous measurements of density
using C$^{34}$S and the mm transitions of \formaldehyde\
\citep{Mangum1993Formaldehyde}, although both of those measurements are
consistent with the density in at least one velocity bin.   The measurements
are consistent despite the high temperature ($T_K=230 $K) measured by
\citep{Mangum1993Formaldehyde}.
}
{fig:g34.26+0.15spectrum}
{1}

\FigureTwo{f31}{f32}
{ {\it Left:}  Same description as Figure \ref{fig:specexample}. 
  {\it Right:}  For panel descriptions, see Figure \ref{fig:g3280densspec}.
  Identified as a single gaussian line.  The line peak could be optically thick
  based on morphological comparison to G69.54-0.98.  Since the carbon RRLs are aligned with
  the \formaldehyde\ column peak and both are blueshifted from the hydrogen RRLs, this spectrum
  is consistent with Scenario 4, in which a layer of gas is being pushed towards the observer
  and compressed.  }
{fig:g35.20-1.74spectrum}
{1}

\FigureTwo{f33}{f34}
{{\it Left:}  Same description as Figure \ref{fig:specexample}.  One of the two off
  positions contained an absorption line at 45 \kms; the position was coincident
  with an IRDC.
  {\it Right:}  For panel descriptions, see Figure \ref{fig:g3280densspec}.  The
  densities measured are very high ($n(\hh) > 10^6\percc$) and lower limits
  are all at similar levels.  The match between the \formaldehyde\ and RRL velocities
  and the slight blueshift of the lower frequency RRLs suggest that this source is
  consistent with Scenario 1, though the lack of a carbon RRL detection suggests that
  the \uchii\ and molecular gas are not interacting.
}
{fig:g35.57-0.03spectrum}
{1}

\OneColFigureTwo{f35}{f36}
{ {\it Left:}  Same description as Figure \ref{fig:specexample}.
  {\it Right:}  For panel descriptions, see Figure \ref{fig:g3280densspec}.
  Identified as a blue gradient because of the low density region centered
  around 53 \kms.  However, in this source, the low density region is also
  substantially lower column \perkmspc, so it may be more appropriate to 
  classify as an envelope.  Because the peak density is slightly redshifted
  from the RRL velocities, this source is consistent with Scenario 2, but the
  lack of carbon RRLs and the lower limits on density at many velocities makes the
  match imperfect.
  }
{fig:g35.58+0.07spectrum}
{1}

\FigureTwo{f37}{f38}
{ {\it Left:}  Same description as Figure \ref{fig:specexample}.
{\it Right:}  Plot of the derived parameters per velocity bin for G37.87-0.40.  This source
is an example with high column and density associated with the line core (55-63
\kms) and declining columns at the edge of the line core (50-55, 63-67 \kms)
with additional low density ($n<10^{4.8}$ \percc) components along the line of sight.
A modest increase in density towards bluer velocities is observed, but the
complexity of the line morphology suggests that this gradient may represent
unassociated line of sight clouds.  The relative velocities of the carbon and
hydrogen recombination lines are consistent with the very simple Scenario 1,
although the \formaldehyde\ line suggests a more complex situation.
}{fig:g37spectrum}
{1}

\FigureTwo{f39}{f40}
{ {\it Left:}  Same description as Figure \ref{fig:specexample} 
  {\it Right:}  Panels described in Figure \ref{fig:g3280densspec}.  The S/N is only
  adequate to place upper limits on density at most velocities.   Because of the low
  S/N, we do not attempt to associate G41.74 with any of the physical scenarios from section
  \ref{sec:scenarios}.}
{fig:g41.74+0.10spectrum}
{1}

\FigureTwo{f41}{f42}
{{\it Left:}  Same description as Figure \ref{fig:specexample} 
 {\it Right:}  Plot of the derived parameters per velocity bin for G43.89-0.78.  
For panel descriptions, see Figure \ref{fig:g3280densspec}.
 The line
core (55 \kms) is very deep in the \oneone\ line but the \oneone/\twotwo\ ratio,
and therefore the spatial density, is substantially higher in the blue tail
(48-52 \kms).  Since the column densities are lower in the blue tail, it
probably represents a relatively small patch of highly turbulent, high-density
material as in Scenario 5.  It is also possible that the blue tail represents an 
outflow as in Scenario 3, but the high density measured is not expected in an outflow.
It is possible that there is a signficant contribution to the 55 \kms\
\oneone\ line from surrounding low-density gas in the beam absorbing against the CMB, 
creating a Type 3 error (Table \ref{tab:systematics}).
}
{fig:g43.89-0.78spectrum}
{1}

\FigureTwo{f43}{f44}
{{\it Left:}  Same panel description as Figure \ref{fig:specexample}.
The H75$\alpha$ line cannot be fit by a single gaussian.
{\it Right:} For panel descriptions, see Figure \ref{fig:g3280densspec}.
The velocity difference between the RRLs and the \formaldehyde\ lines
is consistent with Scenario 2, but no carbon RRL is observed so there
is no direct evidence of interaction between the \uchii\ region and the
molecular gas.
}
{fig:g45.07+0.13spectrum}
{1}

\FigureTwo{f45}{f46}
{ {\it Left:} Same description as Figure \ref{fig:specexample} 
{\it Right:} For panel descriptions, see Figure \ref{fig:g3280densspec}.  The
line is classified as {\it other} because of the very wide, flat line.  The relative velocities
of the recombination lines and their alignment with the \formaldehyde\ line are
consistent with Scenario 1.}
{fig:g45.12+0.13spectrum}
{1}

\FigureTwo{f47}{f48}
{ {\it Left:} Same description as Figure \ref{fig:specexample} 
  {\it Right:} For panel descriptions, see Figure \ref{fig:g3280densspec}.  Identified
  as a blue gradient because density clearly increases towards the blue, while only
  upper limits are measured towards the red.  Based on the relative locations of
  the RRLs and the peak \formaldehyde\ column, the spectrum appears to be consistent with
  Scenario 2.}
{fig:g45.45+0.06spectrum}
{1}

\FigureTwo{f49}{f50}
{ {\it Left:} Same description as Figure \ref{fig:specexample} 
  {\it Right:} Panels described in Figure \ref{fig:g3280densspec}. Identified as a red
  gradient.  The hydrogen recombination line velocities are consistent with Scenario 1.  However, 
  the highest density gas is the most redshifted, which is more consistent with Scenario 2.  
  Additionally, there is a low-density blueshifted tail that is suggestive of an outflow 
  as in Scenario 3. }
{fig:g45.47+0.05spectrum}
{1}

\FigureTwo{f51}{f52}
{ {\it Left:} Same description as Figure \ref{fig:specexample} 
  {\it Right:} Panels described in Figure \ref{fig:g3280densspec}.
  Identified as a red gradient.  The RRL locations and density profile are consistent
  with Scenario 2, but the modest S/N leaves open the possibility of other explanations.}
{fig:g48.61+0.02spectrum}
{1}

\FloatBarrier
\OneColFigure{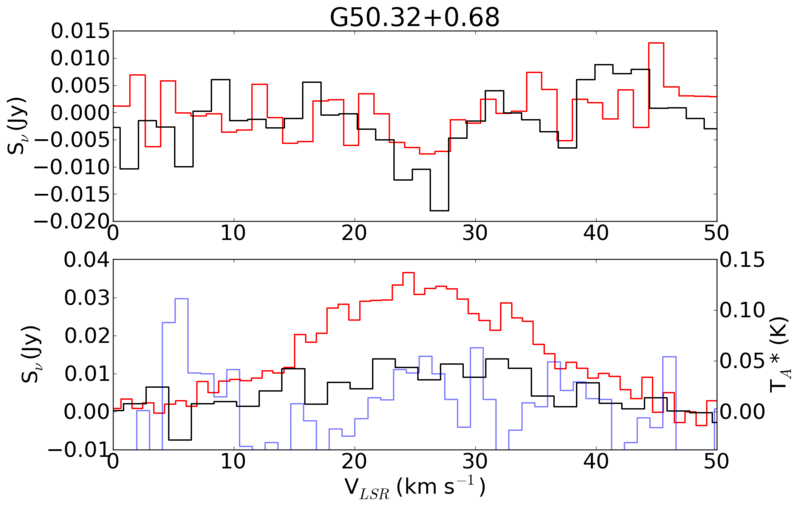}
{ Same description as Figure \ref{fig:specexample}.  There were not enough adjacent high S/N
velocity bins to perform the velocity-density analysis.  However, the velocity
coincidence of the detected H RRLs and \formaldehyde\ lines suggests that this
source is an example of Scenario 1, but no carbon RRL is detected.}
{fig:g50.32+0.68spectrum}
{0.25}{0}

\FigureTwo{f54}{f55}
{ {\it Left:} Same description as Figure \ref{fig:specexample} 
  {\it Right:} Panels described in Figure \ref{fig:g3280densspec}.  It was only 
  possible to derive upper limits on density at most velocities.  
  The locations of the RRLs and \formaldehyde\ lines are consistent with Scenario 2.}
{fig:g60.88-0.13spectrum}
{1}

\FigureTwo{f56}{f57}
{ {\it Left:} Same description as Figure \ref{fig:specexample} 
  {\it Right:} Panels described in Figure \ref{fig:g3280densspec}.
  Identified as a single gaussian morphology because of the peak in 
  density and column.  Because both the \formaldehyde\ and carbon RRLs are 
  blueshifted from the hydrogen RRLs, this source is an excellent example of 
  Scenario 4.}
{fig:g61.48+0.09spectrum}
{1}

\FigureTwo{f58}{f59}
{ {\it Left:} Same description as Figure \ref{fig:specexample} 
  {\it Right:} Panels described in Figure \ref{fig:g3280densspec}.  
  The flat-bottomed line profile suggests that the absorption in this source is
  optically thick.
  The 
  blank region around 8 \kms\ is where there is no allowed parameter space
  in the models.  The rest of the measurements are likely to be incorrect
  because they are biased by optically thick components contributing to
  the line.
  Because of the added uncertainty in the density measurement from the high
  line optical depth, it is not clear what scenario, if any, this source's 
  line profile is consistent with.  However, the RRL at 16.4 \kms\ (off the right
  side of the spectrum) would be consistent with Scenarios 4 and 5.
  }
{fig:g69.54-0.98spectrum}
{1}

\FigureTwo{f60}{f61}
{{\it Left:}  Same description as Figure \ref{fig:specexample} 
 {\it Right:}  Plot of the derived parameters per velocity bin for G70.29+1.60.  
For panel descriptions, see Figure \ref{fig:g3280densspec}.  The velocity
offset between the \oneone\ and \twotwo\ line centers results in a strong blue
density gradient.  Since the density peaks at the same velocity as the \uchii\
region, but there is high column gas redshifted of the density peak, the
spectrum is not directly consistent with Scenario 2, to which it is otherwise
similar.    
}
{fig:g70.29+1.60spectrum}
{1}

\FigureTwo{f62}{f63}
{{\it Left:}  Same description as Figure \ref{fig:specexample} 
 {\it Right:}  Plot of the derived parameters per velocity bin for G70.33+1.59. 
For panel descriptions, see Figure
\ref{fig:g3280densspec}.
Unlike G70.29+1.60, there is no offset between the \oneone\ and \twotwo\
line centers, so this source is consistent with a single gas component
along the line of sight.  Because carbon recombination lines are detected,
though, it is evident that the \uchii\ region is interacting with the molecular gas.
The recombination line velocities are consistent with Scenario 1, but the lack of
a gradient in the \formaldehyde-derived density is not.  The region could be intermediate
between Scenario 2 (infall) and Scenario 1 (expansion), causing the observed density to be
more evenly spread in velocity even though it varies spatially.
}
{fig:g70.33+1.59spectrum}
{1}

\FigureTwo{f64}{f65}
{ {\it Left:}   Same description as Figure \ref{fig:specexample} 
  {\it Right:}  Panels described in Figure \ref{fig:g3280densspec}.  
  There is no continuum emission or RRL detected towards this source: the absorption is
exclusively against the CMB.  \citet{Urquhart2009} report a HC\ion{H}{2}
detection at 0.23 mJy at 6cm, which is substantially lower than the $\sim20$
mJy absorption detected in the \oneone\ line.  The presence of a faint
HC\ion{H}{2} region and water masers \citep{Sunada2007} suggest that this is an
active but very young region, probably the youngest in our sample.  It is
therefore a good example of what can be expected of a survey selected from
millimeter sources instead of \uchii\ regions.
Because of the lack of an RRL detection and the low S/N, we cannot classify this source in an
\uchii\ scenario.
}
{fig:IRAS_20051+3435spectrum}
{1}

\OneColFigure{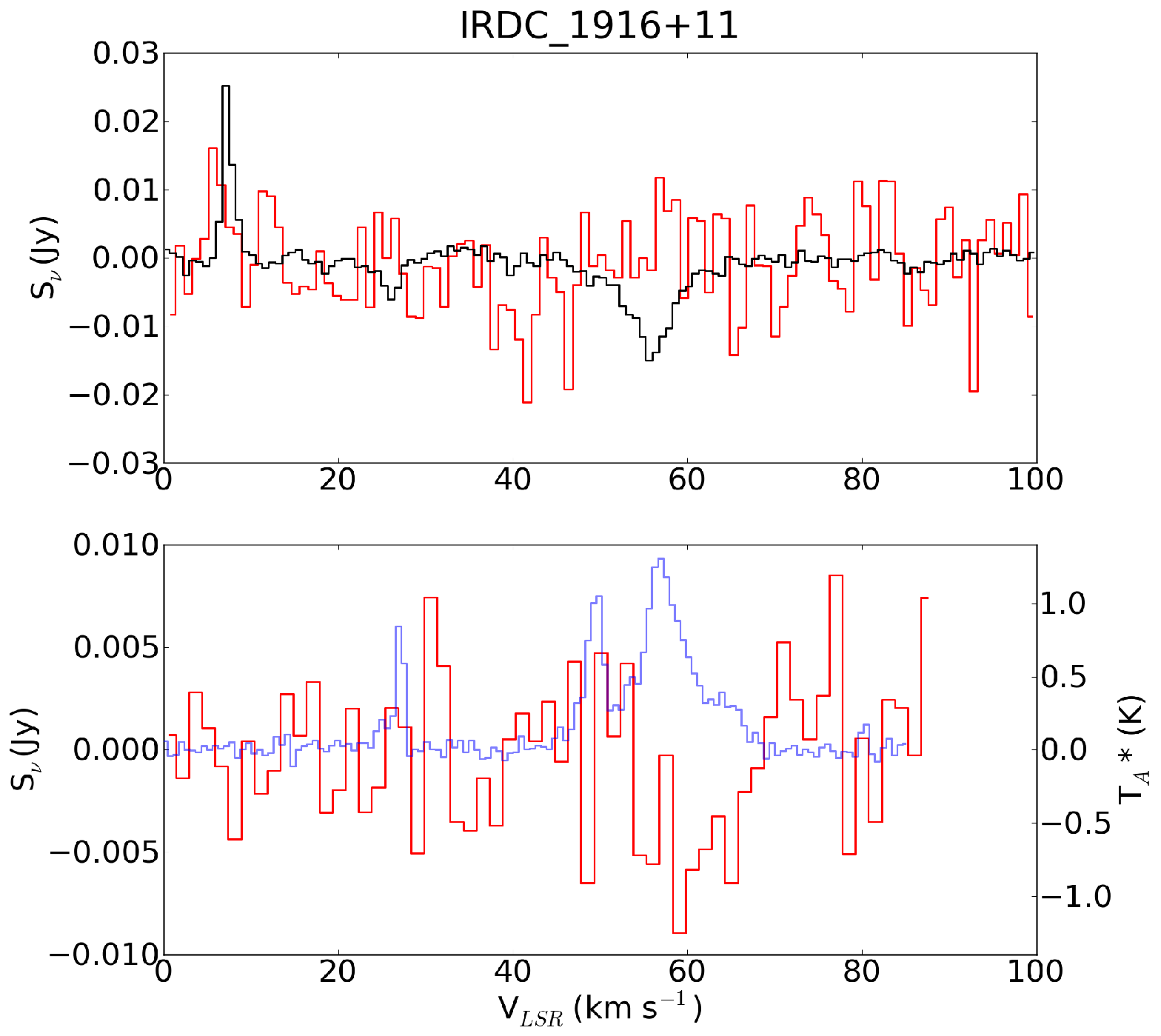}
{ Same description as Figure \ref{fig:specexample}.  The \twotwo\ noise levels
are too great to measure physically interesting limits on the density.
No continuum emission is detected at 2 cm, 6 cm, or 1.1 mm.  
There is also no evidence of an IRDC in the Spitzer 8\um\ image.  This
sightline is probably more accurately characterized as a lack of emission region
rather than a dust-absorbed IRDC.  The three \oneone\ absorption features detected
are associated with \thirteenco\ clouds and only have density upper limits from
our data.  These observations are a strong indication that deep enough
\formaldehyde\ observations in the Galactic plane will always detect
\formaldehyde\ against the CMB or diffuse galactic background, and therefore it
can be used to measure (or constrain) densities of even diffuse GMCs.
}
{fig:IRDC_1916+11spectrum}
{0.25}{0}

\OneColFigure{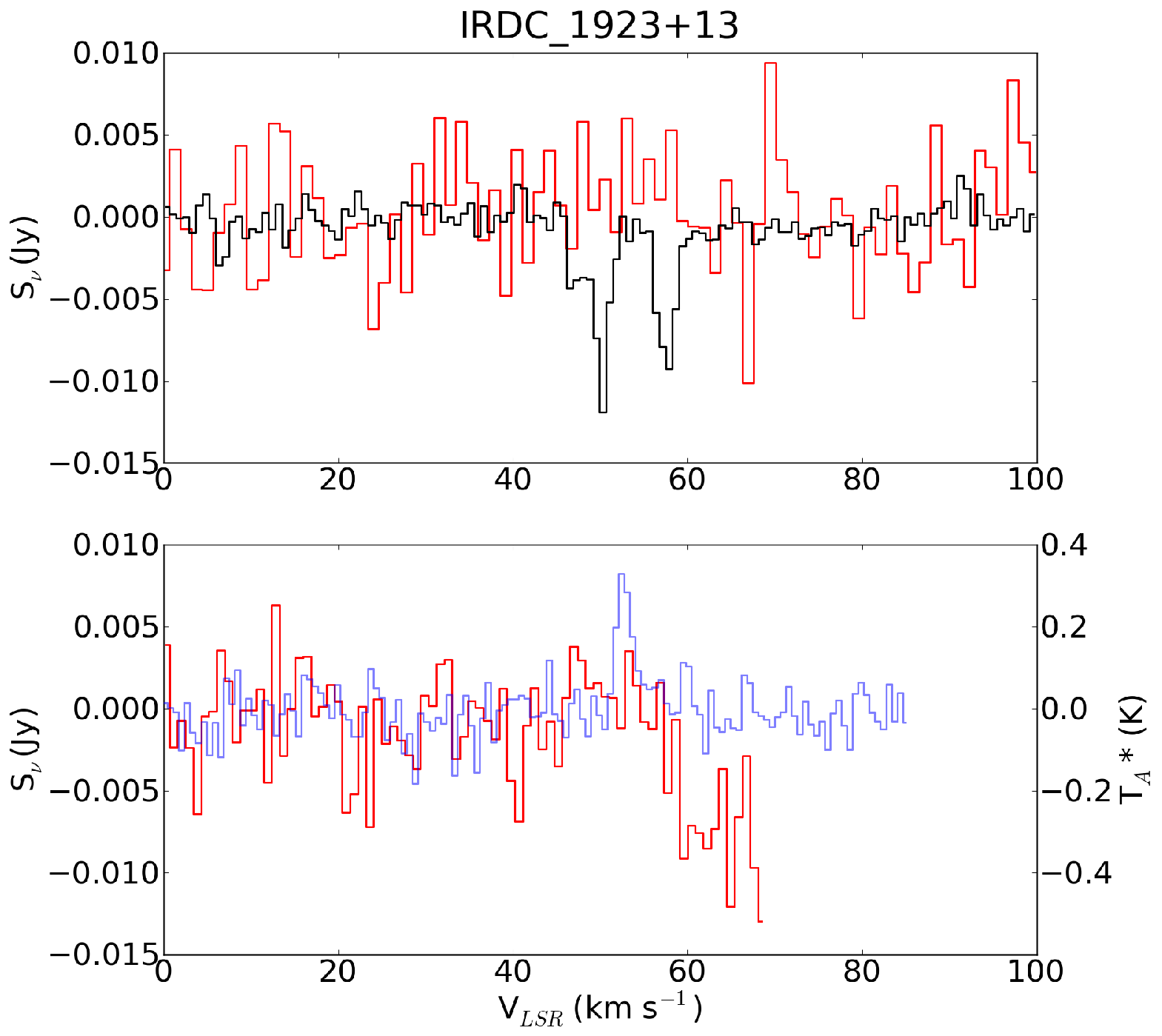}
{ Same description as Figure \ref{fig:specexample}.  The \twotwo\ noise levels
are too great to measure physically interesting limits on the density.}
{fig:IRDC_1923+13spectrum}
{0.25}{0}

\clearpage
\end{document}